\documentclass[11pt,a4paper,UKenglish,section]{article}
\usepackage{jheppub}
\usepackage[biblatex=false,hyperref=false]{atlaspackage}

\usepackage{atlasphysics} 

\usepackage{graphicx} 

\hypersetup{pdftitle={ATLAS draft},pdfauthor={The ATLAS Collaboration}}
\usepackage{preprintcover}

\usepackage{url}
\usepackage{multirow}
\usepackage{subfig}

\PreprintCoverPaperTitle{Analysis of events with $b$-jets and a pair of leptons of the same charge  in $pp$ collisions at $\sqrt{s}=8$~TeV  with the ATLAS detector}

\PreprintIdNumber{CERN-PH-EP-2015-060}

\PreprintJournalName{JHEP}

\PreprintCoverAbstract{
  An analysis is presented of events containing jets including at least one $b$-tagged jet, sizeable missing transverse momentum, and at least two leptons including a pair of the same electric charge, with the scalar sum of the jet and lepton transverse momenta being large.  A data sample with an integrated luminosity of 20.3~\ifb{} of $pp$ collisions at $\sqrt{s} =8$~\TeV{} recorded by the ATLAS detector at the Large Hadron Collider is used.  Standard Model processes rarely produce these final states, but there are several models of physics beyond the Standard Model that predict an enhanced rate of production of such events; the ones considered here are production of vector-like quarks, enhanced four-top-quark production, pair production of chiral $b^\prime$-quarks,  and  production of two positively charged top quarks. 
Eleven signal regions are defined; subsets of these regions are combined when searching for each class of models.  In the three signal regions primarily sensitive to positively charged top quark pair production, the data yield is consistent with the background expectation.  There are more data events than expected from background in the set of eight signal regions defined for searching for vector-like quarks and chiral $b^\prime$-quarks, but the significance of the discrepancy is less than two standard deviations.  The discrepancy reaches  2.5 standard deviations in the set of five signal regions defined for searching for four-top-quark production.  The results  are used to set 95\% CL limits on various models. }

\begin{document}

\title{Analysis of events with $b$-jets and a pair of leptons of the same charge  in $pp$ collisions at $\sqrt{s}=8$~TeV  with the ATLAS detector}

\author{The ATLAS Collaboration}


\abstract{
 	  An analysis is presented of events containing jets including at least one $b$-tagged jet, sizeable missing transverse momentum, and at least two leptons including a pair of the same electric charge, with the scalar sum of the jet and lepton transverse momenta being large.  A data sample with an integrated luminosity of 20.3~\ifb{} of $pp$ collisions at $\sqrt{s} =8$~\TeV{} recorded by the ATLAS detector at the Large Hadron Collider is used.  Standard Model processes rarely produce these final states, but there are several models of physics beyond the Standard Model that predict an enhanced rate of production of such events; the ones considered here are production of vector-like quarks, enhanced four-top-quark production, pair production of chiral $b^\prime$-quarks,  and  production of two positively charged top quarks. 
Eleven signal regions are defined; subsets of these regions are combined when searching for each class of models.  In the three signal regions primarily sensitive to positively charged top quark pair production, the data yield is consistent with the background expectation.  There are more data events than expected from background in the set of eight signal regions defined for searching for vector-like quarks and chiral $b^\prime$-quarks, but the significance of the discrepancy is less than two standard deviations.  The discrepancy reaches  2.5 standard deviations in the set of five signal regions defined for searching for four-top-quark production.  The results  are used to set 95\% CL limits on various models. 
 }

\maketitle

\section{Introduction}

The Standard Model (SM) has been repeatedly confirmed experimentally.
Nonetheless there is a need for physics beyond the SM (BSM) at about the \TeV{} scale, with additional features that explain the baryon asymmetry of the universe, specify the nature of dark matter, and provide a mechanism to naturally stabilize the Higgs boson mass at its observed value of approximately 125~\GeV{}~\cite{Aad:2012tfa,Chatrchyan:2012ufa}.   This paper reports on a search for BSM physics resulting in pairs of isolated high-transverse-momentum (high-\pt) leptons\footnote{Only electrons and muons are considered in the search.  Tau leptons are not explicitly reconstructed, but electrons and muons from $\tau$ decay may enter the selected samples.} with the same electric charge, hereafter denoted as same-sign leptons, (or three or more leptons of any charge) missing transverse momentum, and $b$-jets.  This is a promising search channel since the SM yields of such events are small, and several types of BSM physics  may contribute.

Among the models that predict enhanced same-sign lepton production are those that postulate the existence of vector-like quarks, an enhancement of the four-top-quark production cross section, the existence of a fourth generation of chiral quarks, or production of two positively charged top quarks.  A common data sample is used to search for each of these signatures, but separate final selection criteria are defined based on the characteristics of each signal model.  

Several extensions to the SM that regulate the Higgs boson mass in a natural way require the existence of vector-like quarks (VLQ)~\cite{Dobrescu:1997nm,Chivukula:1998wd,He:2001fz,Hill:2002ap,Contino:2006qr,Anastasiou:2009rv,Kong:2011aa,Carmona:2012jk,Gillioz:2012se,PhysRevLett.60.1813,PhysRevD.41.1286,Grinstein:2010ve,Guadagnoli:2011id,ArkaniHamed:2002qy,Han:2003wu,Perelstein:2003wd,Schmaltz:2005ky,Carena:2006jx,Matsumoto:2008fq}, where `vector-like' means that the left- and right-handed components transform identically under the SU(2)$_{\mathrm{L}}$ weak isospin gauge symmetry.  Since quarks with this structure do not require a Yukawa coupling to the Higgs field to attain mass, their existence would not enhance the Higgs boson production cross section, and thus the motivation persists for a direct search~\cite{Aguilar-Saavedra:2013qpa}.  There are several possible varieties of VLQ; those having the same electric charge as the SM $b$- and $t$-quarks are called $B$ and $T$.  In addition the exotic charge states $T_{5/3}$ and $B_{-4/3}$ may occur, where the subscripts indicate the electric charge.
 Vector-like quarks may exist as  isospin singlets, doublets, or triplets.  Arguments based on naturalness suggest that VLQ may not interact strongly with light SM quarks~\cite{AguilarSaavedra:2002kr,Okada:2012gy}.  Thus it is assumed for this analysis that VLQ decay predominantly to third-generation SM quarks.  For the $B$ and $T$ quarks, charged- and neutral-current decays may both occur ($B \ra Wt, Zb, \hbox{ or } Hb$; $T \ra Wb, Zt, \hbox{ or } Ht$), providing many paths for same-sign lepton production for events with $B\bar{B}$ or $T\bar{T}$ pairs.  

   The branching fractions to each allowed final state are model-dependent, and the ones occurring in models where the $B$ and $T$ exist as singlets or as a $(T,B)$ doublet~\cite{AguilarSaavedra:2009es} are used as a reference.  These branching fractions vary with the $B$ or $T$ mass, and  values for some masses are given in table~\ref{tab:VLQBR}. 
Since the pair production of heavy quarks is mediated by the strong interaction, the cross section is identical for vector-like quarks and $b^\prime$ quarks (described below) of a given mass.  The next-to-next-to-leading-order (NNLO) cross sections from {\sc top++}~v2.0~\cite{Czakon:2013goa,Czakon:2011xx} are used in this paper.
The $T_{5/3}$ quark must decay to $W^+t$, and therefore both single and pair production of this quark can result in same-sign lepton pairs, and both sources are considered.
 
\begin{table}[tt]
  \begin{center}
    \begin{tabular}{l|ccc|ccc}
      \hline\hline
      & \multicolumn{3}{c}{$B$} &\multicolumn{3}{|c}{$T$}  \\
      \hline
	 Mass (model)         &  $Wt$ & $Zb$ & $Hb$ & $Wb$ & $Zt$ & $Ht$ \\ \hline
0.50 \TeV{} (singlet) & 42 & 31 & 27 & 50 & 17 & 33 \\
0.50 \TeV{} (doublet) & 100 & 0 & 0 & 0 & 34 & 66 \\
0.55 \TeV{} (singlet) & 43 & 30 & 27 & 49 & 18 & 32 \\	
0.55 \TeV{} (doublet) & 100 & 0 & 0 & 0 & 37 & 63 \\
0.60 \TeV{} (singlet) & 44 & 29 & 26 &  49 & 19 & 31\\
0.60 \TeV{} (doublet) & 100 & 0 & 0 & 0 & 38 & 62 \\
0.65 \TeV{} (singlet) & 45 & 29 & 26 & 49 & 20 & 30 \\        
0.65 \TeV{} (doublet) & 100 & 0 & 0 & 0 & 40 & 60 \\ \hline
    \end{tabular}
  \end{center}
    \caption{$B$ and $T$ quark branching fractions (in percent), assuming the singlet and $(T,B)$ doublet models of ref.~\cite{AguilarSaavedra:2009es}. In the doublet case it is assumed that the mixing of the $T$ quark with the Standard Model bottom quark is much smaller than the mixing of the $B$ quark with the top quark. \label{tab:VLQBR}}
\end{table}
  
  Same-sign lepton pairs may also arise from the production of four top quarks ($t\bar{t}t\bar{t}$).  The SM rate for this production is small ($\approx 1$ fb~\cite{PhysRevD.44.1987,Barger:2010uw}), but there are several BSM physics models that can enhance the rate, such as top compositeness models~\cite{1126-6708-2008-04-087,PhysRevD.78.074026,Degrande:2010kt}
or Randall-Sundrum models with SM fields in the bulk~\cite{Guchait:2007jd}.  These can generically be described in terms of a four-fermion contact interaction with coupling strength $C_{4t}/\Lambda^2$, where $C_{4t}$ is the coupling constant and $\Lambda $ is the scale of the BSM physics~\cite{Degrande:2010kt}.  The Lagrangian for this interaction is
  
  \begin{equation}
\label{eq:lagrangian4tops}
{\cal L}_{4t} = \frac{C_{4t}}{\Lambda^2} \left(\bar{t}_{\rm R}
\gamma^\mu t_{\rm R}\right)\left(\bar{t}_{\rm R} \gamma_\mu t_{\rm R}\right)
\end{equation}
 where $t_R$ is the right handed top spinor and the $\gamma_\mu$ are the Dirac matrices.  Two specific models are also considered.  The first is  sgluon pair production, where sgluons are colour-adjoint scalars that appear in several extensions to the SM~\cite{Plehn:2008ae,Choi:2008ub,Kilic:2009mi,Kilic:2008pm,Burdman:2006gy,Calvet:2012rk}.   
 If the sgluon mass is above the top quark pair-production threshold, the dominant decay is to $t\bar{t}$, resulting in four top quarks in the final state\footnote{The decays predominantly to $t\bar{t}$ is model-dependent, as discussed in ref.~\cite{Beck:2015cga}.} ($t\bar{t}t\bar{t}$). The cross sections considered in this paper are rescaled to the next-to-leading order (NLO) prediction of Ref.~\cite{GoncalvesNetto:2012nt}. The second model is one with two universal extra dimensions under the real projective plane geometry (2UED/RPP)~\cite{Lyon09}.  The compactification of the extra dimensions leads to discretization of the momenta along their directions. The model is parameterized by the radii $R_4$ and $R_5$ of the extra dimensions or, equivalently, by $m_{\mathrm{KK}} = 1/R_4$ and $\xi = R_4/R_5$. This model predicts the pair production of tier\footnote{A tier of the Kaluza--Klein towers is labelled by two integers, corresponding to the two extra dimensions.} $(1,1)$ Kaluza--Klein excitations of the photon ($A_\mu^{(1,1)}$) with mass approximately $\sqrt{2}m_{\mathrm{KK}}$ that decay to $t\bar{t}$ with a branching fraction assumed to be 100\%. The model also predicts a four-top-quark signal from tiers $(2,0)$ and $(0,2)$. Cosmological observations constrain $m_{\mathrm{KK}}$ in this model to lie approximately between 600~\GeV{} and 1200~\GeV{}~\cite{Lyon12}.
  
 A fourth generation of SM-like quarks includes a charge $-1/3$ quark, called the $b^\prime$~\cite{Frampton:1999xi,Holdom:2006mr,Hung:1997zj,Hou:2008xd}.  Under the assumption that the $b^\prime$-quark decays predominantly to $Wt$, $b^\prime$ pair production 
 results in four $W$ bosons in the final state. If two $W$ bosons with the same electric charge decay leptonically, there will be a same-sign lepton pair in the final state.  If the $b^\prime$-quark can also decay to $Wq$, where $q$ is a light ($u$ or $c$) quark, some $b^\prime$ pairs would also result in same-sign lepton pairs or trileptons (provided that at least one $b^\prime$-quark decays to $Wt$), and therefore the possibility of such decays is explored as well. The existence of additional chiral quark generations  greatly enhances the Higgs boson production cross section, so if the new boson observed at the LHC is a manifestation of a minimal Higgs sector, additional quark generations are ruled out~\cite{Eberhardt:2012gv,Djouadi:2012ae,Eberhardt:2012sb,Eberhardt:2012ck,Denner:2011vt,Passarino:2011kv}. However, a more complex Higgs sector, as in some Two-Higgs-Doublet models~\cite{BarShalom:2012ms}, allows a fourth generation of chiral quarks.

  Production of two positively charged top quarks via $uu \rightarrow tt$ can also result in an excess of same-sign lepton pairs.  This process may be mediated via $s$- or $t$-channel exchange of a heavy particle~\cite{AguilarSaavedra:2011zy,Berger:2011ua}. In the $t$-channel exchange case, the process must include a vertex with a flavour-changing neutral current (FCNC). The neutral particle that is exchanged may be a vector, $Z$-like, particle or a scalar, Higgs-like, particle. Past searches for a new $Z^\prime$ boson have already put strong constraints on this possibility, thus only the scalar case is considered, with the following generic model Lagrangian~\cite{TFCNC}:
  \begin{equation}
  \label{eq:tt_FCNC}
    {\cal L}_{\mathrm {FCNC}} = \kappa_{utH} \bar{t}Hu + \kappa_{ctH} \bar{t}Hc +\mathrm{h.c.}
  \end{equation}
where $H$ is a Higgs-like particle with mass $m_H$ and $\kappa_{utH}$ and $\kappa_{ctH}$ denote the flavour-changing couplings of $H$ to up-type quarks. Two scenarios are tested, one corresponding to a possible FCNC coupling of the newly discovered Higgs boson ($m_H = 125$ \GeV{}) and the other to a second scalar boson with a mass in the range $[250,750]$ \GeV{}.
If the mass of the mediating particle is much greater than the electroweak symmetry breaking scale, an effective four-fermion contact interaction can describe the process, thus extending the
search to non-scalar particles. The corresponding Lagrangian contains separate operators for the different 
initial-state chiralities~\cite{Aad:2012bb}:
\begin{equation}
  \label{eq:lagrangiantt}
\renewcommand{\arraystretch}{1.3}
\begin{array}{lllll}
{\cal L}_{tt}&=&\frac{1}{2}\frac{C_{\mathrm{LL}}}{\Lambda^2}(\bar{u}_{\mathrm{L}}\gamma^\mu t_{\mathrm{L}})(\bar{u}_{\mathrm{L}}\gamma_\mu t_{\mathrm{L}})
&+&\frac{1}{2}\frac{C_{\mathrm{RR}}}{\Lambda^2}(\bar{u}_{\mathrm{R}}\gamma^\mu t_{\mathrm{R}})(\bar{u}_{\mathrm{R}}\gamma_\mu t_{\mathrm{R}})\\
&-&\frac{1}{2}\frac{C_{\mathrm{LR}}}{\Lambda^2}(\bar{u}_{\mathrm{L}}\gamma^\mu t_{\mathrm{L}})(\bar{u}_{\mathrm{R}}\gamma_\mu t_{\mathrm{R}})
&-&\frac{1}{2}\frac{C^{\prime}_{\mathrm{LR}}}{\Lambda^2}(\bar{u}_{{\mathrm{L}}a}\gamma^\mu t_{{\mathrm{L}}b})(\bar{u}_{{\mathrm{R}}b}\gamma_\mu t_{{\mathrm{R}}a})+\mathrm{h.c.}\\
\end{array}\end{equation}
where $C_{\mathrm{LL}}$, $C^{(')}_{\mathrm{LR}}$ and $C_{\mathrm{RR}}$ are the coefficients of effective operators corresponding to each chirality configuration and $\Lambda$ is the scale of the BSM physics.
The $C_{\mathrm{LR}}$ and $C^\prime_{\mathrm{LR}}$ terms lead to kinematically equivalent events, hence only one term is considered in this paper.

\begin{figure}
  \begin{center}
    \subfloat[]{\includegraphics[width=.28\textwidth]{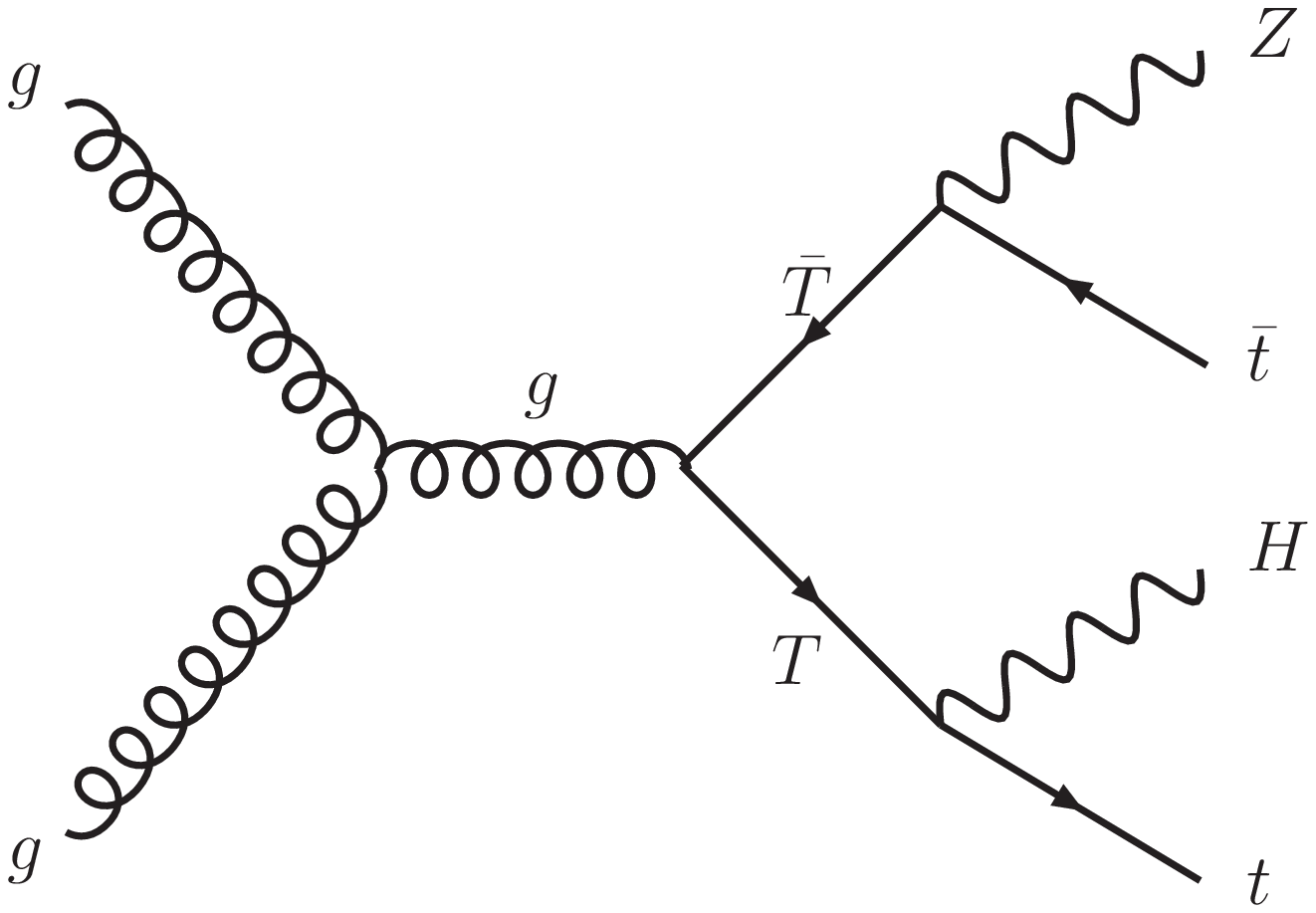}}
    \subfloat[]{\includegraphics[width=.28\textwidth]{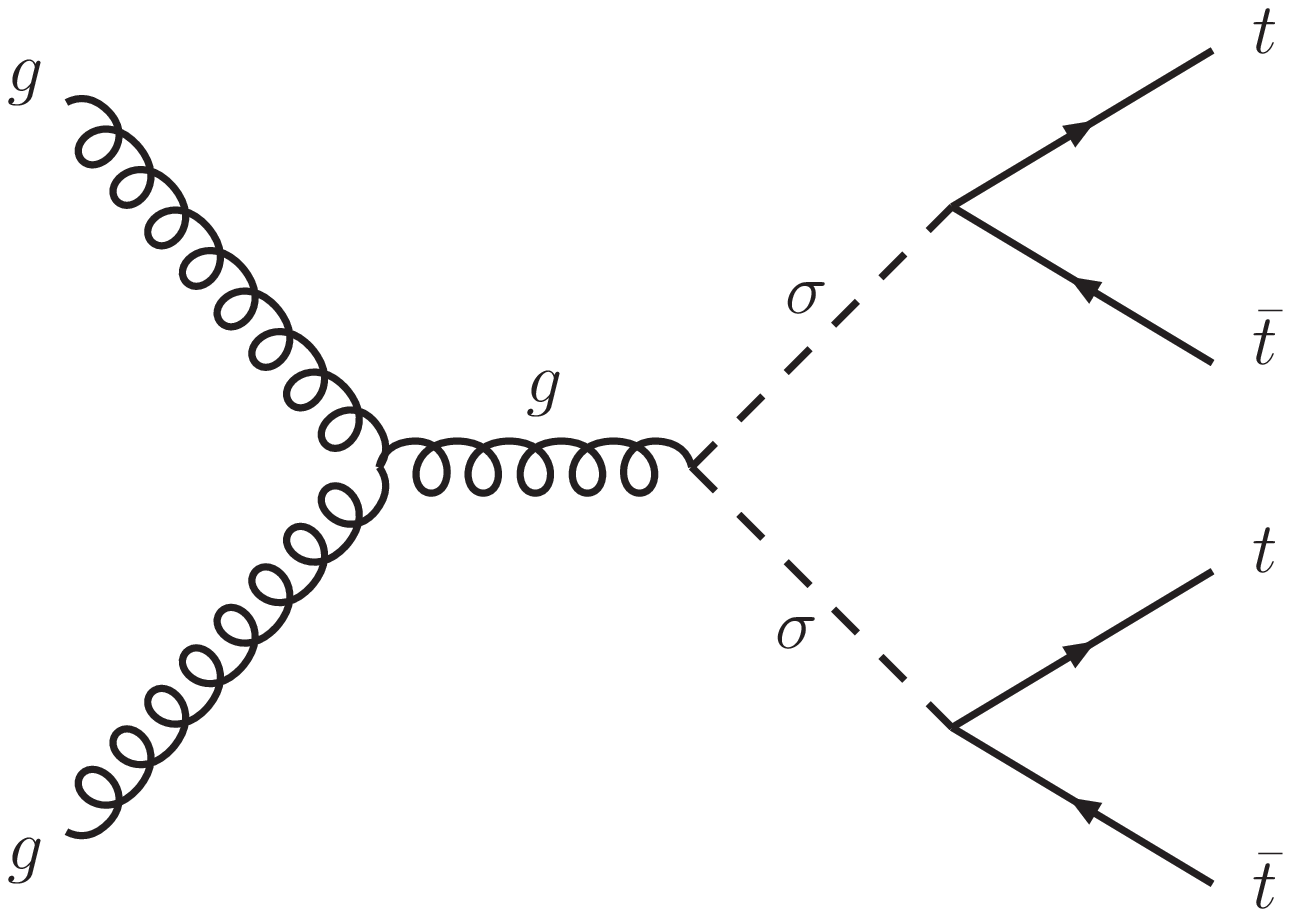}}
    \subfloat[]{\includegraphics[width=.28\textwidth]{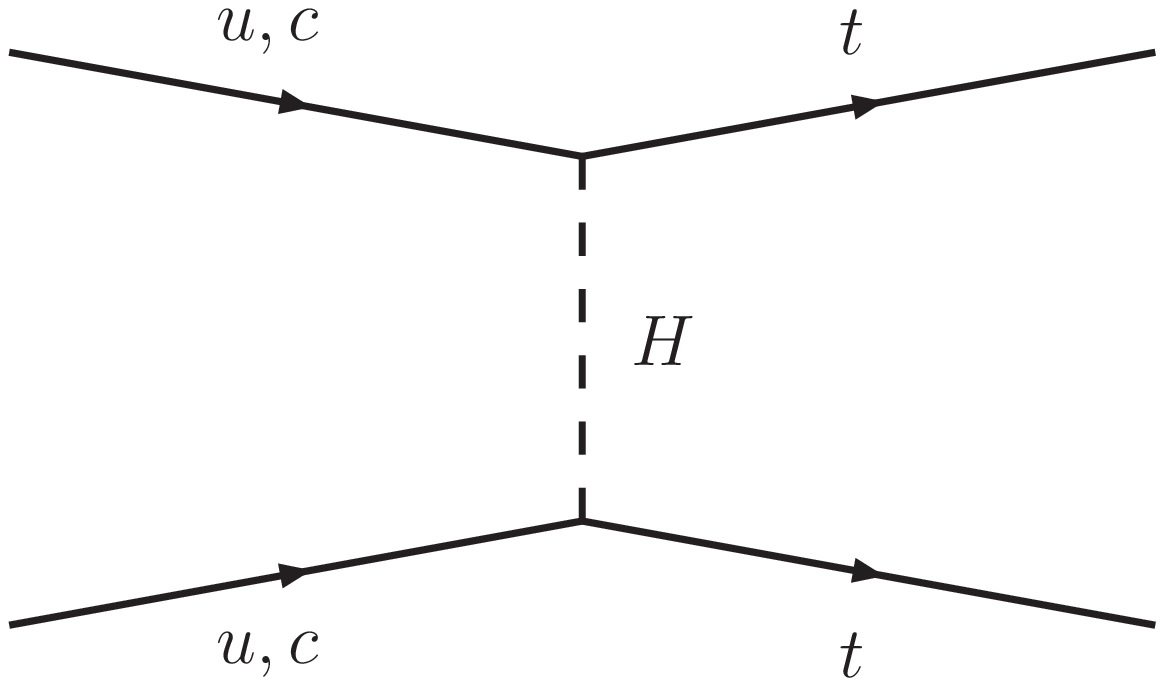}}
  \end{center}
  \caption{Leading-order diagrams for (a) vector-like top quark pair production, (b) sgluon pair production, and (c) same-sign top quark pair production through a BSM flavour-changing Higgs coupling.}
  \label{fig:diagrams}
\end{figure}

Leading-order Feynman diagrams for the production in $pp$ collisions of some of the signals searched for in this analysis are presented in figure~\ref{fig:diagrams}.

Previous searches by the ATLAS collaboration~\cite{Aad:2012bb} using an integrated luminosity of 1.04~\ifb{} of $pp$ collisions at a centre-of-mass energy $\sqrt{s} = 7$ TeV and
the CMS collaboration~\cite{Chatrchyan:2013fea}, using  an integrated luminosity of 19.5~\ifb{} of $pp$ collisions at $\sqrt{s} = 8$ TeV, did not observe a significant excess of same-sign dilepton production.  The ATLAS result was used to set a lower limit of 450 \GeV{} on the mass of a heavy down-type quark, under the assumption that the branching ratio to $Wt$ is 100\%, while 
 the CMS result set upper limits on the four-top-quark production cross section of 49~fb,\footnote{Unless stated otherwise, all limits reported are at the 95\% confidence level (CL).} on the sum of the $tt$ and $\bar{t}\bar{t}$ production cross sections of 720~fb, and on the $tt$ production cross section of 370~fb.  The CMS collaboration  used the same-sign lepton signature to search for $T_{5/3}$ quarks~\cite{Chatrchyan:2013wfa}, ruling out such quarks with mass below 0.80~\TeV{}, and  as part of a broader search for vector-like $T$ quarks~\cite{Chatrchyan:2013uxa}, ruling out  such quarks with mass less than 0.69~\TeV{}.  A more recent search by the  ATLAS collaboration~\cite{Aad:2014pda} using an integrated luminosity of 20.3~\ifb{} of $pp$ collisions at $\sqrt{s} = 8$ TeV with similar final states to those reported here was interpreted in the context of supersymmetric models.  The present analysis improves upon the $\sqrt{s} = 7$ TeV  ATLAS analysis by using a larger data set recorded at a higher centre-of-mass energy, having a higher signal acceptance, and expanding the range of BSM models considered.

\section {Data and Monte Carlo simulations}

The data were recorded by the ATLAS detector~\cite{Aad:2008zzm} in LHC $pp$ collisions at $\sqrt{s} = 8$ \TeV{} between April and December 2012, corresponding  to an integrated luminosity of 20.3~\ifb{}.  The ATLAS detector consists of an inner tracking system surrounded by a superconducting solenoid that provides a 2~T magnetic field, electromagnetic (EM) and hadronic calorimeters, and a muon spectrometer.  The inner detector provides tracking information from pixel and silicon microstrip detectors within pseudorapidity\footnote{ATLAS uses a right-handed coordinate system with its origin at the nominal interaction point (IP) in the centre of the
detector and the $z$-axis coinciding with the axis of the beam pipe. The $x$-axis points from the IP to the centre of the LHC ring,
and the $y$-axis points upward. Cylindrical coordinates ($r$,$\phi$) are used in the transverse plane, $\phi$ being the azimuthal angle around
the beam pipe. The pseudorapidity is defined in terms of the polar angle $\theta$ as $\eta =-\ln \tan({\theta/2})$. For the purpose of the fiducial
selection, this is calculated relative to the geometric centre of the detector; otherwise, it is relative to the reconstructed primary
vertex of each event.} $|\eta| < 2.5$, and from a  transition radiation tracker that covers $|\eta| < 2.0$.   The EM sampling calorimeter uses lead as absorber and liquid argon (LAr) as the active medium, and is divided into a barrel region that covers $|\eta| < 1.475$ and endcap regions that cover $1.375 < |\eta| < 3.2$.  The hadronic calorimeter consists of either LAr or scintillator tile as the active medium, and either steel, copper, or tungsten as the absorber, and covers $|\eta | < 4.9$.  The muon spectrometer covers $|\eta | < 2.7$, and uses multiple layers of high-precision tracking chambers to measure the deflection of muons as they traverse a toroidal field of approximately 0.5 (1.0) T in the central (endcap) regions of the detector.  A three-level trigger system selects events to be recorded for offline analysis.

Signal and the background sources that contain prompt same-sign leptons or trileptons are modelled using Monte Carlo (MC) simulations.  The remaining background sources are determined from the data, as described in section~\ref{sec:bkg}. $B$ and $T$ pair production is modelled using the {\sc protos} v2.2~\cite{AguilarSaavedra:2009es} generator using the MSTW2008LO~\cite{MSTW2008LO} parton distribution functions (PDFs), with {\sc pythia} v6.4~\cite{SJO-0601} used to model extra gluon emission and hadronization. $T_{5/3}$ production (both single and pair) is modelled with {\sc madgraph v5.1~\cite{Alwall:2011uj}} using the CTEQ6L1~\cite{cteq6} PDFs, with {\sc pythia} v8.1~\cite{Sjostrand:2007gs} used for hadronization.  Production of four top quarks is modelled under four scenarios: $i$) Standard Model, $ii$)  contact interaction, $iii$)  sgluon pair, and $iv$) 2UED/RPP.  The sgluon case is generated with {\sc pythia} v6.4 using the CTEQ6L1 PDFs; the other three models are generated with {\sc madgraph} using the MSTW2008LO PDFs  followed by {\sc pythia} v8.1; in the case of 2UED/RPP the {\sc bridge} generator~\cite{Meade:2007js} is used to decay the pair-produced excitations from {\sc madgraph} to \ttbar.  The simulated 2UED/RPP samples correspond to the tier (1,1) for the symmetric ($R_4 = R_5$) case, with $m_{\mathrm{KK}}$ ranging from 600 to 1200~\GeV{}. Constraints on the asymmetric ($R_4 > R_5$) case are derived by an extrapolation that uses kinematical considerations~\cite{Cacciapaglia:2011hx}.  These considerations also permit the extrapolation to signals arising from tiers (2,0) and (0,2) from the generated tier (1,1) signal. Pair production of $b^\prime$ events is modelled with the {\sc pythia} v8.1 generator for $b^\prime$ masses ranging from 400 to 1000 \GeV{}, using the MSTW2008LO PDFs. Production of two positively charged top quarks via a contact interaction
is also modelled using {\sc protos}~\cite{AguilarSaavedra:2010zi} and {\sc pythia} v6.4, with three
different chirality configurations of the contact interaction operator; production via the FCNC exchange of a Higgs-like particle is modelled with  {\sc madgraph} with {\sc pythia} v8.1 used for showering and hadronization.  The MSTW2008LO PDFs are used in simulating both types of $tt$ production.

	 The background contributions from $\ttbar W$ and $\ttbar Z$ (abbreviated as $\ttbar W/Z$ hereafter) and $\ttbar W^+W^-$  are modelled with {\sc madgraph} followed by {\sc pythia} v6.4, while $WZ$ and $ZZ$ plus jet production and $W^\pm W^\pm jj$ production are modelled using {\sc sherpa} v1.4~\cite{GLE-0901}. Background from the production of three vector bosons is modelled using {\sc madgraph} and  {\sc pythia} v6.4, and backgrounds from $t\bar{t}H$, $WH$ and $ZH$ production are modelled using   {\sc pythia} v8.1. The CTEQ6L1 PDFs are used for the $\ttbar W/Z$, three-vector-boson,  $WH$ and $ZH$ samples, the CT10~\cite{Lai:2010vv} PDFs are used for the $WZ$, $ZZ$,  $W^\pm W^\pm jj$ and  $t\bar{t}H$ samples, and the MSTW2008LO PDFs are used for the $\ttbar W^+W^-$ sample.  In most cases (excluding background contributions that are negligibly small) the cross sections are scaled to match next-to-leading-order calculations.

A variable number of additional $pp$ interactions are overlaid on simulated events to model the effect of multiple collisions during a single bunch crossing, and also the effect of the detector response to collisions from bunch crossings before or after the one containing the hard interaction.  Events are then
weighted to reproduce the distribution of the number
of collisions per bunch crossing observed in data.  The detector response is modelled  using either a {\sc geant4}~\cite{Agostinelli:2002hh,Aad:2010ah} simulation of the entire detector or a  {\sc geant4} simulation of the inner tracker and of the muon spectrometer combined with a fast simulation of shower development in the calorimeter~\cite{ATLAS:1300517}. Some samples are generated with both types of simulation, to allow direct comparison between the two, and agreement was found within the systematic uncertainty assigned to the efficiency estimate.  In all cases the simulated events were reconstructed using the same algorithms that were applied to the collision data.

\section{Event selection}
\label{sec:sel}

The final states considered in this search require the presence of two leptons with the same electric charge in the event (events with additional leptons beyond the same-sign pair are also accepted). In addition, two or more jets are required, at least one of which is consistent with origination from a $b$-quark, and sizeable missing transverse momentum \met{} is also required, indicating the presence of neutrinos coming from $W$ boson decays. The criteria used for each of these objects are given below.

Each event is required to pass either an electron trigger (where the chosen triggers require either an isolated electron with $\pt > 24$~\GeV{} or an electron with $\pt > 60$~\GeV{} with no isolation requirement) or a muon trigger (where the triggers chosen require either an isolated muon with $\pt > 24$~\GeV{} or a muon with $\pt > 36$~\GeV{} with no isolation requirement). The trigger efficiency for electrons is $\approx 95$\% while for muons it is $\approx 75$\%, resulting in trigger efficiencies that range from $\approx 95$\% for events with two muons to $>99$\% for events with two electrons.  In addition, events are required to have at least one reconstructed vertex, which must be formed from at least five tracks with $\pt > 0.4 \GeV{}$.  If multiple vertices are reconstructed, the vertex with the largest sum of the squared transverse momenta of its associated tracks is taken as the primary vertex.  Since the events used in this analysis tend to have vertices with many associated tracks, the correct vertex is selected in more than 99\% of the events.

Electrons are identified by requiring a track to match an electromagnetic calorimeter energy cluster, subject to several criteria on the shape of the shower and the consistency between the shower and track.  The selection requirements are varied with the $\eta$ and \pt\ of the electron candidate to optimize the signal efficiency and background rejection~\cite{ATLAS-CONF-2014-032}. The track is required to be within 2~mm in $z$ of the reconstructed primary vertex of the event. A hit in the innermost layer of the inner detector is required to reject photon conversions. Energy clusters in the calorimeter associated with an electron are required to have transverse energy $\et > 25$~\GeV{} and $|\eta| < 2.47$, with the barrel/endcap transition region $1.37 < |\eta| < 1.52$ excluded. The candidate is required to be isolated from  additional tracks within a cone of variable  $\Delta R \equiv \sqrt{(\Delta\eta)^2 + (\Delta\phi)^2} = 10 \GeV{}/p_{\rm T}$~\cite{Rehermann:2010vq}, such that the sum of the transverse momenta of the tracks within that cone must be less than 5\% of the electron $p_{\rm T}$.  In addition, electrons are required to be separated from any jet by at least $\Delta R = 0.4$.

Muons~\cite{Aad:2014rra} are identified from hits in the muon system matched to a central track, where the track must be within 2~mm in $z$ of the primary vertex, and are required to have an impact parameter in the transverse plane that differs from the beam position by less than three impact parameter standard deviations. Requirements are placed on the number of hits in various layers of the muon system, and on the maximum number of layers where hits are missing.  Muon pairs that are consistent with the passage of a cosmic ray are discarded.  Muons are subject to the same track-based isolation requirement as electrons.  Muons are also required to be separated from any jet by $\Delta R = 0.04 + 10 \GeV{}/p_{\rm T}$, and to have $\pt > 25$~\GeV{} and $|\eta| < 2.5$.  Events with a muon within  $\Delta \phi \times \Delta \theta = 0.005 \times 0.005$ of any electron are rejected.  At least one of the selected leptons is required to match a lepton identified by the trigger.

Jets are reconstructed from energy clusters in the calorimeter using an anti-$k_t$ algorithm~\cite{Cacciari:2008gp,Cacciari200657,Cacciari:2011ma} with radius parameter 0.4.  If one or more jets are within $\Delta R=0.2$ of an electron, the jet closest to the electron is discarded (i.e.\ the  cluster of energy in the calorimeter is treated as an electron rather than a jet).  To suppress jets that do not originate from the primary vertex in the event, the jet vertex fraction (JVF) is defined by considering all tracks with $\pt > 0.5 \GeV{}$ within the jet, and finding the fraction of the summed \pt\ from tracks that originate from the primary vertex.  Jets with \pt\ $< 50$ \GeV{} and $|\eta| < 2.4$ that are matched to at least one track are required to have JVF greater than 0.5.  All jets are required to have \pt\ greater than 25 \GeV{} (after energy calibration~\cite{Aad:2011he}) and $|\eta|$ less than 2.5.

A multivariate algorithm~\cite{ATLAS:2011qia} is used to test if each jet is consistent with having arisen from a $b$-quark, based on the properties of the tracks associated with the jet.  A requirement is placed on the output of the discriminant such that $\approx 70$\% of $b$-quark jets and $\approx 1$\% of light-quark or gluon jets pass in inclusive simulated $t\bar{t}$ events.  All jets that meet this criterion are called `$b$-tagged' jets.

The missing transverse momentum is calculated as the negative of the vector \et\ sum from all calorimeter energy clusters, with jet and electron energy calibrations applied to clusters associated with those objects, and corrected for the energy carried away by identified muons.  Energy scale corrections applied to electrons and jets are also propagated to \met.  Events are required to have $\met > 40$ \GeV{}.

If the same-sign leptons are both electrons, their invariant mass $m_{ee}$ is required to be greater than 15~\GeV{} and to satisfy $|m_{ee}-m_Z (=91 \GeV{})| > 10$ \GeV{}. These requirements reject events from known resonances decaying to an electron--positron pair where the charge of either the  electron or positron is misidentified. Finally, the scalar sum of all jet and lepton transverse momenta (\HT), is required to be greater than 400 \GeV{}, since the signals considered here produce a high number of particles with high transverse momenta. These  preselection criteria are applied to all searches; some of them are tightened when optimizing the selection for each signal model (see section~\ref{sec:opt}).   Figure~\ref{fig:preselplots} shows the distributions of \HT\ and \met\ after applying this selection (except for the requirements on \HT\ and \met\ themselves).

\begin{figure}
  \begin{center}
    \subfloat[]{\includegraphics[width=.4\textwidth]{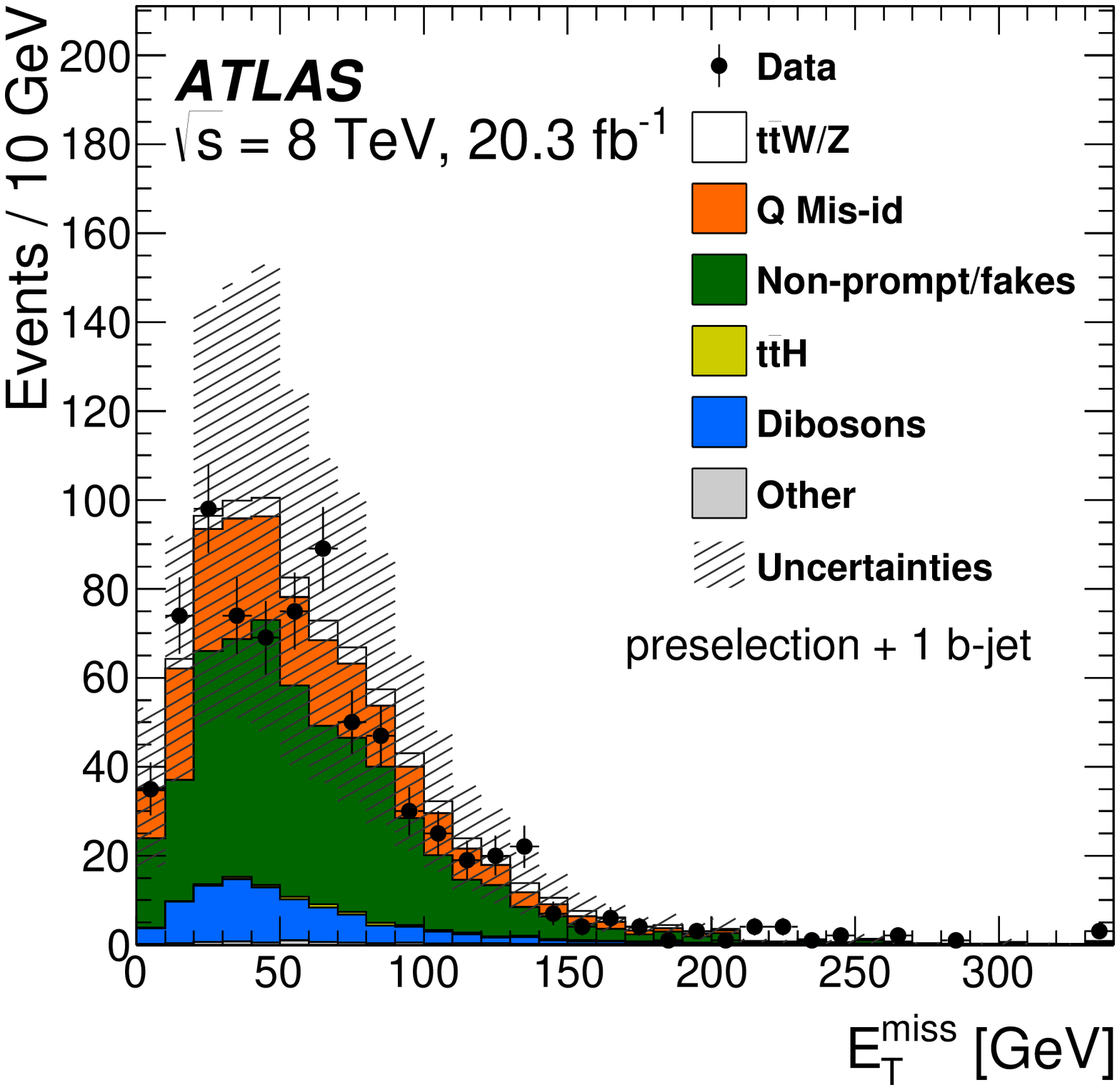}\label{fig:preselplot:MET1tag}}
    \subfloat[]{\includegraphics[width=.4\textwidth]{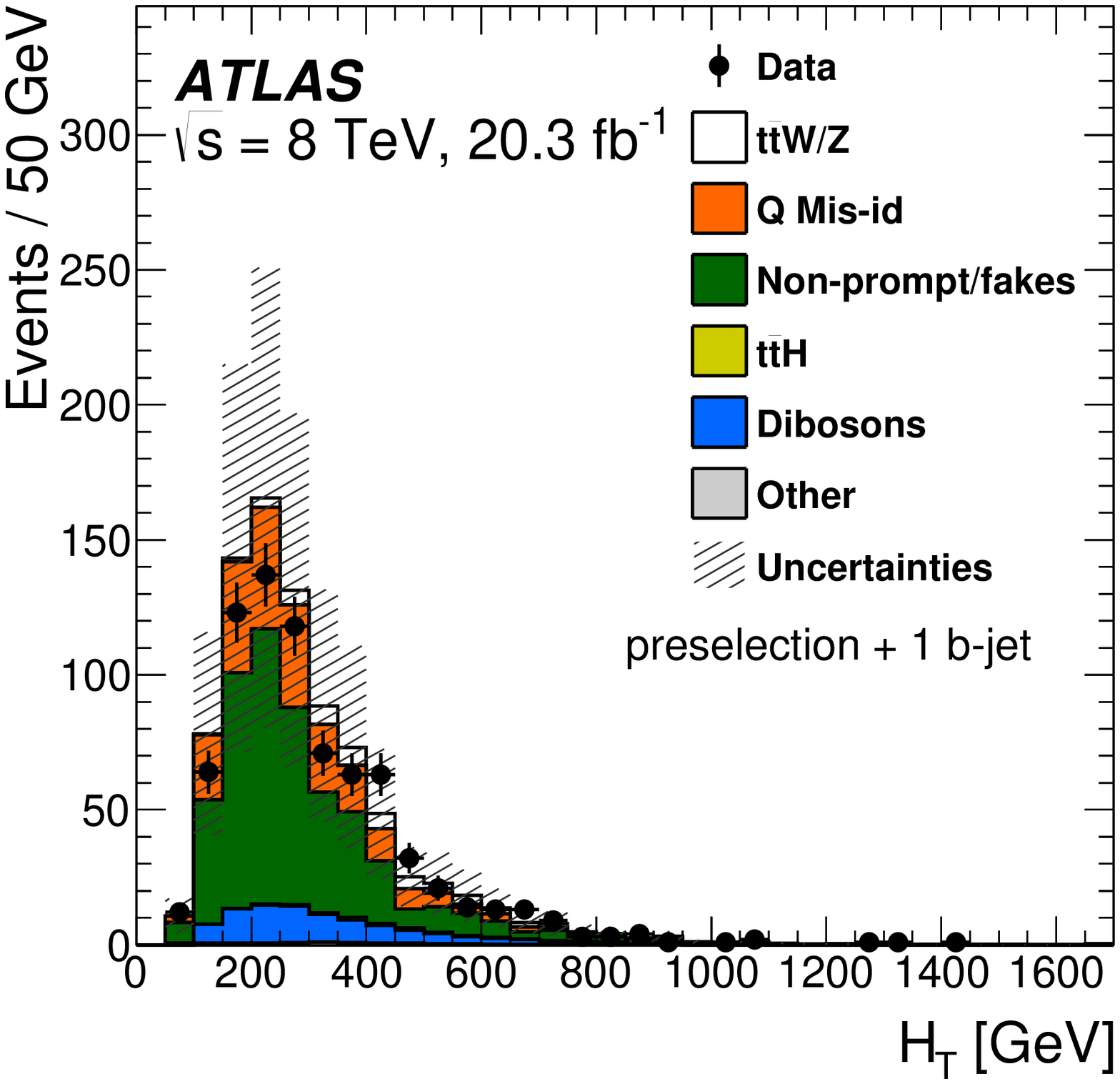}\label{fig:preselplot:HT1tag}}\\
    \subfloat[]{\includegraphics[width=.4\textwidth]{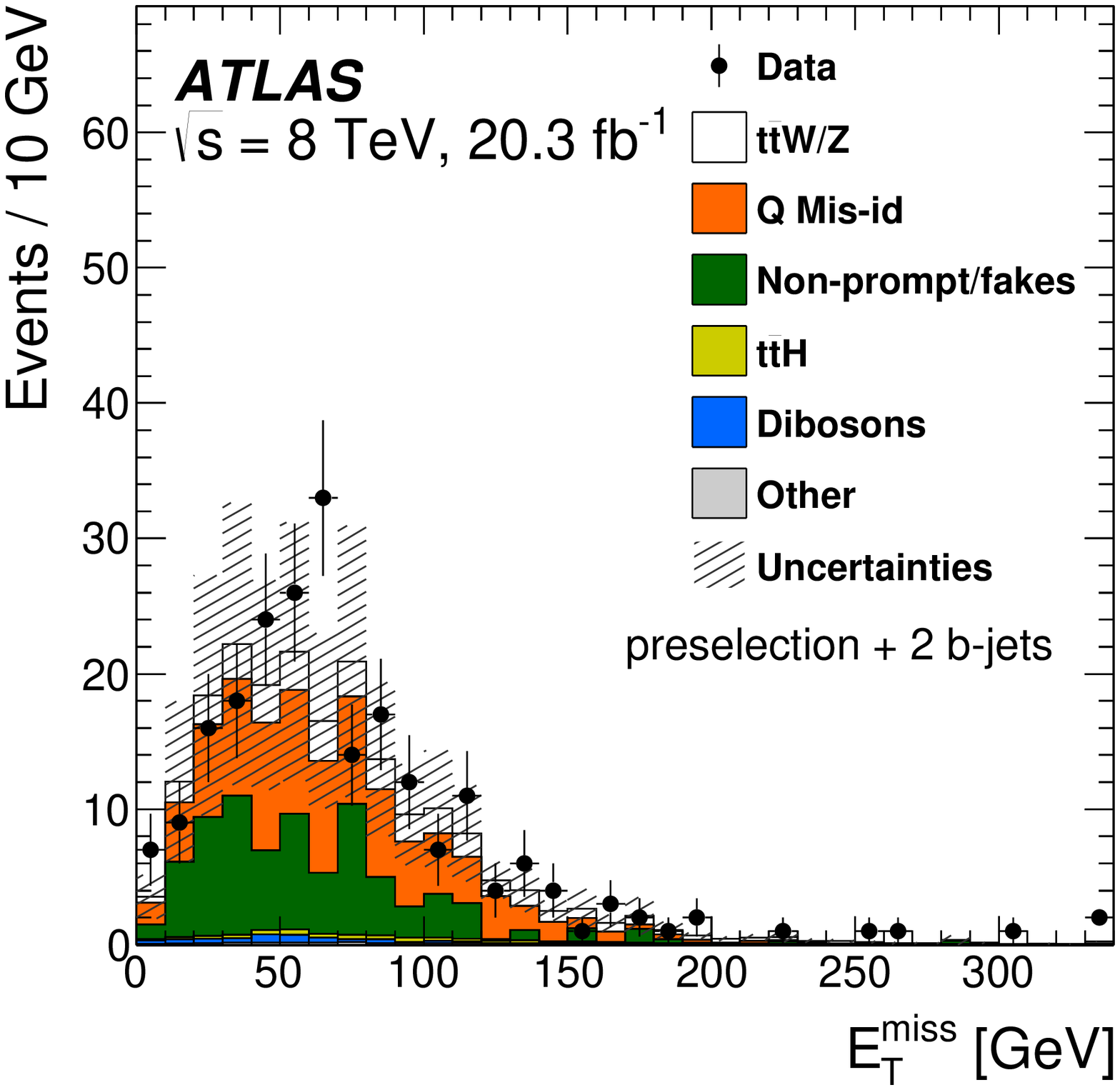}\label{fig:preselplot:MET2tag}}
    \subfloat[]{\includegraphics[width=.4\textwidth]{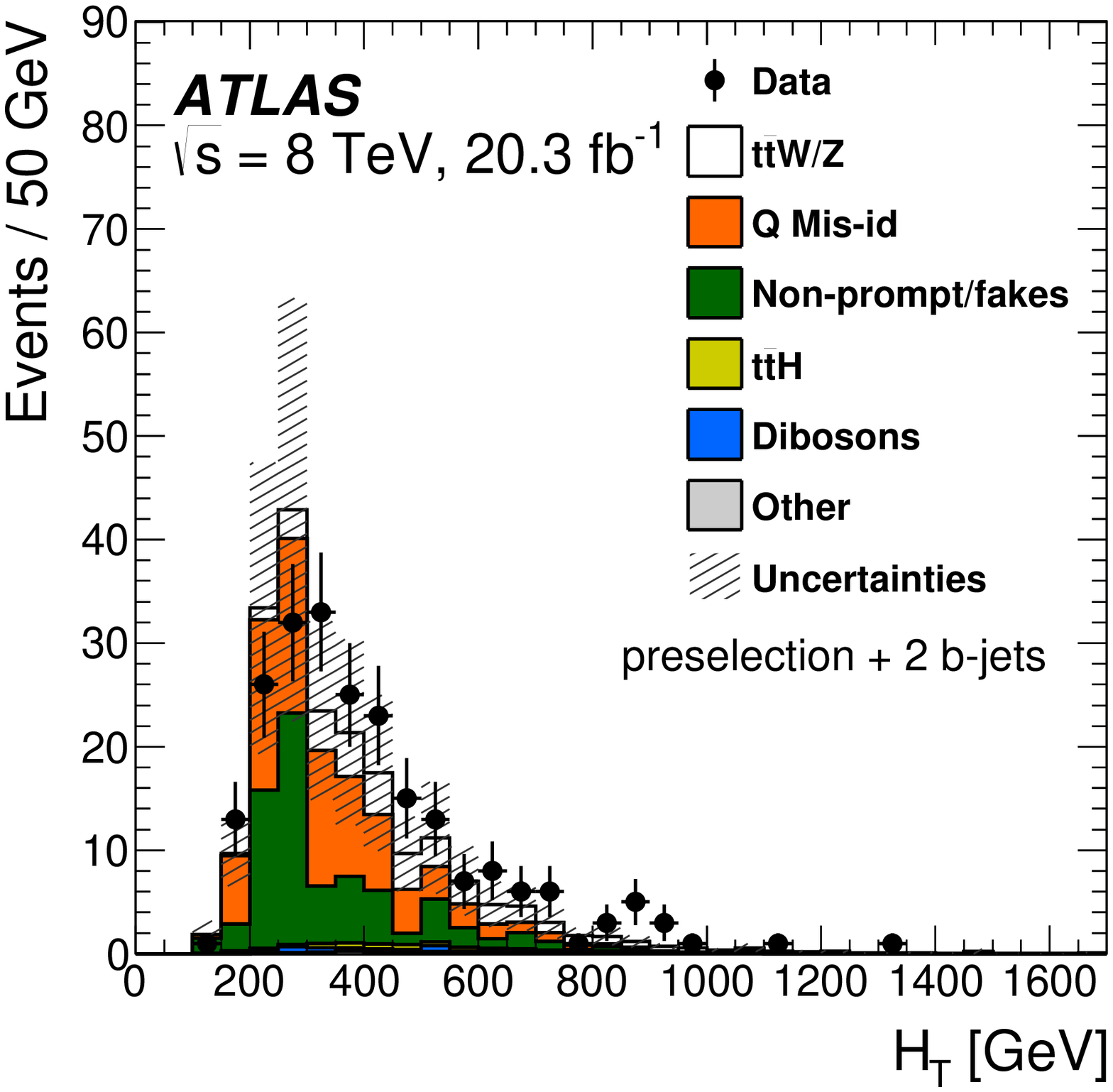}\label{fig:preselplot:HT2tag}}\\
    \subfloat[]{\includegraphics[width=.4\textwidth]{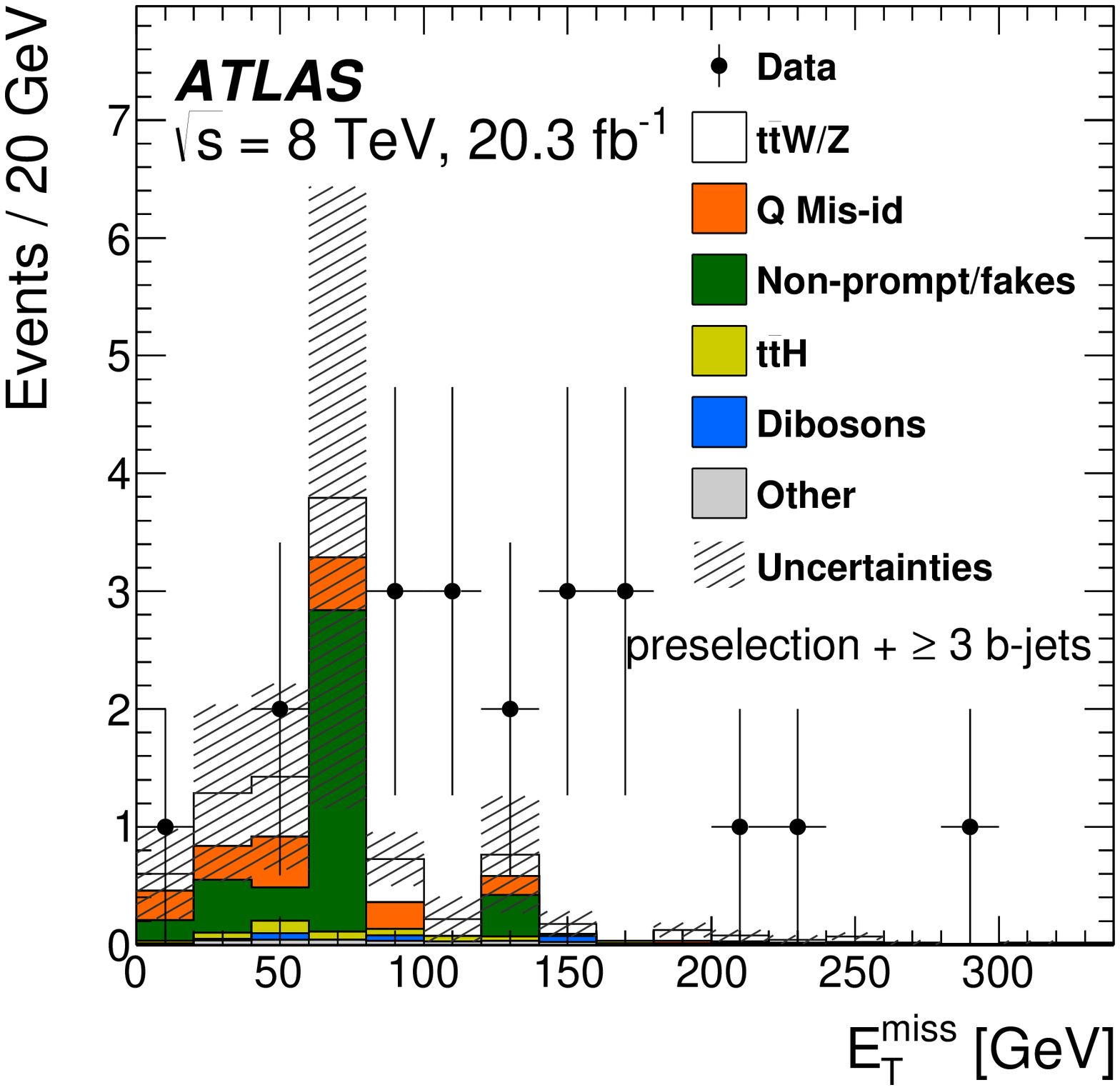}\label{fig:preselplot:METge3tag}}
    \subfloat[]{\includegraphics[width=.4\textwidth]{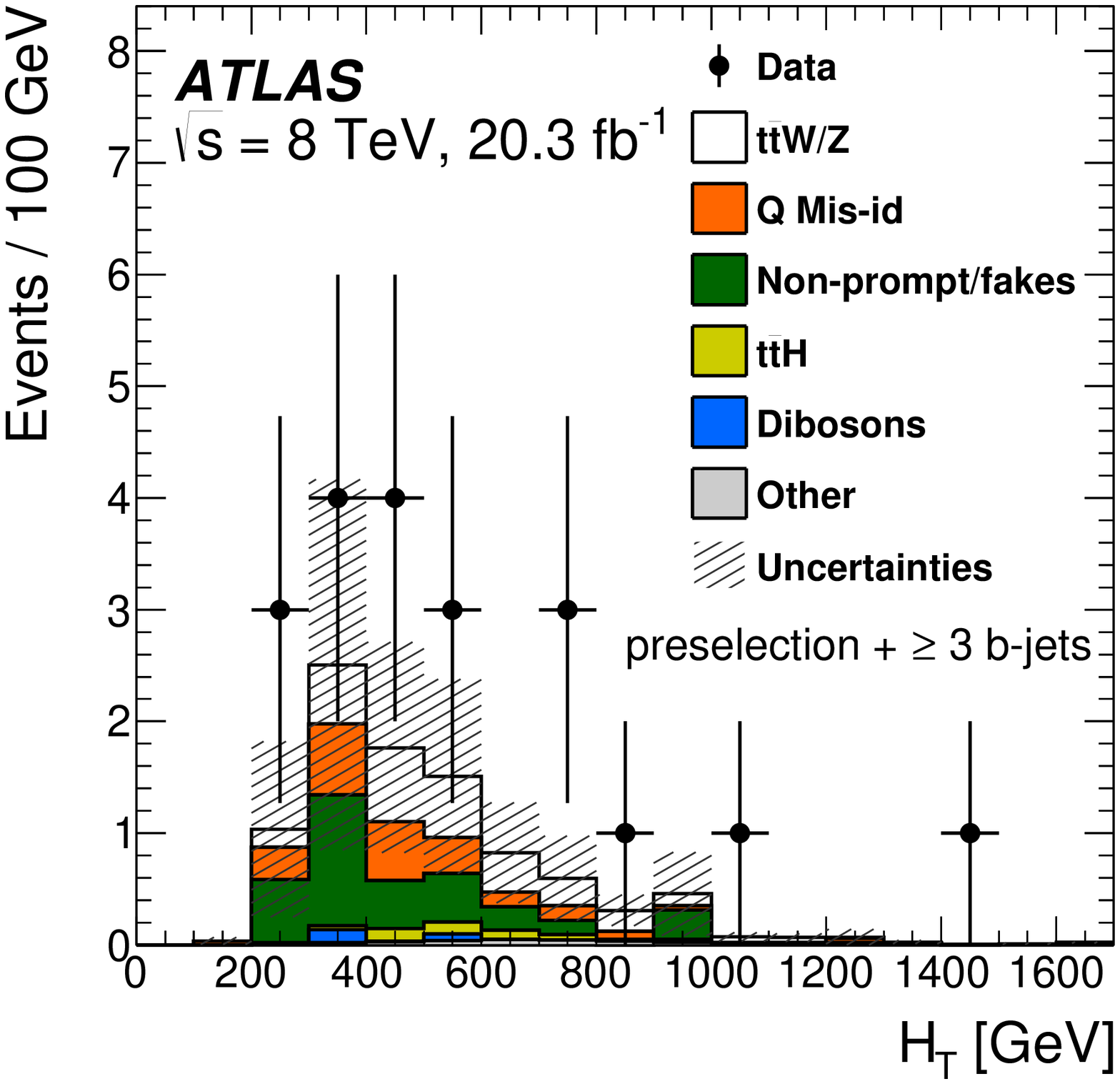}\label{fig:preselplot:HTge3tag}}\\
  \end{center}
  \caption{Distributions of the missing transverse momentum \met\  and scalar sum of jet and lepton transverse momenta \HT\ after applying the preselection criteria, for events with one, two, or more than two $b$-tagged jets.  The points with error bars are the data, the stacked histograms show the expectation from background, and the shaded band is the total uncertainty on the background expectation.}
  \label{fig:preselplots}
\end{figure}

\section {Background estimation}
\label{sec:bkg}

Background arises from two distinct sources: SM processes that result in same-sign lepton pairs, and instrumental backgrounds where objects are misidentified or misreconstructed such that events appear to have the required set of leptons.  The former category includes production of $W^{\pm}W^{\pm}jj$, $\ttbar W/Z$, $\ttbar W^+W^-$,  $t\bar{t}H$, $WH$, $ZH$, $tWZ$, $tH$, $WZ$ and $ZZ$ with a heavy-flavour jet, or three vector bosons.  In addition, the four-top-quark production predicted in the SM is included as a background to all searches for other signals, though its contribution is small due to the small cross section.  All of these processes have small cross sections, and their expected yields are computed using simulation. Known differences between the lepton selection and $b$-tag efficiencies between the MC simulation and data control samples are taken into account when computing the expected yields.
 
Instrumental backgrounds have contributions from two categories: $i$) events where one or more jets are misidentified as leptons, or which contain non-prompt leptons, and $ii$) events that contain two leptons of opposite charge, where one of the charges is mismeasured.  The `matrix method' is used to estimate the contribution from events with misidentified (fake or non-prompt) leptons.  In this method, the default (`tight') lepton identification criteria (section~\ref{sec:sel}) are relaxed to form a `loose' sample.  Lepton isolation requirements are not imposed, and therefore the loose sample contains a larger fraction of fake/non-prompt leptons than the tight sample.  The fraction of real leptons (meaning prompt leptons from the decay of a $W$, $Z$, or $H$ boson) passing the loose criteria that also pass the tight criteria is referred to as $r$. Similarly, the fraction of fake/non-prompt leptons passing the loose cuts that also pass the tight cuts is referred to as $f$.   Using measured values of $r$ and $f$, one can construct a matrix that relates the observed yields of dilepton events in the categories loose--loose, loose--tight, and tight--tight to the real--real, real--fake, and fake--fake yields. An analogous procedure is applied to three lepton events starting with categories for all possible combinations of three loose or tight leptons, resulting in an estimate of the  number of events with one or more misidentified leptons in the selected sample ($N^\mathrm{tt}_\mathrm{fake}$).

Single-lepton events are used to measure $r$ and $f$. The criteria used to select these events are different for electrons and muons  due to the differences in the sources of fake/non-prompt leptons for each flavour.  For electrons, $r$ is measured using events with \met\ $> 150$ GeV, where the dominant contribution is from $W\rightarrow e\nu$, and $f$ is measured using events with the transverse mass of the \met\ and electron\footnote{The transverse mass of a lepton and \met\ is defined as $m_{\rm T}(W) \equiv \sqrt{2p_{{\rm T}\ell}  \met(1-\cos\Delta \phi)}$ where $p_{{\rm T}\ell}$ is the lepton \pt and $\Delta \phi$ is the azimuthal angle between the lepton and the direction of the missing transverse momentum vector.} $m_{\rm T}(W)< 20$ GeV and $\met + m_{\rm T}(W) < 60$ GeV, where the dominant contribution is from multijet production where one or more jets is misidentified as an electron.  For muons, $r$ is measured using events with $m_{\rm T}(W) > 100$ GeV, a sample dominated by $W\rightarrow \mu\nu$, and $f$ is measured using events where the impact parameter of the muon with respect to the primary vertex is more than five standard deviations from zero, consistent with muons arising from heavy-flavour hadron decays.  The small contribution of real leptons to the control samples used to measure $f$ is estimated from simulation, and this contribution is subtracted from the sample.
  The values of $r$ and $f$ are parameterized with respect to properties of the leptons (e.g. $| \eta |$ and \pt) and of the event (e.g. the number of $b$-tagged jets).  Typical values are $r = 0.90$ and $f = $ 0.20 -- 0.40 for electrons, and  $r=0.95 $-- 1.00 and $f= $ 0.12 --0.30 for muons.
 
The triggers used for low-\pt\ leptons require isolation;  since the  tight and loose lepton  criteria differ in their isolation requirements, fake/non-prompt leptons in events where only the  low-\pt\ triggers fired are more isolated, on average, than those from an unbiased trigger, meaning that  $f$ for these leptons is substantially higher.  Therefore, $r$ and $f$ are measured separately for samples collected with the different triggers, and  the appropriate values are applied based on the lepton triggers that fired in each event.
A further complication may arise due to the small number of events in the loose sample, which can lead to a calculated value of $N^\mathrm{tt}_\mathrm{fake}$ that is negative or very close to zero.  In the case of negative values $N^\mathrm{tt}_\mathrm{fake}$ is set to zero when computing limits.  To properly estimate the statistical uncertainty on the fake/non-prompt lepton contribution given the small number of events, a Poisson likelihood for the estimate from the matrix method is used, and the standard deviation of the probability density function (p.d.f.) from this likelihood is used to set the uncertainty.  In cases where the prediction from the matrix method is less than or near zero, the standard deviation is computed relative to zero rather than to the mean of the p.d.f.

Charge misidentification (`Q mis-Id') is negligible for muons due to the small probability for muons to radiate photons, the long lever arm to the muon system and the fact that the charge is measured in both the inner detector and the muon spectrometer.  For electrons, the rate of charge misidentification is calculated from a sample of $Z \ra ee$ events, selected with no requirement placed on the charge of the two electron tracks. It is assumed that the rate at which the charge of an electron is misidentified varies with the $|\eta|$ and \pt\ of each electron but is uncorrelated between the two electrons in each event.  Further assuming that the sample consists entirely of opposite-sign electron pairs, the number of measured same-sign events $N_{\rm ss}^{ij}$ where one electron is in the $i^{\mathrm{th}}$ ($|\eta|$,\pt) bin and the other in the $j^{\mathrm{th}}$ bin is expected to be

\begin{equation}
\label{eq:n_ss}
N_{\rm ss}^{ij} \approx N^{ij}(\varepsilon_i + \varepsilon_j)
\end{equation}
\\
where $N^{ij}$ is the total number of events in the $i$-$j$ $|\eta|$--$\pt$ bin, and $\varepsilon$ is the rate of charge mismeasurement.  The value of $\varepsilon$ in each $|\eta|$--$\pt$ bin is then extracted by maximizing the Poisson likelihood for the observed number of same-sign pairs in each $|\eta|$--$\pt$ bin to be consistent with the expectation from equation~\ref{eq:n_ss}.  One limitation of this estimate is that electrons from $Z$ decay only rarely have large \pt, rendering the uncertainty on the charge misidentification rate for high-\pt electrons large.  To reduce this uncertainty, the rate of charge misidentification is 
estimated using simulated $t\bar{t}$ events as a function of the electron \pt.  This rate is scaled to match  the 
rate observed in data for the \pt\ range covered by the $Z$ events, and the rate for electrons 
with larger \pt\ is extrapolated according to the scaled prediction from simulation. Closure tests comparing the number of events in the same-sign $Z$ peak to the expectation based on the opposite-sign $Z$ peak and the charge mismeasurement rates were performed in data and simulation and
show good agreement.

To determine the number of events expected from charge mismeasurement in the signal region, a sample is selected using the same criteria as for the analysis selection, except that an opposite-sign rather than same-sign $ee$ or $e\mu$ pair is required.  The measured $\varepsilon$ values are then applied to each electron in this sample to determine the expected number of mismeasured same-sign events in the analysis sample. One source of charge mismeasurement is from `trident' electrons, where the electron emits a hard photon that subsequently produces an electron--positron pair, resulting in three tracks with small spatial separation.  If the wrong track is matched to the EM cluster, the charge may be incorrect.  However, such electrons would also appear to be isolated far less frequently than electrons that do not emit hard radiation. Therefore the value of $r$ for trident electrons is lower than for electrons that have a correctly measured charge, meaning that they also contribute to the fake/non-prompt electron estimate from the matrix method.  To avoid double-counting events with trident electrons in the background estimate, the charge mismeasurement rate is measured in a data sample where the non-prompt/fake contribution, estimated using the matrix method, has been removed.  

Simulation was used to estimate the sources of events in the signal regions that have fake/non-prompt leptons and/or electrons with mismeasured charge, and it is found that $t\bar{t}$ events provide the dominant contribution.

The background estimates are validated using samples where one or more of the  preselection criteria are vetoed so that the samples are statistically independent and the expected yield from signal events is small.
One such validation
region, called the `low  \HT{}$+1b$' region, is defined by applying the preselection criteria, except that the requirement on \HT\ is modified to $100\GeV{}<~\HT~<~400$~\GeV{}.  This validation region is particularly useful because the background composition is similar to that of the preselection (including the fact that $t\bar{t}$ events are the dominant source of the fake/non-prompt lepton and charge mismeasurement background contributions).  The predicted and
observed yields in this validation region are given in tables~\ref{ctrl:lowht_1b_ss_yield} and~\ref{ctrl:lowht_1b_3l_yield}.  Events with three leptons are considered explicitly in table~\ref{ctrl:lowht_1b_3l_yield} since the fake/non-prompt lepton background contribution from trilepton events is not negligible, and it is important to check that this component of the background is well understood.  The ``other bkg.'' category includes $WWW$, $WWZ$, $WH$, $ZH$, $t\bar{t}WW$, SM four-top-quark, and single-top-quark production.  Similar agreement between the data yield and background expectation is observed in validation regions where no requirement on $b$-tagged jets is imposed, and where  \met\ is required to be less than 40~\GeV{} or \HT\ is required to be in the range 100--400~\GeV{}.

\begin{table}
  \begin{center}
\begin{tabular}{l *{3}{c}}
\hline\hline
Sample & $ee$ & $e\mu$ & $\mu\mu$ \\
\hline
Q mis-Id & $136 \pm 2 \pm 41$ & $118 \pm 1 \pm 35$ &  ---  \\
Fake/Non-prompt & $153 \pm 11 \pm 107$ & $225 \pm 11 \pm 158$ & $29 \pm 3 \pm 20$ \\
$t\bar{t}W/Z$ & $4.57 \pm 0.19 \pm 1.88$ & $14.2 \pm 0.3 \pm 5.8$ & $8.43 \pm 0.27 \pm 3.56$ \\ 
$t\bar{t}H$  & $0.39 \pm 0.04 \pm 0.04$ & $1.31 \pm 0.08 \pm 0.13$ & $ 0.76 \pm 0.06 \pm 0.07$ \\
Dibosons & $5.57 \pm 0.45 \pm 1.08$ & $15.9 \pm 0.8 \pm 2.9$ & $9.00 \pm 0.58 \pm 1.79$ \\
Other bkg. & $0.32 \pm 0.11 \pm 0.11$ & $ 0.75 \pm 0.20 \pm 0.20$ & $0.27 \pm 0.06 \pm 0.06$ \\ \hline
Total bkg. & $299 \pm 11 \pm 115$ & $375 \pm 11 \pm 162$ & $47 \pm 3 \pm 20$ \\
\hline
Data & 271 & 307 & 52 \\
\hline

\end{tabular}
\end{center}
    \caption{Observed and expected numbers of events in the low-\HT{}$+1b$ validation 
      region for the same-sign dilepton channels.  The first uncertainty is statistical and the second is systematic (systematic uncertainties are described in section~\ref{sec:syst}).
      \label{ctrl:lowht_1b_ss_yield}}
\end{table}

\begin{table*}
  \begin{center}
 \begin{tabular}{l *{4}{c}}
\hline\hline
Sample & $eee$ & $ee\mu$\\
\hline
Fake/Non-prompt & $8.0 \pm 2.3 \pm 5.6$ & $13.2 \pm 2.4 \pm 9.2$ \\
$t\bar{t}W/Z$ & $1.20 \pm 0.09 \pm 0.46$ & $2.55 \pm 0.13 \pm 0.87$ \\
$t\bar{t}H$  & $0.07 \pm 0.02 \pm 0.01$ & $0.28 \pm 0.03 \pm 0.03$ \\
Dibosons & $5.78 \pm 0.51 \pm 1.14$ & $6.78 \pm 0.57 \pm 1.33$  \\
Other bkg. & $0.04 \pm 0.02 \pm 0.02$ & $0.11 \pm 0.02 \pm 0.02$ \\ \hline
Total bkg. & $15.1 \pm 2.4 \pm 5.7$ & $22.9 \pm 2.5 \pm 9.4$  \\
\hline
Data & 15 & 18  \\
\hline

\\
\hline\hline
Sample & $e\mu\mu$ & $\mu\mu\mu$ \\
\hline
Fake/Non-prompt &  $17.9 \pm 2.8 \pm 12.5$ & $1.34 \pm 0.55 \pm 0.94$ \\
$t\bar{t}W/Z$ &$ 3.38 \pm 0.16 \pm 1.15$ & $2.70 \pm 0.14 \pm 1.00$ \\
$t\bar{t}H$ & $0.32 \pm 0.03 \pm 0.03$ & $0.14 \pm 0.02 \pm 0.01$ \\
Dibosons & $8.42 \pm 0.57 \pm 1.78$ & $9.23 \pm 0.65 \pm 1.82$ \\
Other bkg. & $0.12 \pm 0.02 \pm 0.02$ & $0.15 \pm 0.03 \pm 0.03$ \\ \hline
 \hline
Total bkg. & $30.1 \pm 2.8 \pm 12.7$ & $13.6 \pm 0.9 \pm 2.4$ \\
\hline
Data & 36 & 14 \\

\hline

\end{tabular}
  \end{center}
   \caption{Observed and expected numbers of events  in the low-\HT{}$+1b$ validation 
      region for the trilepton channels.  The first uncertainty is statistical and the second is systematic (systematic uncertainties are described in section~\ref{sec:syst}).}\label{ctrl:lowht_1b_3l_yield}

\end{table*}

\section{Selection optimization}
\label{sec:opt}
The selection is defined to optimize the expected limit on signals. 
Since many BSM physics models (each of them dependent on mass and/or coupling parameters) could result in anomalous production of the sort sought in this analysis, defining a selection that is sensitive to all of them is a challenge. As a first step toward a solution, the same-sign top signal is considered separately from the others, as it has unique characteristics: contributions are expected dominantly from positively charged lepton pairs\footnote{Production of $\bar{u}\bar{u}\rightarrow\bar{t}\bar{t}$ is also possible through these processes but the production cross section in $pp$ collisions for this process is two orders of magnitude lower than that for $uu\rightarrow tt$.  Therefore, considering only the $tt$ final state reduces the background by a factor of two while having only a small impact on the signal.}, and the jet multiplicity tends to be lower.  The selection for same-sign top events is optimized with respect to \HT, \MET, and the number of $b$-tagged jets $N_b$, with contributions from the $ee$, $e\mu$, and $\mu\mu$ channels considered separately.

The remaining signals share a similar final-state topology, but the distribution of events differs between them in several variables.
Therefore several event categories are defined, based on features of the events such as  \HT, \MET, and $N_b$, as shown in table~\ref{tab:allSR}.  Splitting the sample in this manner 
provides good overall efficiency for signal events, while allowing events that are least likely to arise from background (i.e.\ events with large values of \HT, \MET, or $N_b$) to be treated separately in the analysis, thereby enhancing the sensitivity to BSM physics.  The boundaries between categories in \HT\ and \MET\ were chosen to optimize the sensitivity to four-top-quark signals; these values are close to optimal for the other signals considered as well.

All of the categories are considered when searching for vector-like quarks or chiral $b^\prime$-quarks, while only the categories that require at least two $b$-tagged jets are considered when searching for the production of four top quarks.  One consequence of defining several signal categories is that the data-driven background estimates are subject to large statistical fluctuations.  To mitigate this, all lepton flavours are summed within each category.  The signal regions are defined based on the expected yields of signal and background, taking into account statistical and systematic effects, without considering the distribution of data.

 \begin{table*}[tb]
	\begin{center}
	\begin{tabular}{ c  | c | c | c | c}
	\hline
	\hline
	\multicolumn{3}{c|}{Definition} & \multicolumn{2}{c}{Name} \\
	\hline
        \multicolumn{4}{c}{$e^\pm e^\pm+e^\pm\mu^\pm+\mu^\pm\mu^\pm+eee+ee\mu+e\mu\mu+\mu\mu\mu$, $N_j\geq2$} \\
        \hline
	\multirow{3}{*}{$400  < \HT < 700\GeV{}$} 	& $N_b = 1$ & & SRVLQ0 &  \\
	\cline{2-2}\cline{4-5}
	 										& $N_b = 2$& $\met>40$~\GeV & SRVLQ1 & SR4t0 \\
	\cline{2-2}\cline{4-5}
	 										& $N_b \geq 3$ & & SRVLQ2 &SR4t1 \\
	\hline
	\multirow{5}{*}{$\HT \geq 700~\GeV{}$} 				& \multirow{2}{*}{$N_b = 1$}  				& $40 < \hbox{\met} < 100\GeV{}$ 		& SRVLQ3 & \\
	\cline{3-5}
											 &					 				& $\hbox{\met} \geq 100~\GeV{}$					& SRVLQ4 & \\
	\cline{2-5}
				& \multirow{2}{*}{$N_b = 2$}  				& $40< \hbox{\met} < 100\GeV{}$ 		& SRVLQ5 & SR4t2 \\
	\cline{3-5}
											 &					 				& $\hbox{\met}\geq 100~\GeV{}$					& SRVLQ6 & SR4t3 \\
	\cline{2-5}
											&  $N_b \geq 3$								  			 		& $\met>40$~\GeV & SRVLQ7 & SR4t4 \\
	\hline
        \multicolumn{4}{c}{$e^+e^+$, $e^+\mu^+$, $\mu^+\mu^+$, $N_j\in[2,4]$, $\Delta\phi_{\ell\ell}>2.5$} \\
        \hline
        $\HT>450~\GeV$ & $N_b\geq1$ & $\met>40~\GeV$ & \multicolumn{2}{c}{SRtt$ee$, SRtt$e\mu$, SRtt$\mu\mu$} \\
        \hline
	\end{tabular}
	\end{center}
		\caption{Definitions of the different signal regions. $N_j$ is the number of jets that pass the selection requirements, and $\Delta\phi_{\ell\ell}$ is the separation in $\phi$ between the leptons. In regions SRVLQ0--SRVLQ7, contributions from all lepton flavours are summed.\label{tab:allSR}
}

\end{table*}

\section{Systematic uncertainties} 
\label{sec:syst}

 \begin{table*}[tt]
  \begin{center}
   \begin{tabular}{l|cccccccc}
      \hline\hline
       \multirow{2}{*}{Source}  & \multicolumn{8}{c}{VLQ signal region number} \\
       \cline{2-9}
        &0 & 1 & 2 & 3  & 4 & 5 & 6 & 7 \\
      \hline
      Cross section  & $\pm 8.0$ & $\pm 13.6$ & $\pm 15.1$ &  $\pm 11.1 $ &$\pm 12.1 $  & $\pm 16.8$  &  $\pm 25.2$ & $\pm 23.8$\\
      \midrule
       Jet energy scale & $^{+1.7}_{-1.6}$ & $^{+1.2}_{-1.8}$   & $^{+1.4}_{-1.7}$ & $^{+1.8}_{-2.1}$  & $^{+2.6}_{-4.2}$ & $^{+3.8}_{-1.5}$ &  $^{+8.5}_{-4.8}$ & $^{+7.3}_{-2.9}$ \\
      \midrule
      $b$-tagging efficiency & $\pm 1.0$ &$\pm 2.6$ &$^{+5.7}_{-5.5}$   & $^{+1.9}_{-2.0}$ & $^{+1.6}_{-1.7}$ & $^{+3.8}_{-3.7}$ &  $^{+5.1}_{-5.0}$ & $^{+8.3}_{-8.2}$ \\
            \midrule
      Lepton ID efficiency & $\pm 1.3$ & $\pm 1.6$  & $\pm 1.6$ &  $^{+2.1}_{-2.0}$&  $^{+2.1}_{-2.0}$  &$^{+2.2}_{-2.1}$  &  $^{+2.8}_{-2.2}$ & $\pm 2.5$  \\
      	\midrule
            Jet energy resolution & $\pm 0.5$  & $\pm 0.2$ & $\pm 3.1$  & $\pm 1.9$ &  $\pm 0.3$ &$\pm 0.9$ &  $\pm 0.8$ & $\pm 3.4$ \\
        \midrule    
      Luminosity & $\pm 0.9$ &  $\pm 1.1$  &$\pm 1.3$  & $\pm 1.4$ &  $\pm 1.5$ & $\pm 1.5$ & $\pm 2.1$ &  $\pm 1.9$ \\
        \midrule
              Fake/non-prompt leptons      & $\pm 33$ & $\pm 18$ & $\pm 25$ & $\pm 23$ & $\pm 26 $ & $\pm 16$ & $\pm 1.5$& $\pm 3.8$ \\
       \midrule	
      Charge misID &$^{+5.9}_{-5.7}$ & $^{+9.3}_{-9.1}$ & $^{+5.4}_{-5.1}$ & $^{+7.4}_{-6.7}$ & $^{+5.0}_{-4.6}$ & $^{+8.7}_{-8.1}$ & $^{+9.0}_{-8.5}$ & $^{+11.0}_{-10.1}$ \\
            \hline

    \end{tabular}
\end{center}
    \caption{The largest systematic uncertainties (in \%) on the total background yield for the four-top/$b^\prime$/VLQ selection.}\label{tab:syst_bkg}

\end{table*}

 \begin{table*}[tt]
  \begin{center}
    \begin{tabular}{l|cccccccc}
      \hline\hline
      \multirow{2}{*}{Source} & \multicolumn{8}{c}{VLQ signal region number} \\
       \cline{2-9}
           & 0 & 1 & 2 & 3  & 4 & 5 & 6 & 7 \\
      \hline
      Jet energy scale &$^{+11.3}_{-9.0}$  & $^{+11.5}_{-6.3}$   & $^{+28.0}_{-17.3}$ & $^{+3.7}_{-2.1}$ &  $^{+5.4}_{-2.4}$ & $^{+3.9}_{-2.0}$ & $^{+4.5}_{-6.5}$ & $^{+6.6}_{-3.0}$  \\
      \midrule
       $b$-tagging efficiency &$^{+2.5}_{-3.0}$ & $^{+6.3}_{-6.1}$ & $^{+16.4}_{-15.9}$ &  $^{+3.1}_{-3.7}$ & $^{+3.4}_{-4.0}$ & $^{+7.4}_{-7.2}$ &  $^{+7.6}_{-7.4}$ & $^{+12.1}_{-11.9}$ \\
      \midrule
      Lepton ID efficiency & $\pm2.9$ & $\pm2.9$  &$\pm2.8$  & $\pm2.9$  & $^{+3.2}_{-3.1}$ &   $\pm2.9$ &  $^{+3.2}_{-3.1}$ &  $^{+3.0}_{-2.9}$ \\
      \midrule
            Jet energy resolution & $\pm 0.8$ & $\pm 2.5$ & $\pm 3.9$ & $\pm 0.3$ &$\pm 0.7$  &$\pm 0.7$  & $\pm 1.0$ &  $\pm 0.1$ \\
        \midrule
      Luminosity & $\pm 2.8$ & $\pm 2.8$ & $\pm 2.8$ & $\pm 2.8$  &$\pm 2.8$  &$\pm 2.8$ & $\pm 2.8$ & $\pm 2.8$ \\
       \hline
    \end{tabular}
\end{center}
    \caption{The largest systematic uncertainties (in \%) on the yield of a representative signal (600 \GeV{} vector-like $B$ pair production) for the four-top/$b^\prime$/VLQ selection.}\label{tab:syst_signal}
\end{table*}

Tables~\ref{tab:syst_bkg} and~\ref{tab:syst_signal}  show  the sources of systematic uncertainties that contribute more than 1\% uncertainty on the  expected background or signal yield for the four-top/$b^\prime$/VLQ selection.  These uncertainties have similar impact on the expected yields for the other signal models.
For the yields derived from simulation, the largest source of uncertainty is the cross-section calculation.  For the $\ttbar W/Z$ background, this is based on variations in the PDFs, variations of the renormalization and factorization mass scales (varied up and down by a factor of four from the nominal value of 172.5~\GeV{})~\cite{Campbell:2012dh}, and variations in the parameters controlling the initial-state radiation model, resulting in a 43\% uncertainty.  For other background contrbutions, varying the renormalization and factorization scales  results in uncertainties of  30\% for $WZ$ and $ZZ$ production, 25\% for $W^\pm W^\pm jj$ production, $+38\%/-26\%$ for $t\bar{t}W^+W^-$ production, and 10\% for $t\bar{t}H$, $tH$, $WH$, $ZH$,  $tWZ$,  $WWW$ and $ZWW$ production.  These uncertainties, applied to the event yields shown in tables~ \ref{tab:sec:sel:categorisation_yields} and \ref{tab:sec:sel:categorisation_yields_2}, result in the overall cross section uncertainties reported in table~\ref{tab:syst_bkg}. 
 The uncertainty on the integrated luminosity is 2.8\%~\cite{Aad:2013ucp}.  This uncertainty applies only to the backgrounds estimated from simulation, not to the data-driven estimates of the fake/non-prompt lepton and electron charge mismeasurement backgrounds, so the overall contribution of the luminosity uncertainty shown in table~\ref{tab:syst_bkg} is less than 2.8\%.   The largest detector-specific uncertainties  arise from the jet energy scale~\cite{Aad:2011he}, the $b$-tagging efficiency~\cite{ATLAS:2011qia}, and the lepton identification efficiency~\cite{ATLAS-CONF-2014-032,Aad:2014rra}.

Systematic uncertainties on the background contributions estimated from data are evaluated separately.  Six effects are considered when assigning the systematic uncertainty on the predicted yield of events from electron charge mismeasurement:  $i$) the statistical uncertainty on the probability for an electron to have its charge mismeasured, $ii$) the statistical uncertainty on the \pt-dependent scale factor, $iii$) the difference observed in simulated $Z$ boson events between the true charge mismeasurement rate and the rate obtained by applying the same method as is used for the data, $iv$)  the difference in the \pt-dependent scale factor when measured using different  $t\bar{t}$ simulated samples, $v$)  the variation in the result observed when the width of the $Z$ peak region is varied, and $vi$) the statistical uncertainty on the correction for the overlap in the measurement of charge misidentification and fake-electron background estimates.  The magnitudes of these effects depend on the event characteristics, so the uncertainty on the background from electron charge misidentification  varies from $23$ to $40$\% in the signal and control regions, as presented in Tables~\ref{ctrl:lowht_1b_ss_yield} and \ref{finalyieldsstop}-\ref{tab:sec:sel:categorisation_yields_2}. The expected yield of fake/non-prompt leptons is subject to uncertainties in the real and fake/non-prompt lepton efficiencies that arise from $i$) variations in the values of $r$ and $f$ when different control regions are used to measure them, $ii$) the small number of events in those control regions, and $iii$) the MC model used to subtract the real lepton contribution from the fake/non-prompt lepton control region. When assessing effect $i$, the following alternative control regions are used: for electrons, the alternative fake/non-prompt control region requires one loose electron and $\met < 20$~\GeV{}, while for muons, the alternative control region requires one loose muon, $m_{\rm T}(W)  < 20$ GeV and $\met+m_{\rm T}(W)  < 60$~\GeV{}.  In both cases the expected contribution from real leptons in the control region is subtracted using simulation. The alternative control regions for $r$ are formed by increasing the requirement on \met\ from $> 150$~\GeV{} to $> 175$~\GeV{} for electrons and by increasing the requirement on  $m_{\rm T}(W)$ from $> 100$~\GeV{} to $> 110$~\GeV{} for muons. 
 Effects $i)$--$iii)$ sum to a 70\% uncertainty on the predicted yield of fake/non-prompt leptons.
 
\section{Results}

\begin{table*}
\begin{center}
\begin{tabular}{l*{3}{r@{ $\pm$ }r@{ }l}}
\hline\hline
 & \multicolumn{3}{c}{SRtt$ee$} & \multicolumn{3}{c}{SRtt$e\mu$} & \multicolumn{3}{c}{SRtt$\mu\mu$}\\
\hline
$t\bar{t}W/Z$    & $0.58$ & $ 0.06$ & $\pm\ 0.25$ & $1.20$ & $ 0.09$ & $ \pm\ 0.53$ & $0.64$ & $ 0.07$ & $\pm\ 0.28$ \\
$t\bar{t}H$        & $0.05$ & $0.02$ & $\pm\ 0.01$ & $0.12$ & $0.02$ & $\pm\ 0.02$ & $0.03$ & $0.01$ & $\pm\ 0.01$\\
Dibosons              & $0.27$ & $0.14$ & $ \pm\ 0.07$ & $0.38$ & $0.09$ & $ \pm\ 0.10$ & $0.19$ & $0.12$ & $ \pm\ 0.04$\\
Fake/Non-prompt & $0.87$ & $0.79$ & $ \pm\ 0.61$ & $2.92$ & $1.27$ & $ \pm\ 2.04$ & $0.34$ & $0.29$ & $ \pm\ 0.24$\\
Q mis-Id & $2.66$ & $0.25$ & $^{+1.04}_{-0.96}$ & $2.79$ & $0.26$ & $^{+0.96}_{-0.92}$ & \multicolumn{3}{c}{---} \\
Other bkg.  &$ 0.01$ & $0.08$ & $\pm\ 0.00$ & $0.05$ & $0.08$ & $ \pm\ 0.01$ & $0.12$ & $0.11$ & $\pm\ 0.03$ \\
\hline
Total bkg. & $4.5 $ & $ 0.8$ & $ ^{+1.3}_{-1.2} $ & $7.5 $ & $ 1.3$ & $ \pm\ 2.5 $ & $1.3 $ & $ 0.3$ & $ \pm\ 0.4 $\\
\hline
    Data & \multicolumn{3}{c}{6} &  \multicolumn{3}{c}{5} &  \multicolumn{3}{c}{2} \\
 \hline   
    $p$-value & \multicolumn{3}{c}{0.38} & \multicolumn{3}{c}{0.84} & \multicolumn{3}{c}{0.45} \\
      \hline
\end{tabular}
\end{center}
\caption{Observed and expected numbers of events  with 
        statistical (first) and systematic (second) uncertainties for the positively charged top pair signal
        selection. The $p$-values for agreement between the observed yield and the expected background in each signal region are reported. }\label{finalyieldsstop}
\end{table*}

	\begin{table}
	\begin{center}
		\begin{tabular}{l*{3}{r@{ $\pm$ }r@{ }l}}
		\hline\hline
		 & \multicolumn{3}{c}{SRVLQ0} & \multicolumn{3}{c}{SRVLQ1/SR4t0} & \multicolumn{3}{c}{SRVLQ2/SR4t1} \\
		\hline
		$t\bar{t}W/Z$    & $16.2$ & $ 0.3$ & $\pm\ 7.0$ & $12.6$ & $ 0.3$ & $ \pm\ 5.4$ & $1.24$ & $ 0.09$ & $\pm\ 0.53$ \\
		$t\bar{t}H$        & $2.5$ & $0.1$ & $\pm\ 0.3$ & $1.8$ & $0.1$ & $\pm\ 0.2$ & $ 0.26$ & $0.03$ & $\pm\ 0.05$ \\ 
		Dibosons                   & $11.2$ & $0.6$  & $ \pm\ 2.8$    & $0.95$ & $0.19$ & $\pm\ 0.25$ & $0.07$ & $0.12$ & $\pm\ 0.05$ \\
		Fake/Non-prompt & $42.1$ & $5.4$ & $ \pm\ 24.6$ & $8.61$ & $2.34$ & $ \pm\ 5.02$ & $1.17$ & $0.82$ & $ \pm\ 0.68$ \\
		Q mis-Id & $20.8$ & $0.7$ & $ \pm\ 5.2$ & $15.1$ & $0.6$ & $ \pm\ 3.5$ & $0.74$ & $0.11$ & $ \pm\ 0.18$ \\
		Other bkg. &$1.76$ & $0.13$ & $\pm\ 0.17$ & $0.75$ & $0.04$ & $\pm\ 0.10$ & $0.10$ & $0.08$ & $\pm\ 0.03$ \\
		\hline
		Total bkg. & $94.5 $ & $ 5.4$ & $ \pm\ 24.9 $ & $40.0 $ & $ 2.4$ & $ \pm\ 7.3 $ & $3.6 $ & $ 0.9$ & $ \pm\ 0.8$ \\
		\hline
Data & \multicolumn{3}{c}{$107$} & \multicolumn{3}{c}{$54$} & \multicolumn{3}{c}{$6$} \\		
\hline
$p$-value & \multicolumn{3}{c}{0.36}	& \multicolumn{3}{c}{0.12} 	& \multicolumn{3}{c}{0.24}	\\ \hline
\\
 \end{tabular}
 \\
 \begin{tabular}{l*{2}{r@{ $\pm$ }r@{ }l}}
		\hline\hline
		&  \multicolumn{3}{c}{SRVLQ3} & \multicolumn{3}{c}{SRVLQ4}\\
		\hline
		$t\bar{t}W/Z$    & $2.07$ & $ 0.10$ & $\pm\ 0.89$ & $3.14$ & $ 0.13$ & $ \pm\ 1.35$ \\
		$t\bar{t}H$        & $0.40$ & $0.04$ & $\pm\ 0.07$ &  $0.57$ & $0.05$ & $\pm\ 0.07$ \\
		Dibosons                  & $2.36$ & $0.29$ & $\pm\ 0.61$ & $2.03$ & $ 0.25$ & $\pm\ 0.49$ \\ 
		Fake/Non-prompt & $3.09$ & $1.29$ & $ \pm\ 1.80$ & $4.24$ & $1.59$ & $ \pm\ 2.47$\\
		Q mis-Id & $1.72$ & $0.22$ & $ \pm\ 0.63$ & $1.45$ & $0.17$ & $ \pm\ 0.52$\\
		Other bkg. & $0.22$ & $0.08$ & $\pm\ 0.03$ & $0.41$ & $0.10$ & $\pm\ 0.06$ \\
		\hline
		Total bkg. &  $9.87 $ & $ 1.35$ & $ \pm\ 2.10 $ & $11.9 $ & $ 1.6$ & $ \pm\ 2.8 $\\
		\hline
Data  & \multicolumn{3}{c}{$7$} & \multicolumn{3}{c}{$10$}\\		
\hline
$p$-value 	& 	\multicolumn{3}{c}{0.83}	& \multicolumn{3}{c}{0.71} \\
      	\hline

		\end{tabular}
	\end{center}
		\caption{Observed and expected numbers of events with statistical (first) and systematic (second) uncertainties for five of the signal regions defined for VLQ, chiral $b^\prime$-quark and four-top-quark production searches.  The $p$-values for agreement between the observed yield and the expected background in each signal region are reported.\label{tab:sec:sel:categorisation_yields}}
	\end{table}
	
	\begin{table}
	\begin{center}		
		\begin{tabular}{l*{3}{r@{ $\pm$ }r@{ }l}}
		\hline\hline
		 & \multicolumn{3}{c}{SRVLQ5/SR4t2} & \multicolumn{3}{c}{SRVLQ6/SR4t3} & \multicolumn{3}{c}{SRVLQ7/SR4t4}\\
		\hline
		$t\bar{t}W/Z$   & $1.87$ & $ 0.09$ & $\pm\ 0.80$ & $2.46$ & $0.11$ & $\pm\ 1.06$ & $0.57$ & $ 0.05$ & $\pm\ 0.25$ \\
		$t\bar{t}H$       & $0.31$ & $0.04$ & $\pm\ 0.05$ & $0.44$ & $0.04$ & $\pm\ 0.06$ & $0.08$ & $0.02$ & $\pm\ 0.02$ \\	
		Dibosons                  & $0.33$ & $0.14$ & $\pm\ 0.10$ & $0.04$ & $0.12$ & $\pm\ 0.03$ & $0.00$ & $0.12$ & $\pm\ 0.00$ \\
		Fake/Non-prompt & $1.03$ & $0.97$ & $ \pm\ 0.60$ & $0.00$ & $1.02$ & $ \pm\ 0.28$ & $0.04$ & $0.83$ & $ \pm\ 0.24$\\
		Q mis-Id & $1.17$ & $0.16$ & $ \pm\ 0.38$ & $1.09$ & $0.14$ & $ \pm\ 0.34$ & $0.30$ & $0.09$ & $ \pm\ 0.10$\\
		Other bkg. & $0.16$ & $0.08$ & $\pm\ 0.02$ & $0.23$ & $0.08$ & $\pm\ 0.05$ & $0.14$ & $0.08$ & $\pm\ 0.08$ \\
		\hline
		Total bkg. & $4.9 $ & $ 1.0$ & $ \pm\ 1.0 $ & $4.3 $ & $ 1.1$ & $ \pm\ 1.1 $ & $1.1 $ & $ 0.9$ & $ \pm\ 0.4 $\\
		\hline
		Data & \multicolumn{3}{c}{$6$} & \multicolumn{3}{c}{$12$} & \multicolumn{3}{c}{$6$}\\
		\hline
	$p$-value & \multicolumn{3}{c}{0.46} & \multicolumn{3}{c}{0.029} & \multicolumn{3}{c}{0.036} \\

		\hline
		\end{tabular}
	\end{center}
		\caption{Observed and expected numbers of events with statistical (first) and systematic (second) uncertainties for three of the signal regions defined for VLQ, chiral $b^\prime$-quark and four-top-quark production searches. The $p$-values for agreement between the observed yield and the expected background in each signal region are reported. \label{tab:sec:sel:categorisation_yields_2}}
	\end{table}

 \begin{figure} 
        \begin{center}
                \includegraphics[width=\linewidth]{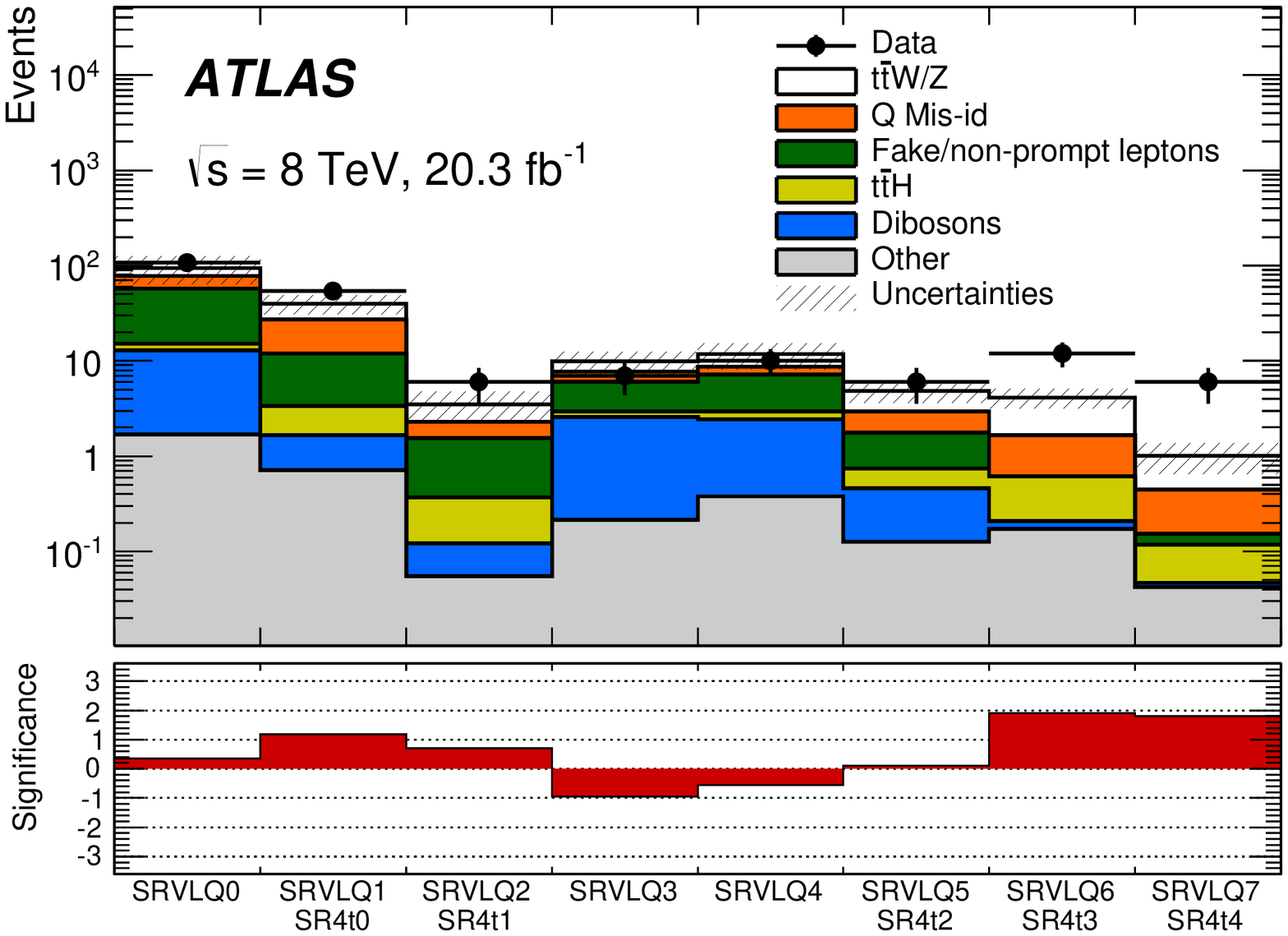}
                \caption{Expected background yields and observed data events in various signal regions. Uncertainties include both the statistical and systematic errors. The difference between data and expectations is 
                        quantified by the means of the significance, computed from the $p$-values in tables~\ref{tab:sec:sel:categorisation_yields} and~\ref{tab:sec:sel:categorisation_yields_2}.}
                \label{fig:yields_categorisation}
        \end{center}
\end{figure}

The observed yields for each signal selection are given in tables~\ref{finalyieldsstop}--\ref{tab:sec:sel:categorisation_yields_2} and figure~\ref{fig:yields_categorisation}.  The $CL_s$ method~\cite{Junk:1999kv,0954-3899-28-10-313} is used to assess the consistency between the observed yields and each potential BSM physics signal, where the log-likelihood ratio $L_{\mathrm{R}}$ is used as the test statistic.  For each model, $L_{\mathrm{R}}$ is defined as 
\begin{equation}
L_{\mathrm{R}} = -2\log\frac{L_{s+b}}{L_b}
\end{equation}
where $L_{s+b}$ ($L_b$) is the Poisson likelihood to observe the data under the signal-plus-background (back\-ground-only) hypothesis. Pseudo-experiments are generated under each hypothesis, taking into account statistical fluctuations of the total predictions according to Poisson statistics, as well as Gaussian fluctuations in the signal and background expectations describing the effect of systematic uncertainties.  The quantities $CL_{s+b}$ and $CL_b$ are defined as the fractions of signal plus background  and background-only pseudo-experiments with $L_{\mathrm{R}}$ larger than the observed value.  Signal cross sections for which $CL_s = CL_{s+b}/CL_b < 0.05$ are deemed excluded at the 95\% CL.  Expected limits assuming the absence of signal are also computed; these are the basis for assessing the intrinsic sensitivity of the analysis.  

In the signal regions defined for searching for positively charged top quark pair production, the observed yields agree well with the expectation from background.  The resulting limits on the cross section for this process are shown in table~\ref{tab:limitObs:sstop} in both the contact interaction and Higgs-like FCNC models. For the special case of the 125~\GeV{} Higgs boson, the limit on the cross section leads to a limit of BR($t\rightarrow uH)<0.01$.  The results can also be expressed as limits on the parameters defined in equations~\ref{eq:tt_FCNC} and~\ref{eq:lagrangiantt}:
\begin{figure}
  \begin{center}
    \subfloat[]{\includegraphics[width=.45\textwidth]{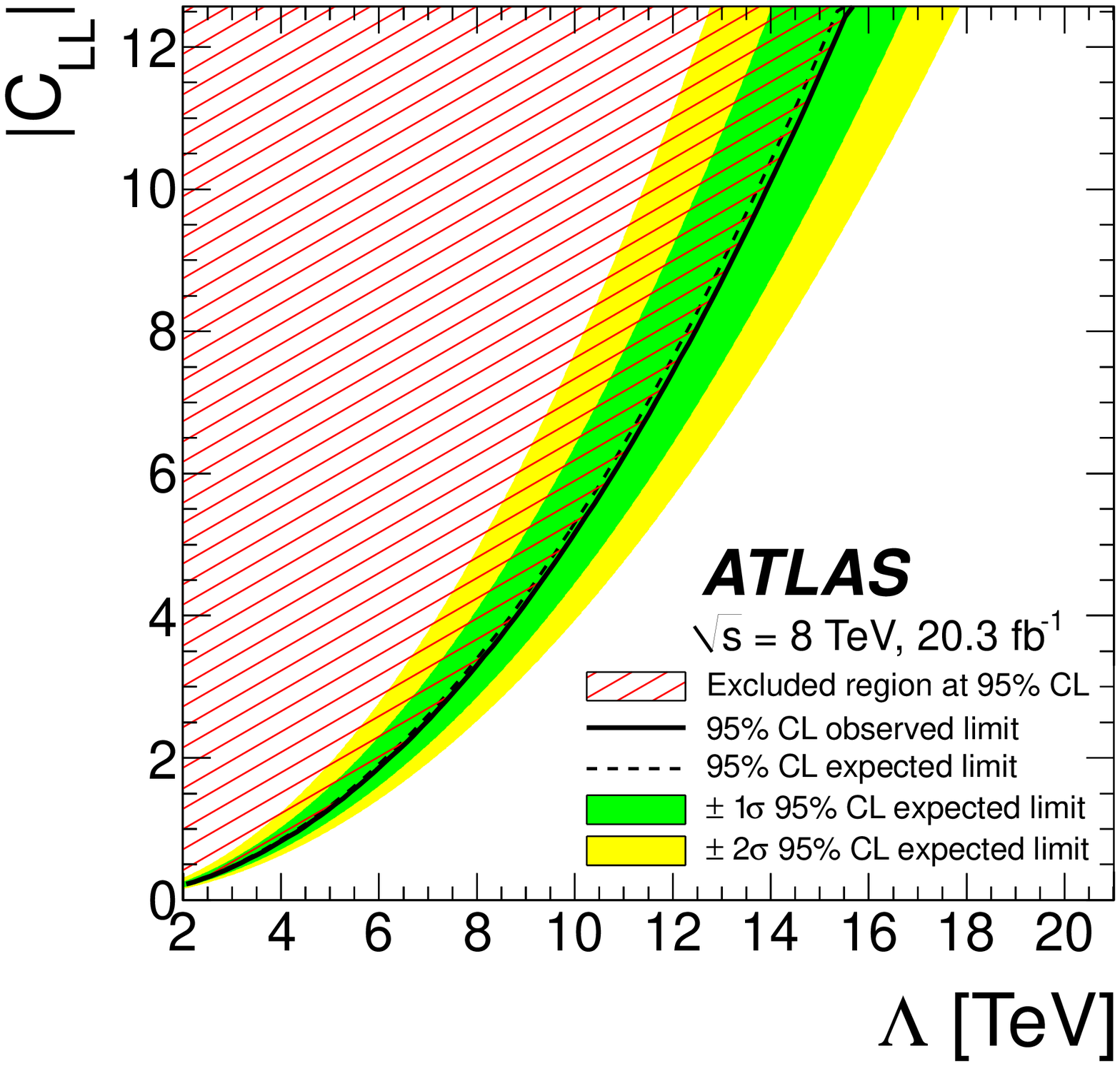}\label{fig:sstopLL:limit}}
    \subfloat[]{\includegraphics[width=.45\textwidth]{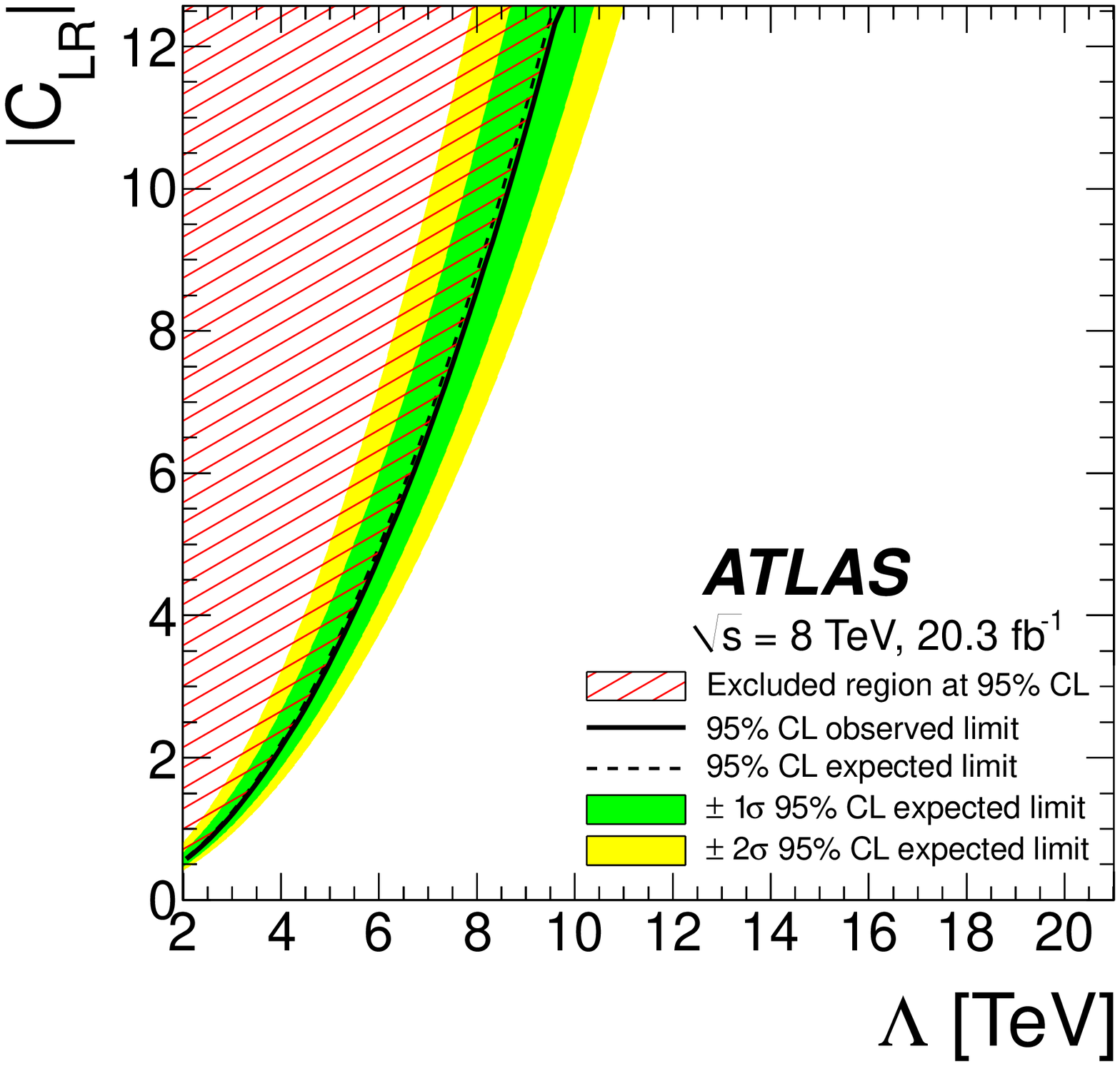}\label{fig:sstopLR:limit}}\\
    \subfloat[]{\includegraphics[width=.45\textwidth]{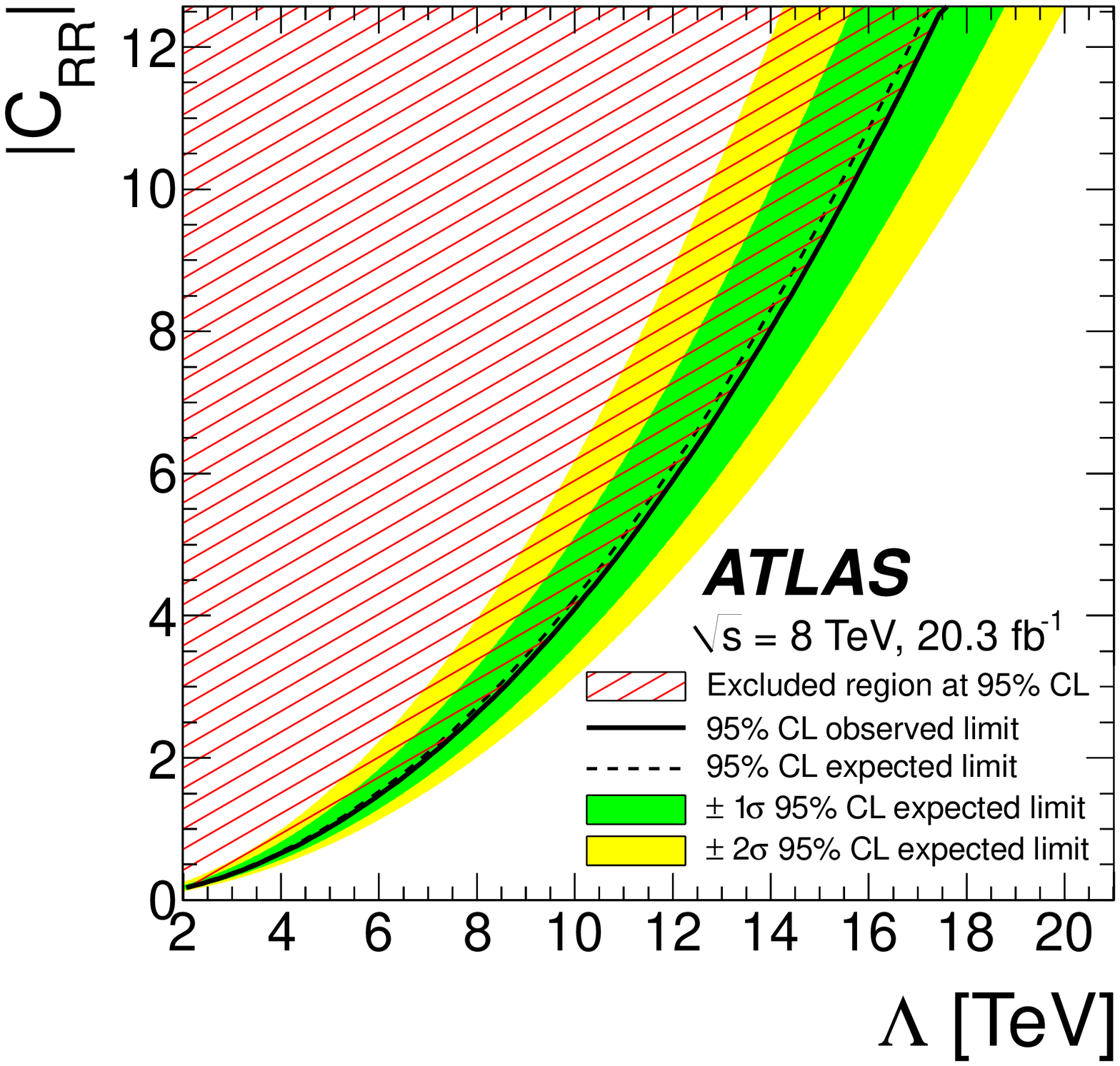}\label{fig:sstopRR:limit}}
    \subfloat[]{\includegraphics[width=.45\textwidth]{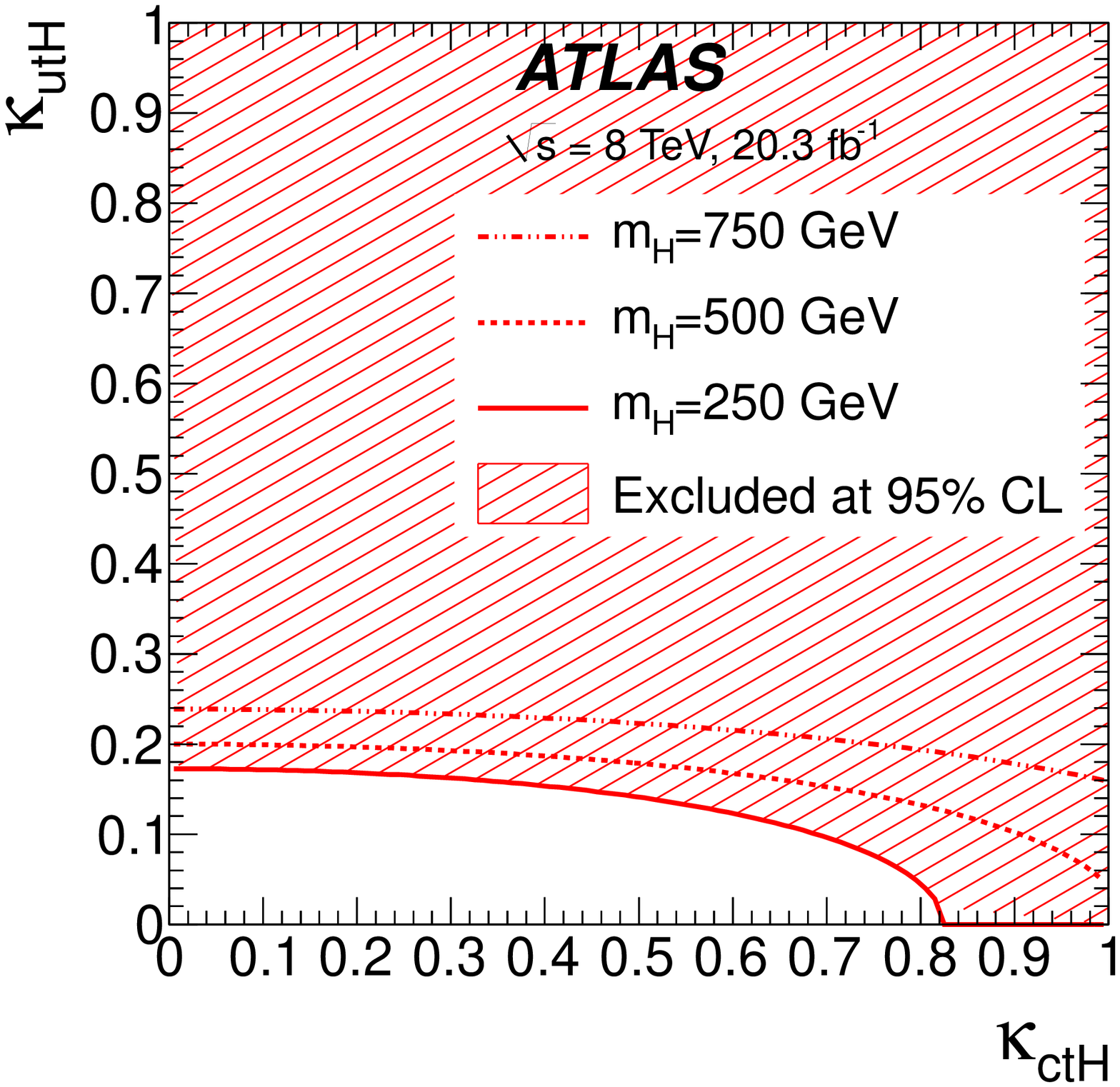}\label{fig:sstopFCNC:limit}}
  \end{center}
  \caption{Expected and observed limits on the coupling constant $|C|$ in the contact interaction model for same-sign top quark pair production as a function
    of the BSM physics energy scale $\Lambda$ for the (a) left--left, 
    (b) left--right and (c) right--right terms.
    Plot (d) shows the observed limits on the Higgs-like 
    exchange with FCNC coupling in the plane ($\kappa_{utH}$,$\kappa_{ctH}$) for
    three different hypotheses on the mass of the heavy Higgs-like particle.}
  \label{fig:limitObs:sstop}
\end{figure}
for each chirality, the upper limit on $C$ as a function of $\Lambda$ is shown in figure~\ref{fig:limitObs:sstop}; the same figure also shows the limits on $\kappa_{utH}$ and  $\kappa_{ctH}$  in the Higgs-like FCNC model.

\begin{table}
  \begin{center}
     \begin{tabular}{l|c|c|c}
      \hline\hline
      Model & \multicolumn{2}{c|}{$\sigma(pp\rightarrow tt)$ [fb]} & Coupling const. \\
      & Exp. & Obs. & Observed \\
      \hline
      \multicolumn{3}{c|}{Contact interaction model} & $|C|/\Lambda^2$ [\TeV$^{-2}$] \\
      \hline
      Left--left & 64 & 62 & 0.053 \\
      Left--right & 53 & 51 & 0.137 \\
      Right--right & 40 & 38 & 0.042 \\
      \hline
      \multicolumn{3}{c|}{Higgs-like FCNC model} & $\kappa_{utH}$ or $\kappa_{ctH}$ \\
      \hline
      $uu\ra tt$ ($m_H=125$~\GeV{}) & 37 & 35 & 0.16 \\
      $uu\ra tt$ ($m_H=250$~\GeV{}) & 21 & 20 & 0.17 \\
      $uu\ra tt$ ($m_H=500$~\GeV{}) & 12 & 11 & 0.20 \\
      $uu\ra tt$ ($m_H=750$~\GeV{}) & 9.3 & 8.4 & 0.24 \\
      $cc\ra tt$ ($m_H=250$~\GeV{}) & 71 & 69 & 0.81 \\
      $cc\ra tt$ ($m_H=500$~\GeV{}) & 37 & 35 & 1.02 \\
      $cc\ra tt$ ($m_H=750$~\GeV{}) & 28 & 27 & 1.29 \\
      \hline
    \end{tabular}
  \end{center}
     \caption{Observed and expected 95\% CL upper limits on the cross section for  same-sign top-quark
      production, and on the coupling constants.}\label{tab:limitObs:sstop}

\end{table}

\begin{figure}
        \begin{center}
                \subfloat[]{\includegraphics[width=0.5\linewidth]{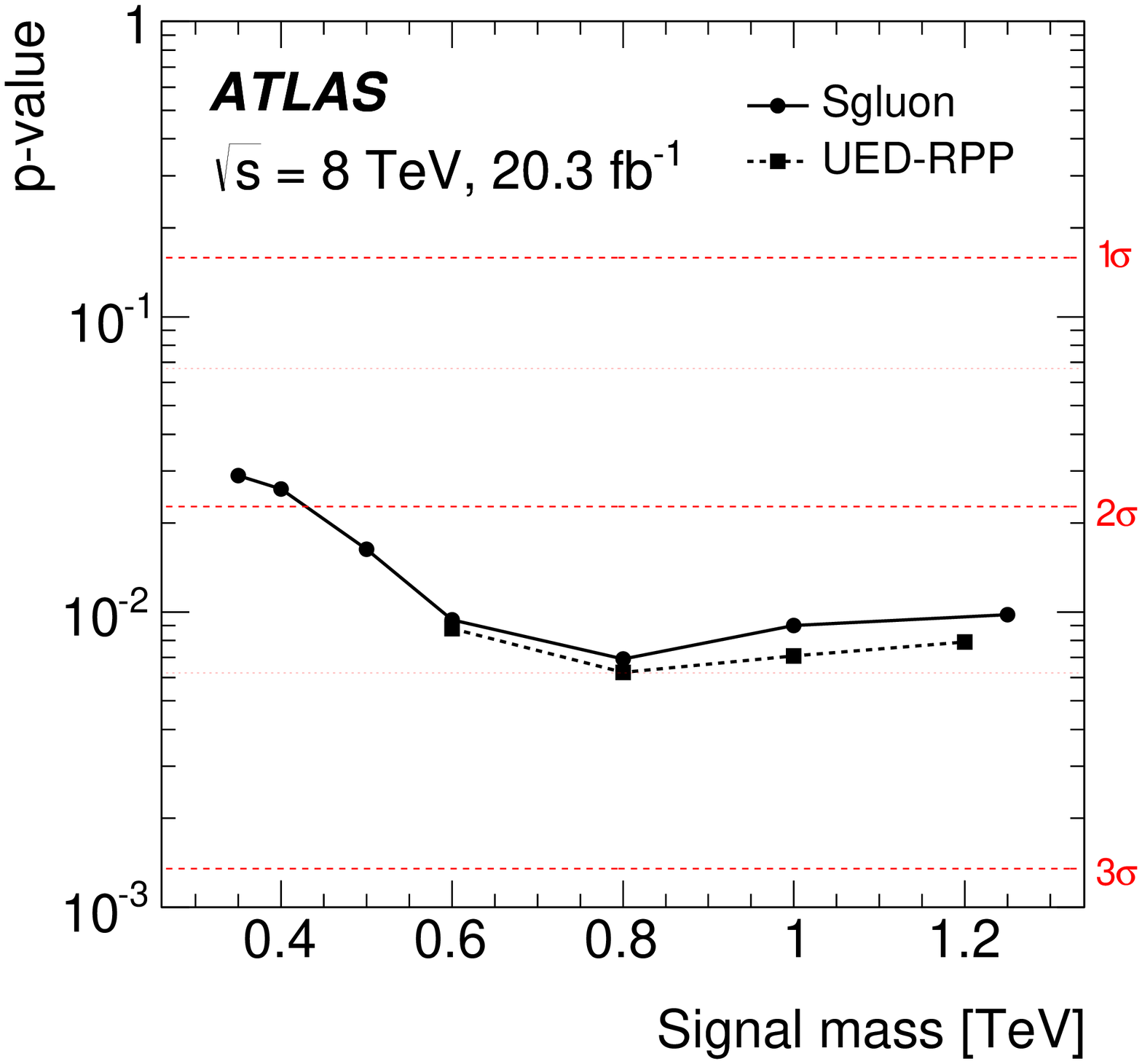}
                 \label{fig:4top_pvalues}}
               \subfloat[]{ \includegraphics[width=0.5\linewidth]{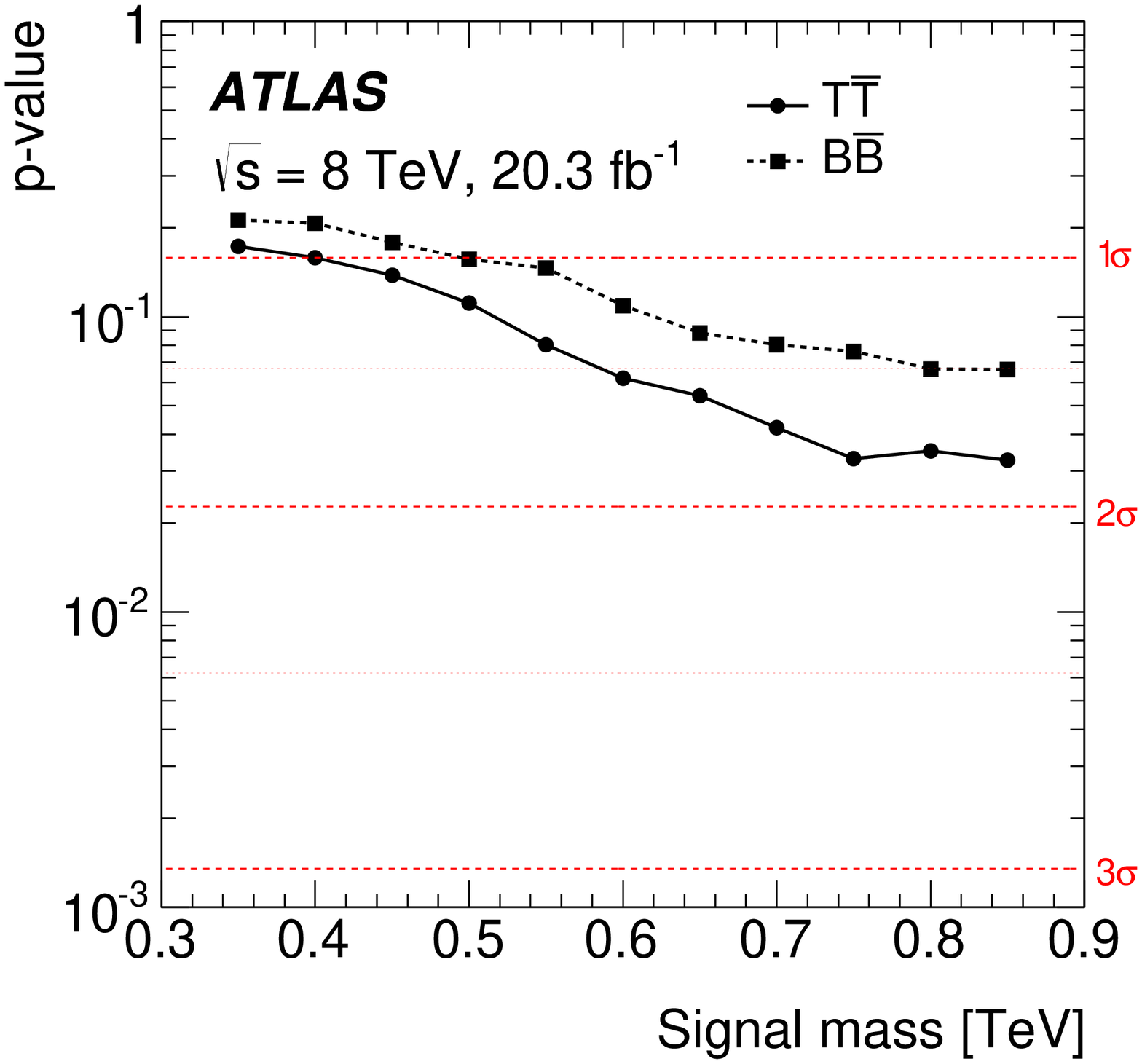}
                 \label{fig:vlq_pvalues}}
                 \caption{(a) Probability for the data in the four-top-quark signal regions (SR4t0--SR4t4) to be consistent with a zero cross section for anomalous four-top-quark production under two model scenarios, as a function of the characteristic mass scale of the models.  (b) Probability for the data in the VLQ/$b^\prime$-quark signal regions (SRVLQ0--SRVLQ7) to be consistent with a zero cross section for various heavy quarks, as a function of the quark mass.}

        \end{center}
\end{figure}

\begin{figure}[t]
  \begin{center}
  \subfloat[]{\includegraphics[width=0.5\linewidth]{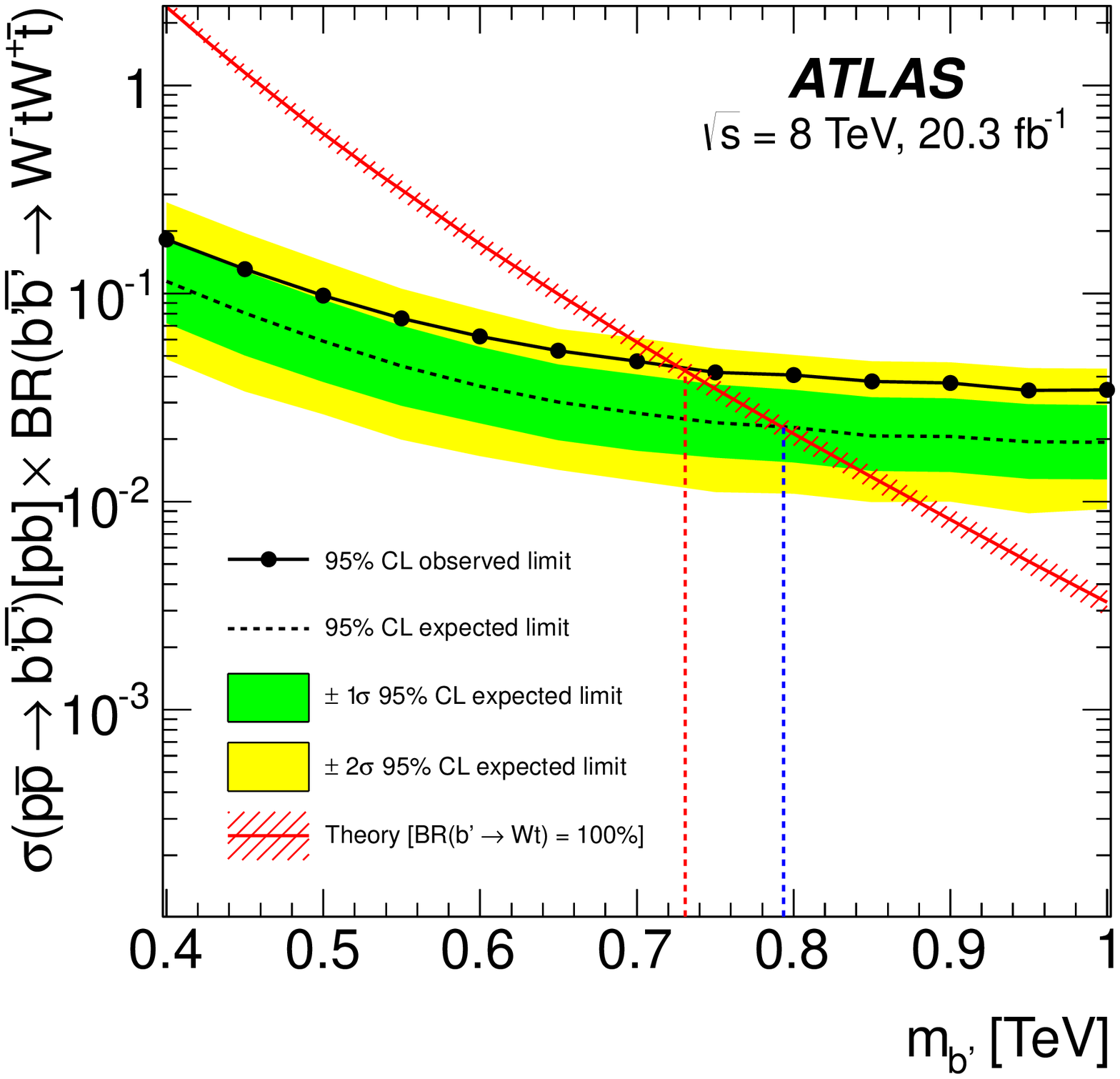}
    \label{limit:bprime}} \\
    \subfloat[]{\includegraphics[scale=0.38]{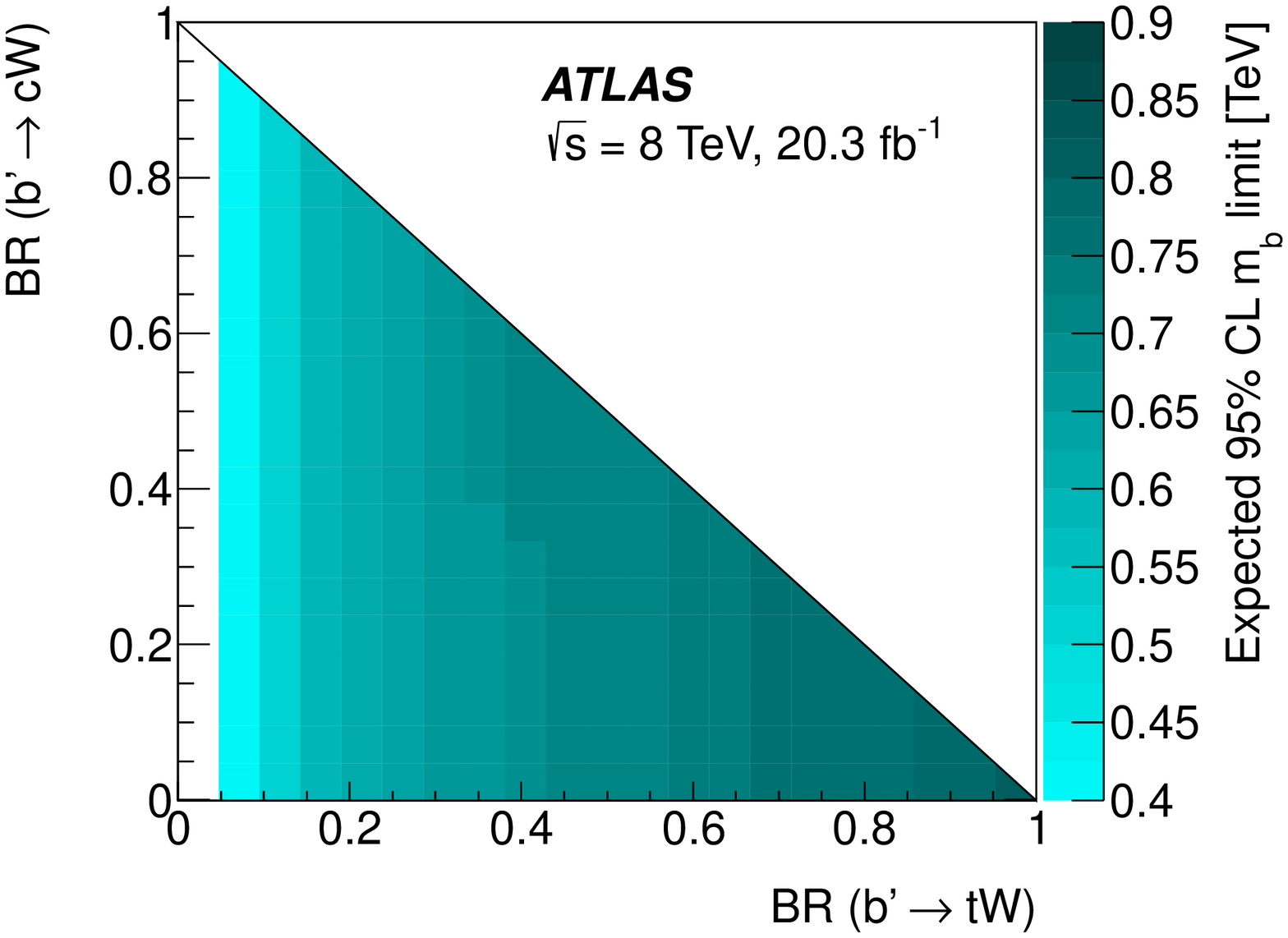}\label{fig:bprime2Dlimitgradient_exp}}
    \subfloat[]{\includegraphics[scale=0.38]{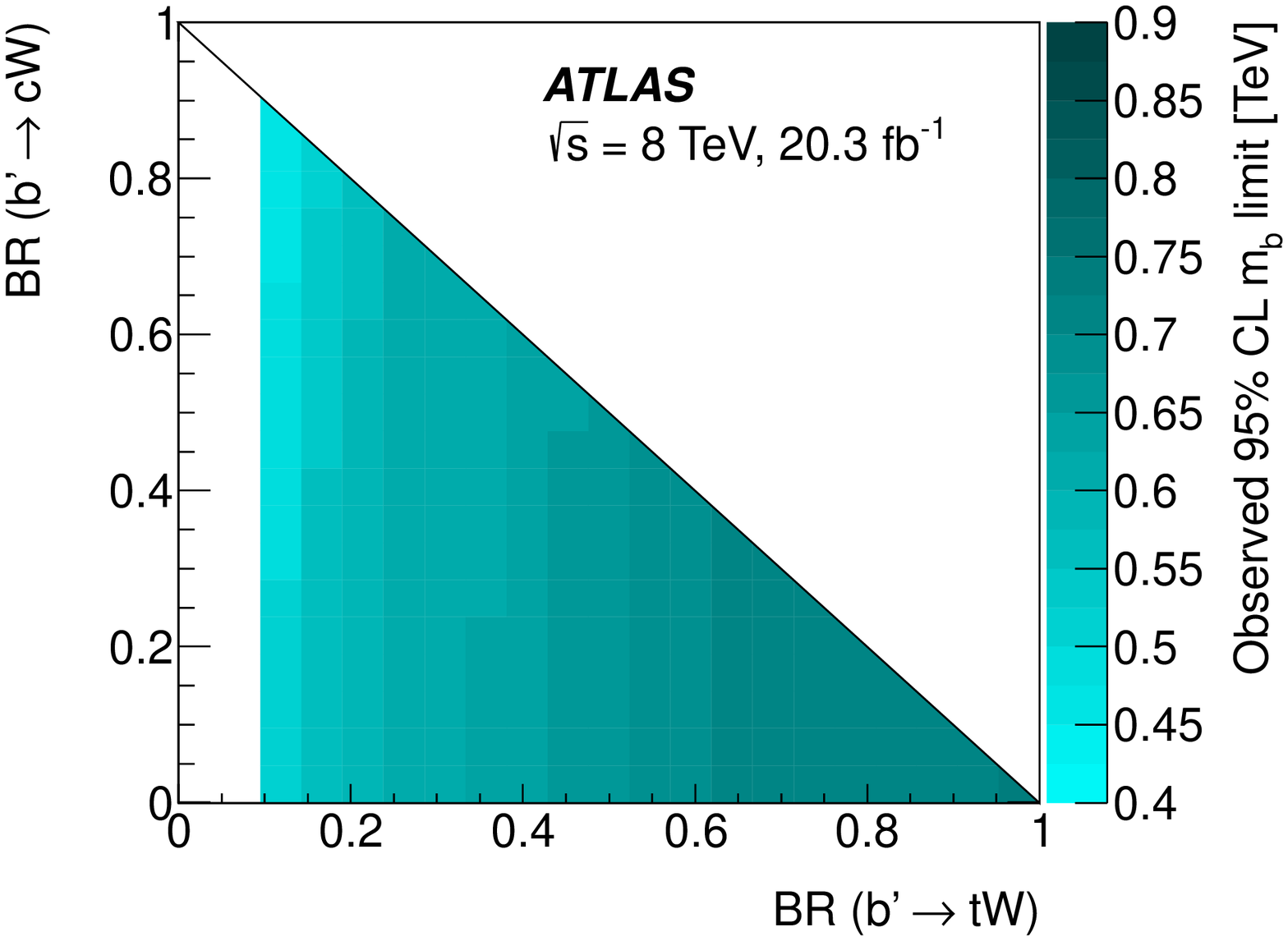}\label{fig:bprime2Dlimitgradient_obs}}
    \caption{Limits on $b^\prime$-quark pair production: (a) Expected and observed upper limits on the $b^\prime$  pair-production cross section times the branching ratio for 
      $b^\prime\bar{b^\prime} \ra W^-tW^+\bar{t}$, as a function of the $b^\prime$-quark mass (the vertical dashed lines indicate the expected and observed limits on the $b^\prime$-quark mass, and the shaded band around the theory cross section indicates the total uncertainty on the calculation); (b) expected and (c) observed exclusion limits on the $b^\prime$-quark mass
       as a function of the assumed branching
      ratios into $c$- and $t$-quarks. 
    }\label{fig:bprime2Dlimitgradient}
  \end{center}
\end{figure}

 In contrast to the same-sign top signal regions, some of the signal regions defined for VLQ, $b^\prime$-quark, and four-top-quark production exhibit an excess over expected background.  The excess is largest in the subset of the signal regions used for the four-top-quark search, where at least two $b$-tagged jets are required.  While it is still of interest to limit the set of models consistent with the data as described above, it is also important in this case to assess the consistency of the data with the background-only hypothesis.  This is done by computing $p \equiv 1-CL_b$.   The resulting $p$-values depend on the signal model and the signal regions considered, as shown in figures~\ref{fig:4top_pvalues} and~\ref{fig:vlq_pvalues}.  For signals where all eight signal regions are considered (as is the case for VLQ and $b^\prime$ models), the significance is above one standard deviation but less than two.  For signals for which only SR4t0--SR4t4 are considered (as is the case for four-top-quark production models) the significance reaches 2.5 standard deviations.  Several checks (detailed in section~\ref{sec:crosschecks}) of the background estimates were performed.  Some features of the events in the signal regions that exhibit the most significant excesses (SR4t3 and SR4t4) are presented in section~\ref{sec:eventfeatures}.

  The excess is not significant enough to support a claim of BSM physics.   Therefore 95\% CL limits (upper limits on cross sections, or lower limits on masses) relevant for each model are calculated.  The observed excess causes these limits to be less restrictive than expected for the background-only hypothesis.
  The data place 95\% CL upper limits on the $b^\prime$-quark pair production cross section that vary with the mass of the $b^\prime$-quark.  Limits obtained assuming a 100\% branching ratio to $Wt$ are presented in figure~\ref{limit:bprime} (expected mass limit at 0.79 \TeV{}, observed at 0.73 \TeV{}), and limits where decays to $u$- or $c$-quarks are also considered are shown in figures~\ref{fig:bprime2Dlimitgradient_exp} and~\ref{fig:bprime2Dlimitgradient_obs}.

Limits on the VLQ pair-production cross section, assuming  the branching fractions to $W$, $Z$, and $H$ modes prescribed by the singlet model, are shown in figure~\ref{fig:limitObs:VLQ}. Comparison with the calculated  cross-section results in  lower limits on the $B$-quark mass of  $0.62$~\TeV{} and on the $T$ quark mass of $0.59$~\TeV{} at 95\% CL.  The expected limits in the absence of a signal contribution are 
0.69~\TeV{} for the $B$-quark mass and 0.66~\TeV{} for the $T$-quark mass.  If the three branching fractions are allowed to vary independently (subject to the constraint that they sum to one), the data can be interpreted as excluding at 95\% CL some of the possible sets of branching ratios for a given $B$- or $T$-quark mass.  These exclusions are shown in figures~\ref{fig:BBS_2DLimits} and \ref{fig:TTS_2DLimits}.

\begin{figure} 
  \begin{center}
     \subfloat[]{\includegraphics[width=.45\textwidth]{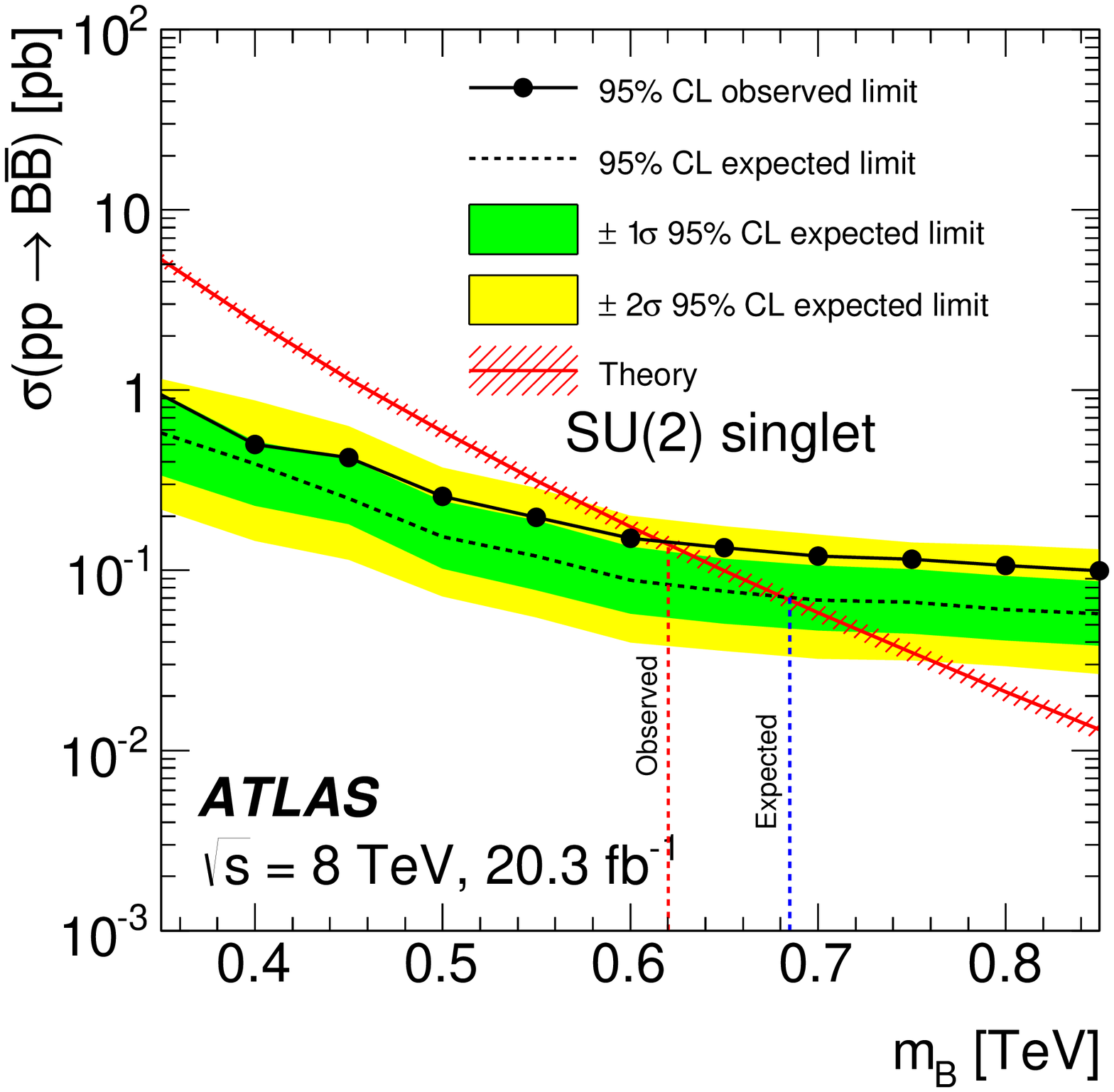}}
   \subfloat[]{\includegraphics[width=.45\textwidth]{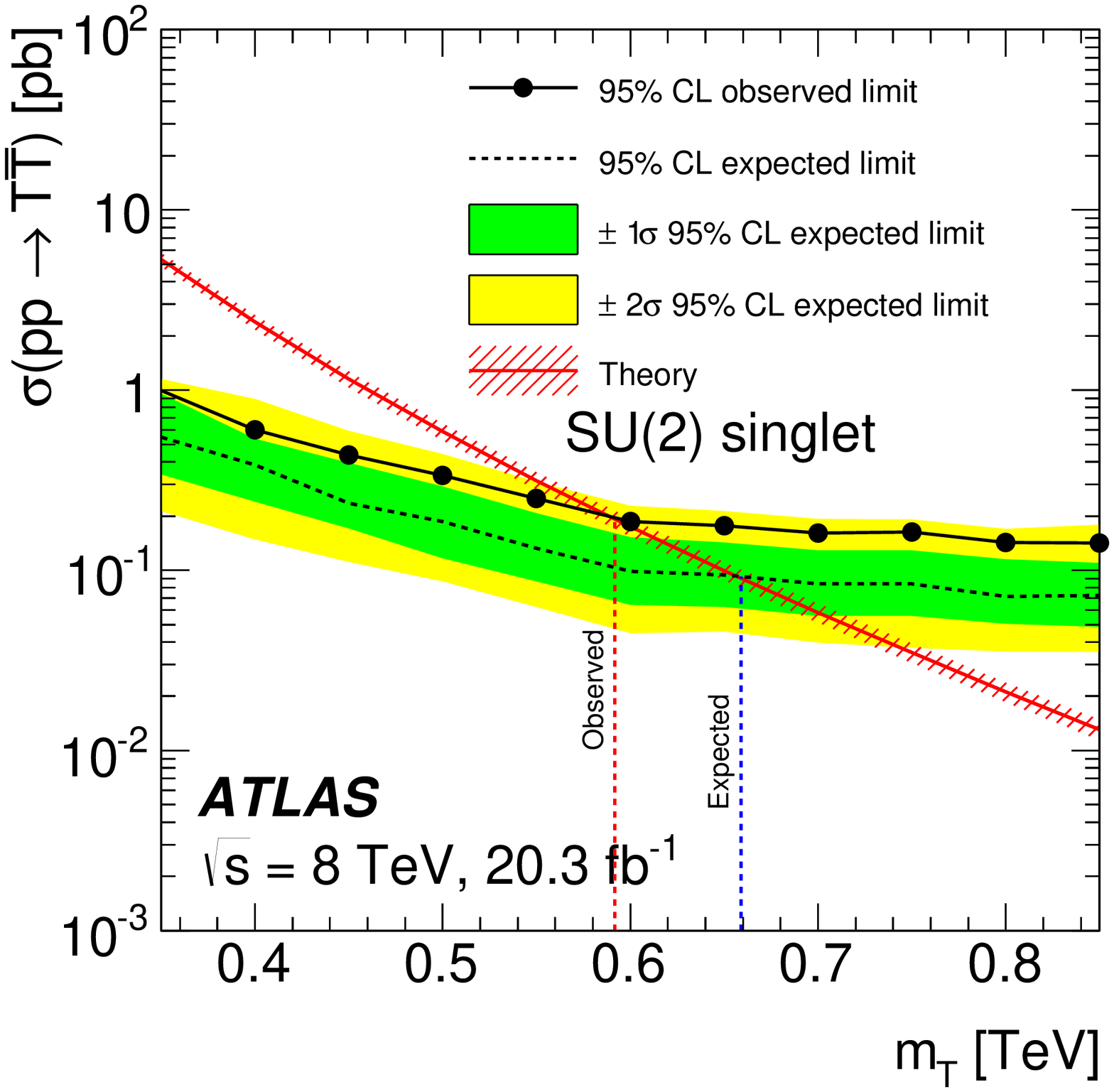}}
    \caption{Expected and observed limits on the pair production cross section as a function of mass for  
      (a) vector-like $B$ and (b) vector-like $T$ quarks.  The vertical dashed lines indicate the expected and observed limits on the vector-like quark mass.  These limits assume branching ratios given by the model where the $B$ and $T$ quarks exist as singlets~\cite{AguilarSaavedra:2009es}.  The shaded band around the theory cross section indicates the total uncertainty on the calculation.}
     \label{fig:limitObs:VLQ}
  \end{center}
\end{figure}

\begin{figure} 
  \begin{center}
     \subfloat[]{\includegraphics[width=.45\columnwidth]{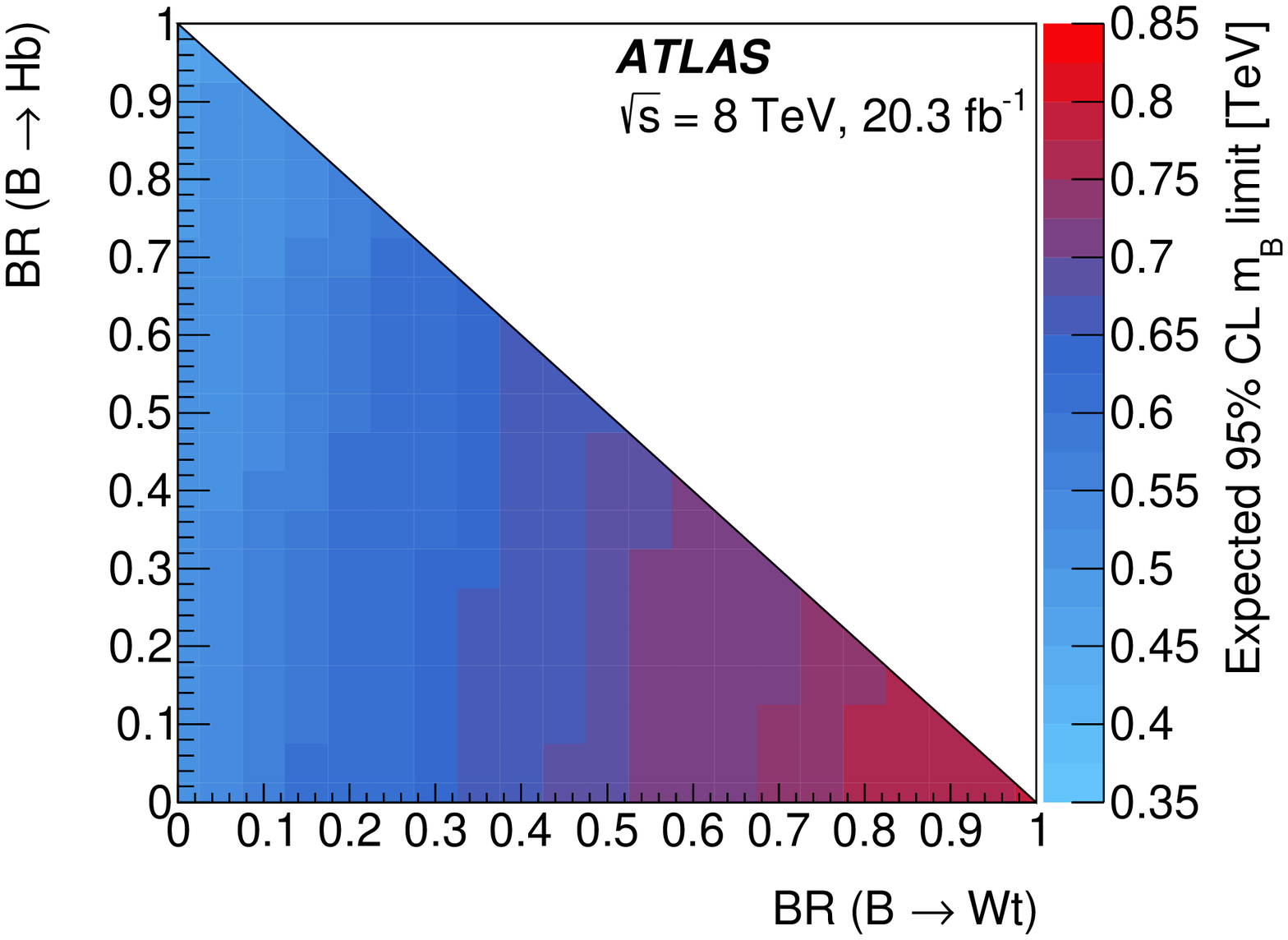}}
     \subfloat[]{\includegraphics[width=.45\columnwidth]{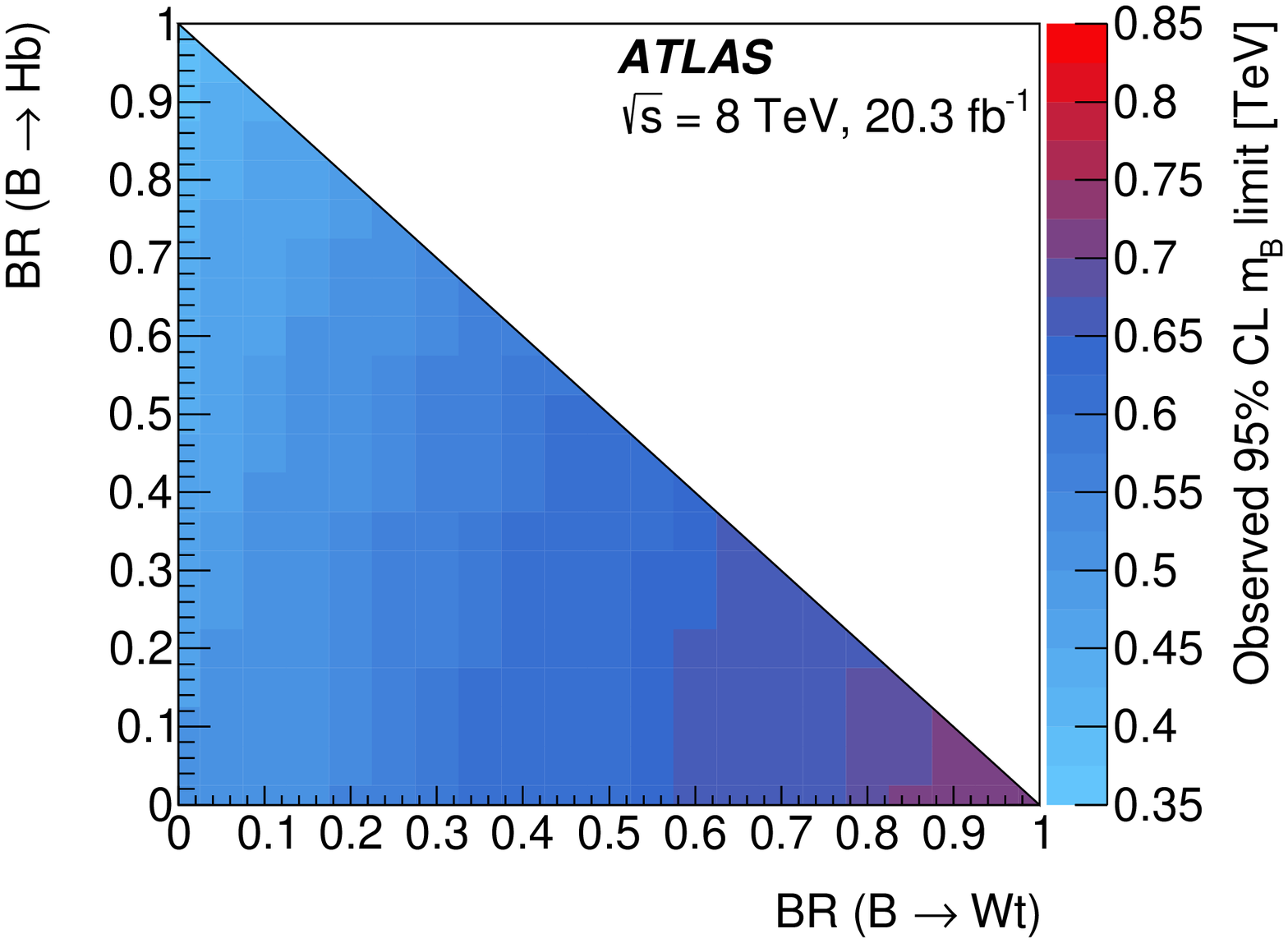}}
    \caption{Expected (a) and observed (b) vector-like $B$ quark mass hypotheses excluded at 95\% CL as a function of the assumed branching ratios.}
\label{fig:BBS_2DLimits}
  \end{center}
\end{figure}

\begin{figure} 
  \begin{center}
   \subfloat[]{\includegraphics[width=.45\columnwidth]{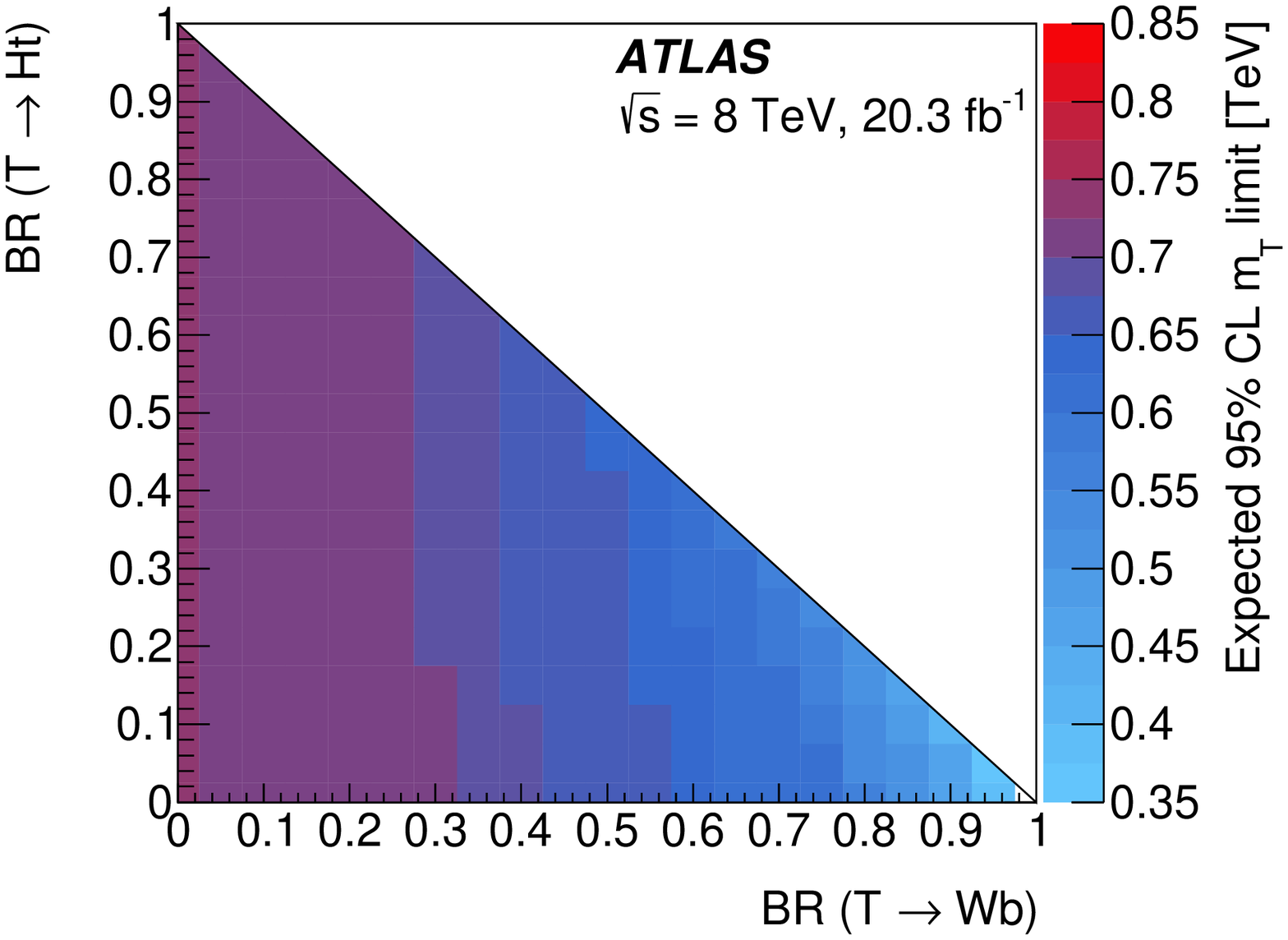}}
   \subfloat[]{\includegraphics[width=.45\columnwidth]{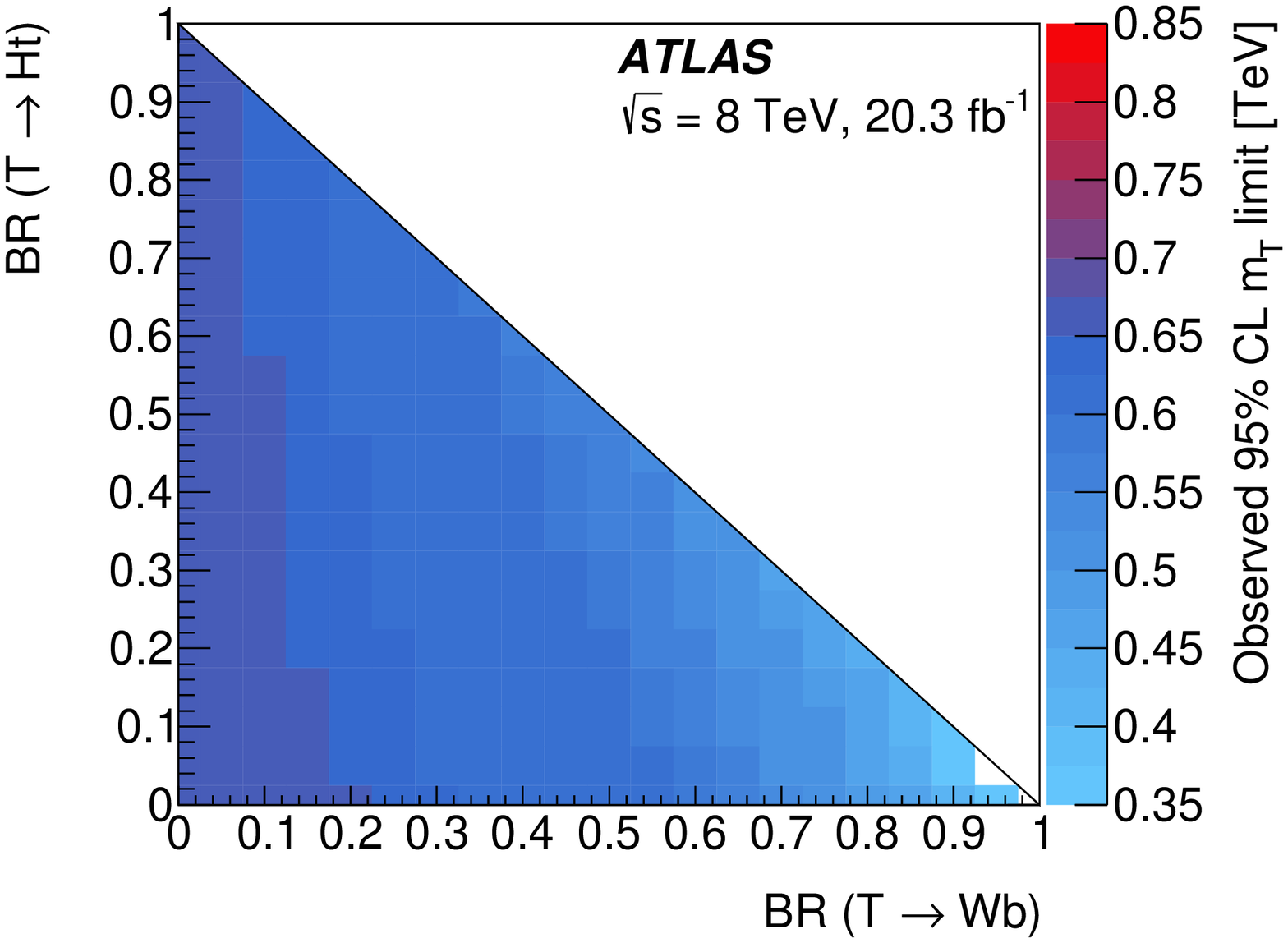}}
    \caption{Expected (a) and observed (b) vector-like
 $T$ quark mass hypotheses excluded at 95\% CL as a function of the assumed branching ratios.}
\label{fig:TTS_2DLimits}
  \end{center}
\end{figure}

Limits on $T_{5/3}$ production are set for  pair production only, and for the sum of pair and single production for two different values of the coupling $\lambda$  of the $T_{5/3}$ to $Wt$ ($\lambda = 0.5$ and 1.0)~\cite{Contino:2008hi}.  This coupling is related to the mixing parameter $g^{*}$ used by the model in refs.~\cite{XTVLQ,Buchkremer:2013bha}:  $\lambda = m_{T_{5/3}} gg^{*}/\ m_{W}\sqrt{2}$.  The pair-production limits are shown in figure~\ref{fig:T53PPlimit}, and correspond to a mass limit of 0.74 \TeV{} (0.81 \TeV{} expected). The limits on pair plus single production with $\lambda = 0.5$ are shown in figure~\ref{fig:T53PPSP05limit}, where the observed mass limit is 0.75 \TeV{} and the expected limit is 0.81 \TeV{}.  Finally, limits on pair plus single production with $\lambda = 1.0$ are shown in figure~\ref{fig:T53PPSP10limit}, where again the observed mass limit is 0.75 \TeV{} and the expected limit is 0.81 \TeV{}.
             
\begin{figure} 
\begin{center}
   \subfloat[]{\includegraphics[width=.5\columnwidth]{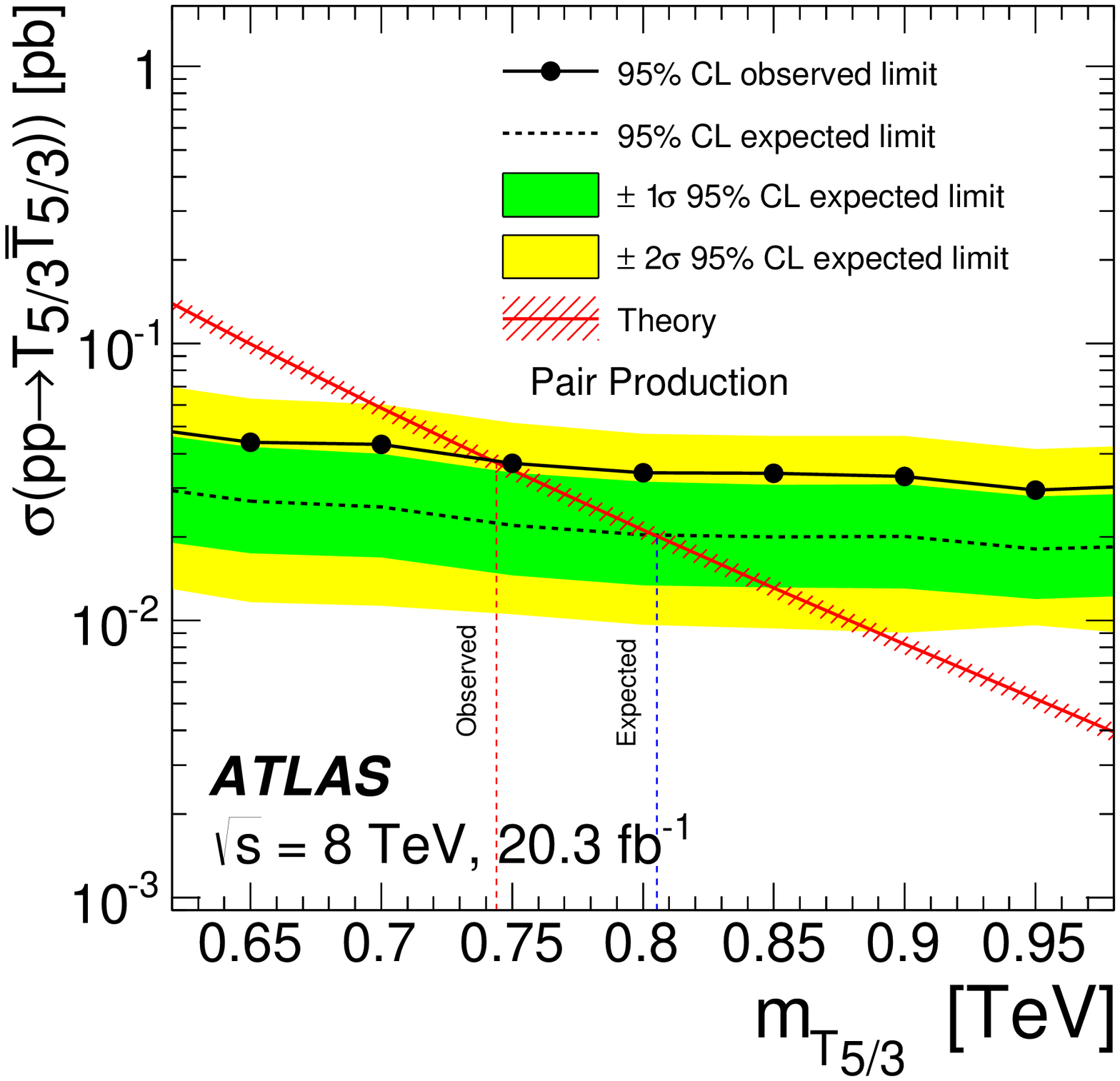}
  \label{fig:T53PPlimit}}
   \subfloat[]{\includegraphics[width=.5\columnwidth]{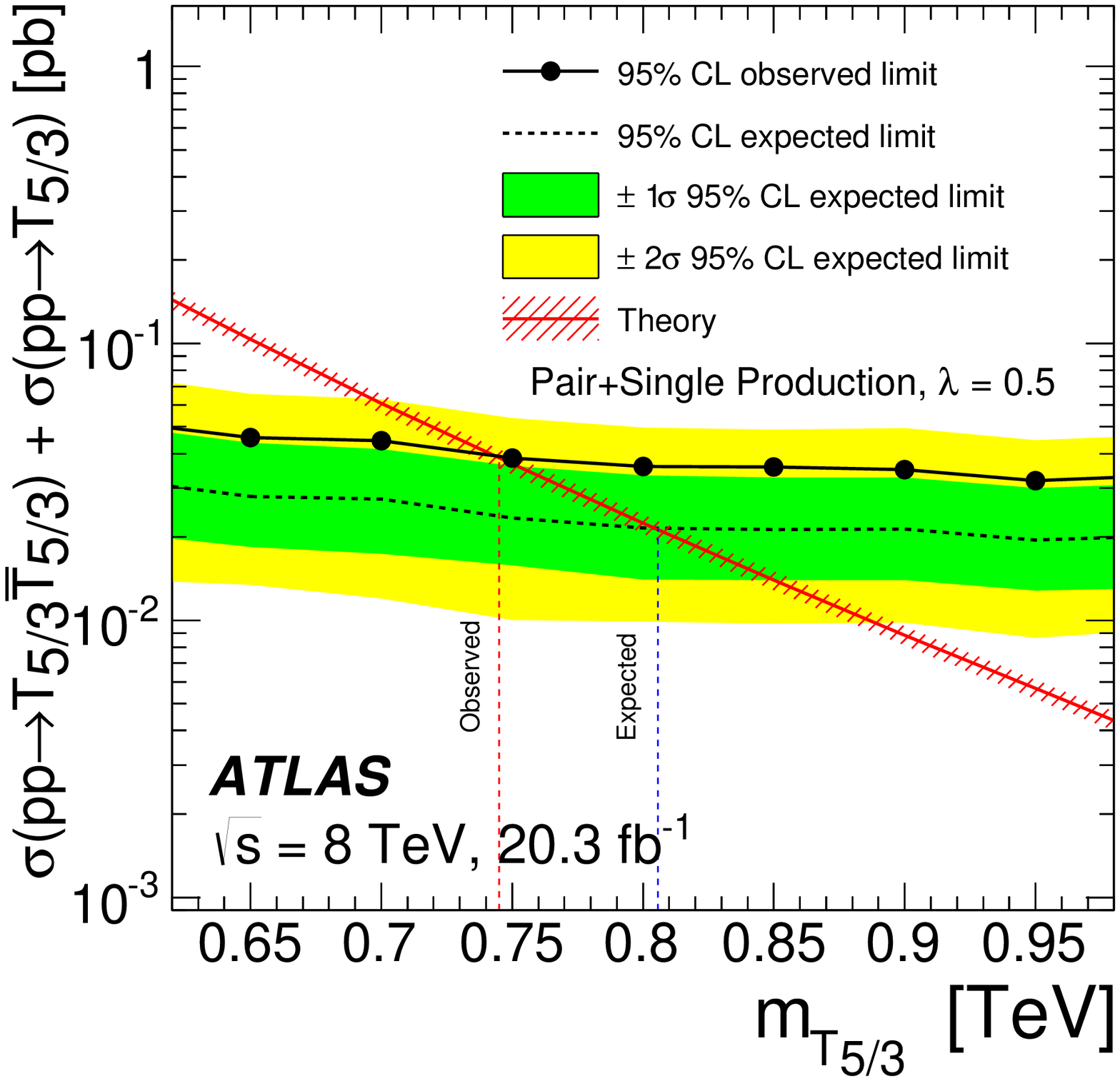} 
  \label{fig:T53PPSP05limit}} \\
  \subfloat[]{\includegraphics[width=.5\columnwidth]{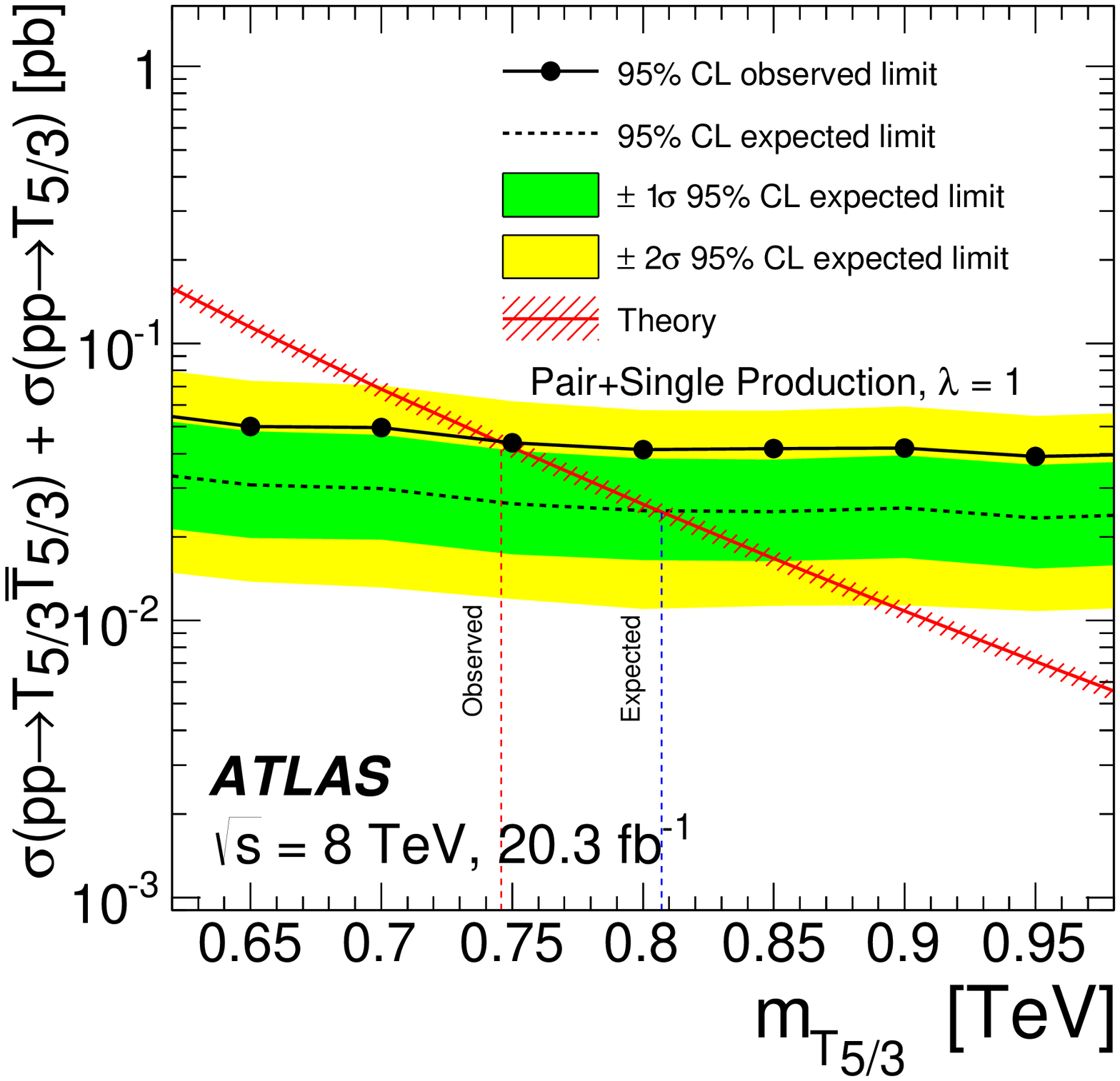}
  \label{fig:T53PPSP10limit}}
  \caption{Expected and observed cross section limits as a function of mass on (a) $T_{5/3}$ pair production, (b) $T_{5/3}$ pair plus single production for coupling $\lambda$ = 0.5, and (c) $T_{5/3}$ pair plus single production  for  $\lambda$ = 1.0.  The vertical dashed lines indicate the expected and observed limits on the $T_{5/3}$ mass, and the shaded band around the theory cross section indicates the total uncertainty on the calculation.}
\end{center}
\end{figure}

\begin{figure} 
  \begin{center}
    \subfloat[]{\includegraphics[width=0.5\linewidth]{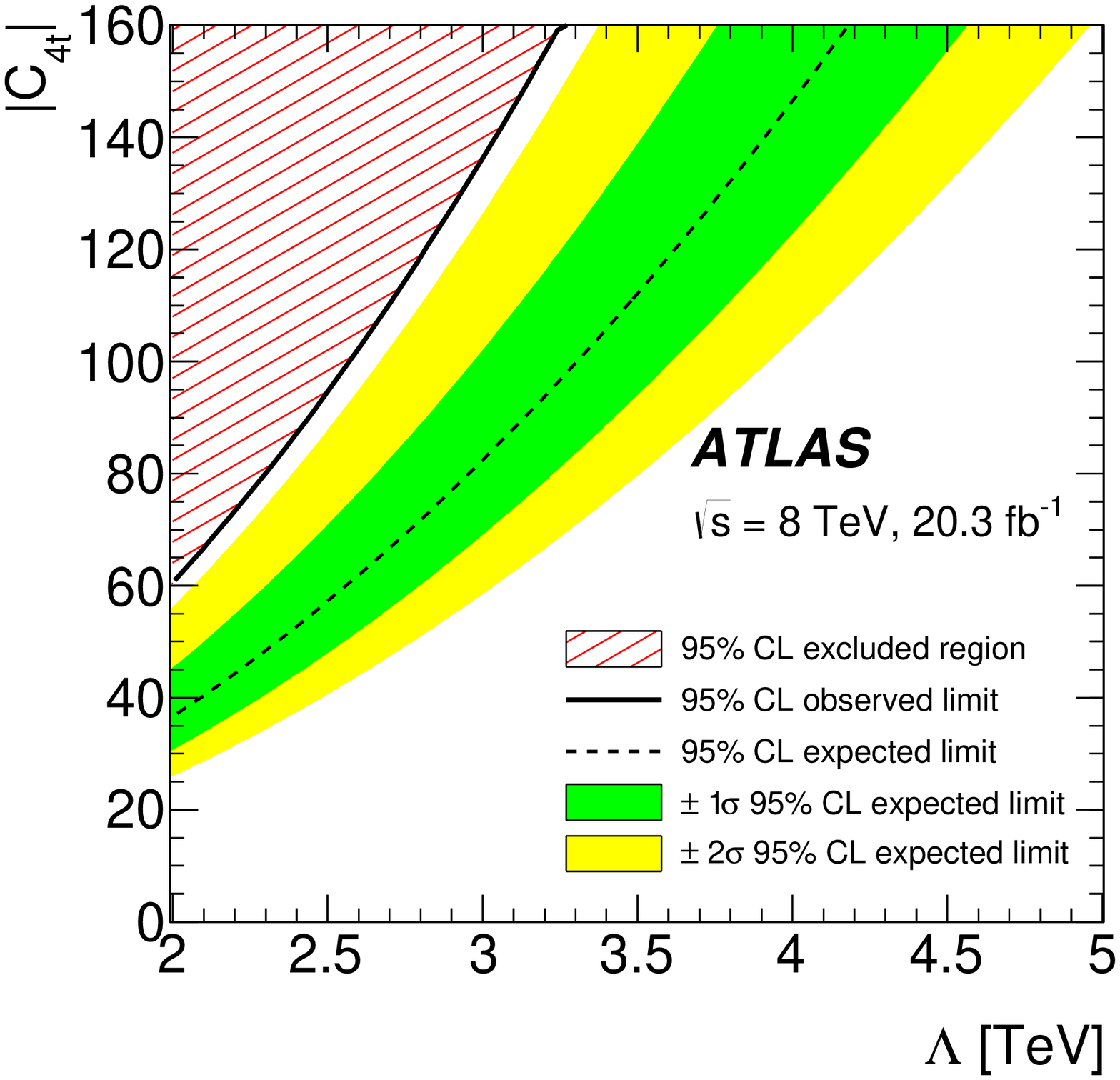}
      \label{fig:limitObs:4tCI}}
	\subfloat[]{\includegraphics[width=0.5\linewidth]{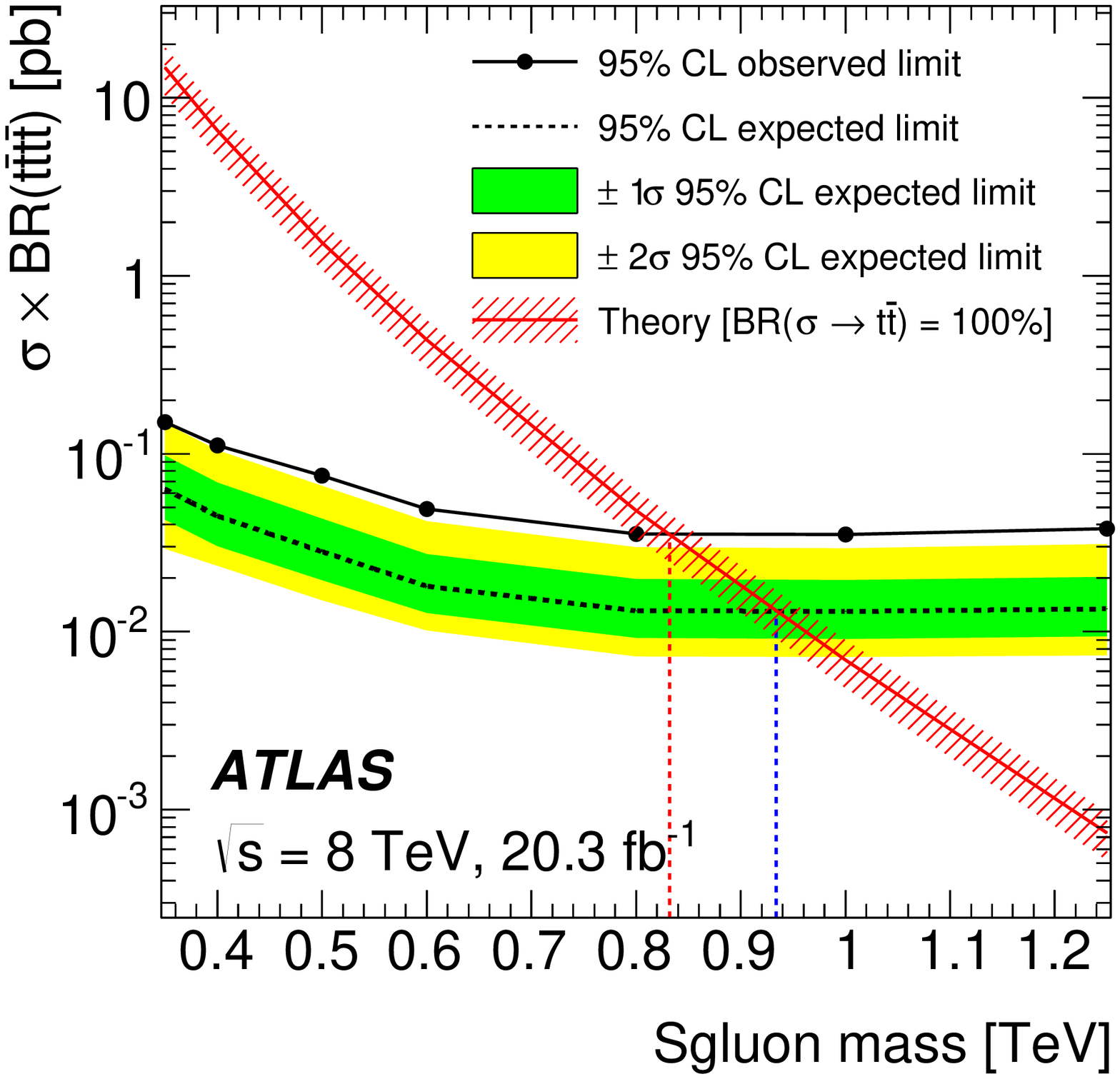}
	\label{fig:obs:limit_sgluon}}\\
	\vspace{-0.15in}
	\subfloat[]{\includegraphics[width=0.5\linewidth]{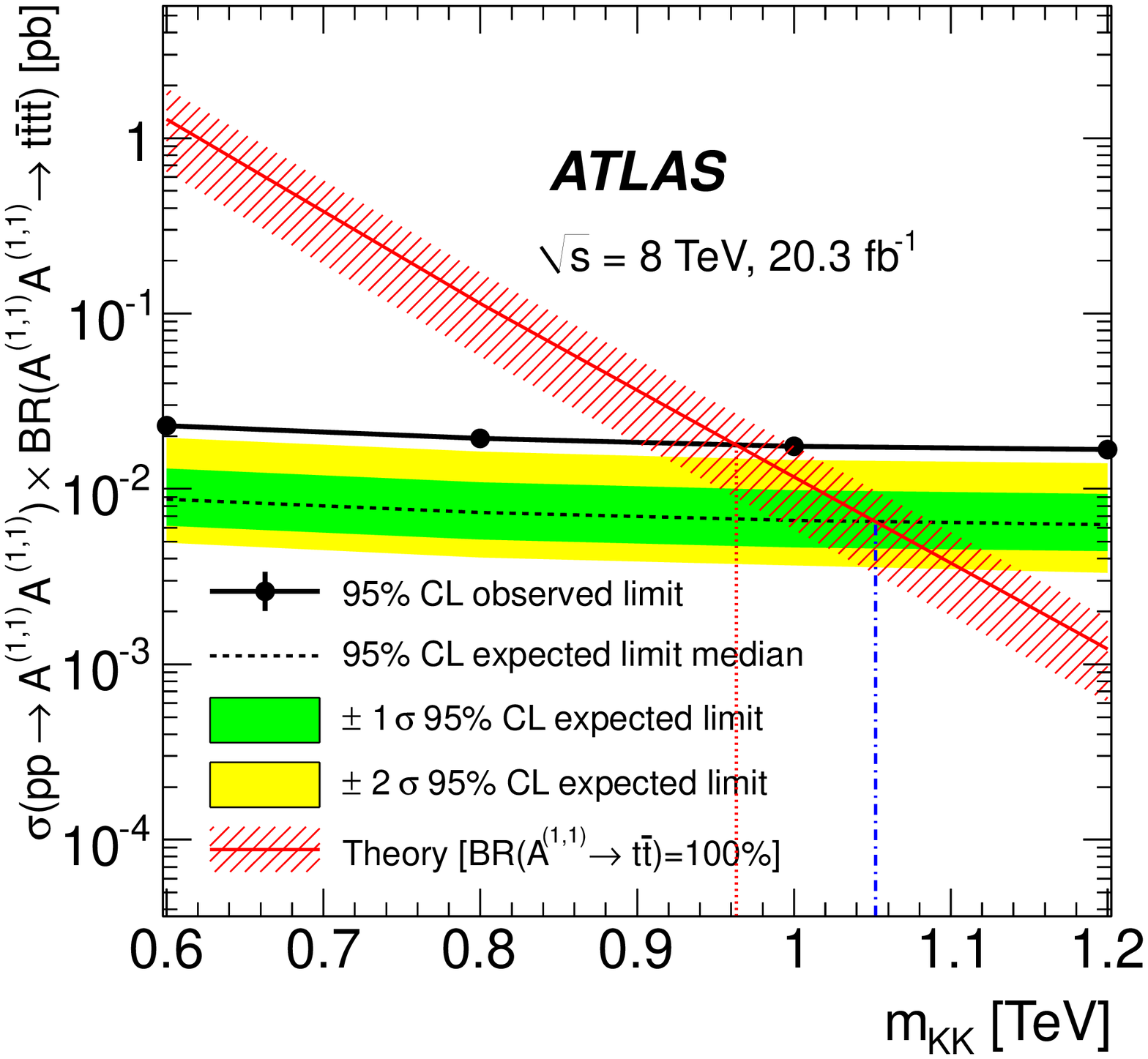}
	\label{fig:rpp:observed_1D11}}
	\subfloat[]{\includegraphics[width=0.5\linewidth]{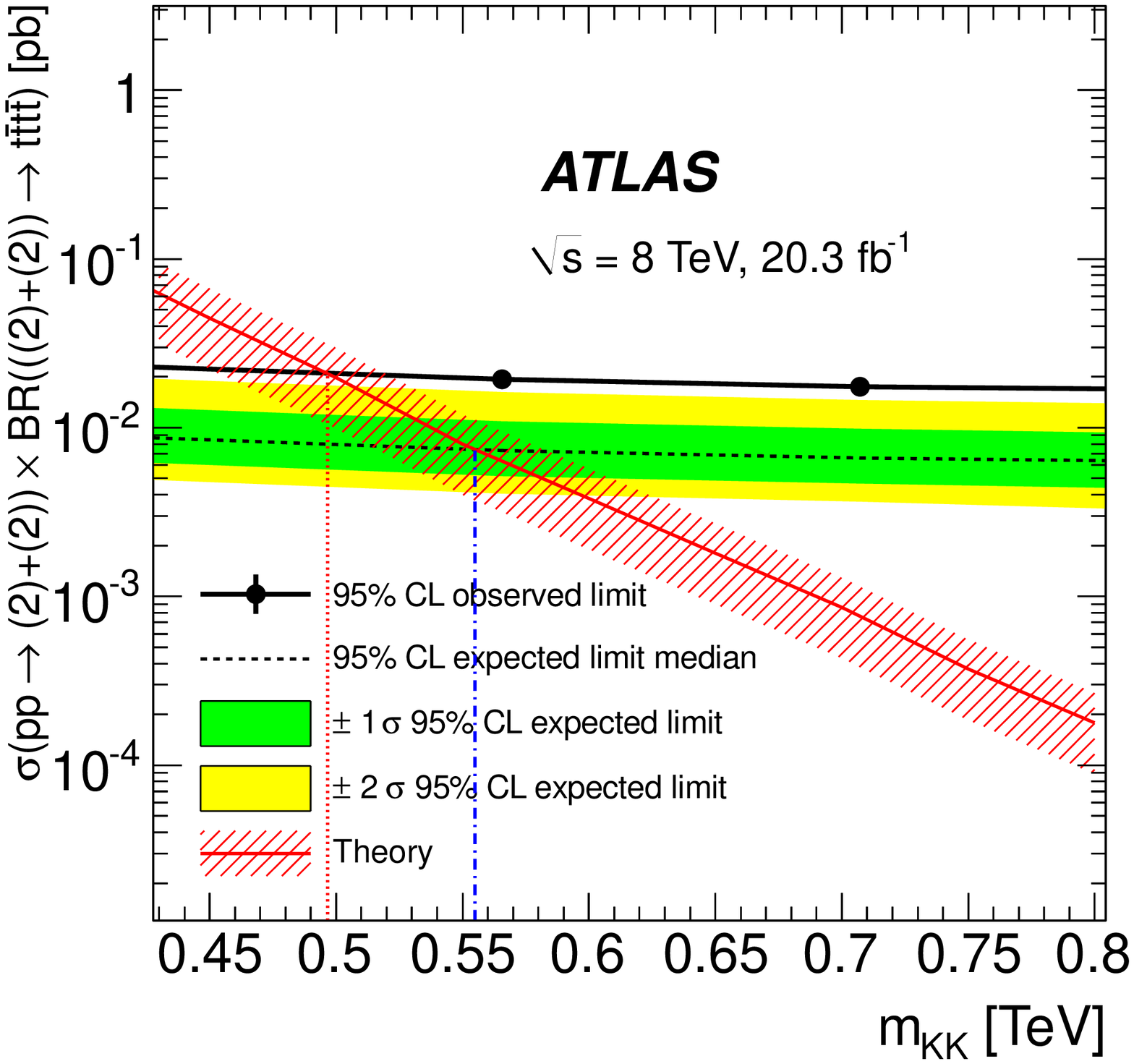}
	\label{fig:rpp:observed_1D20}}
	 \caption{Limits obtained from the search for four-top-quark production.  (a) Expected and observed limits on the coupling constant 
        $|C_{4t}|$ in the contact interaction model for four-top production as a function of the BSM physics energy scale $\Lambda$. The region in the top left corner, corresponding to $|C_{4t}|/\Lambda^2>15.1$~\TeV{}$^{-2}$, is 
        excluded at 95\% CL. (b) Expected and observed limits for the sgluon pair-production cross section times branching ratio to four top quarks  as a function of the sgluon mass. (c) Expected and observed limits on the four-top-quark production rate for the 2UED/RPP model in the symmetric case.  The theory line corresponds to the production of four-top-quark events by tier $(1,1)$ with a branching ratio of $A^{(1,1)}$ to $t\bar{t}t\bar{t}$ of 100\%.  
         (d)  Expected and observed limits  on  the four-top-quark production rate  for the 2UED/RPP model in the symmetric case.  The notation $(2)+(2)$ is a shorthand for $A^{(2,0)}A^{(2,0)}+A^{(0,2)}A^{(0,2)}$. The theory line corresponds to the four-top-quark production by tiers $(2,0)+(0,2)$ alone (BR$(A^{(1,1)}\mapsto t\bar{t}t\bar{t}) = 0)$. The vertical dashed lines in (b), (c), and (d) indicate the expected and observed limits on the sgluon mass or on $m_{\mathrm{KK}}$, and the shaded band around the theory cross section indicates the total uncertainty on the calculation.
}
   \end{center}
   \end{figure}
   
The upper limit on the cross section for four-top-quark production is  70~fb assuming SM kinematics, and 61~fb for production with a BSM-physics contact interaction (expected limits are respectively 27~fb and 22~fb).  The cross-section limit for the contact interaction case is lower than for the SM since the contact interaction tends to result in final-state objects with larger \pt, which increases the selection efficiency.  The limits are also interpreted in the context of specific BSM physics models.  For the contact interaction model, the upper limit on $|C_{4t}|/\Lambda^2$ is 15.1~\TeV{}$^{-2}$, as illustrated in figure~\ref{fig:limitObs:4tCI}.  The lower limit on the sgluon mass is 0.83~\TeV{}, assuming that the sgluons are pair-produced and always decay to \ttbar\ (for an expected limit of 0.94~\TeV{}), as shown in figure~\ref{fig:obs:limit_sgluon}.  The observed limits on the cross section times branching ratio for the 2UED/RPP signal are shown in figures~\ref{fig:rpp:observed_1D11},~\ref{fig:rpp:observed_1D20} and~\ref{fig:rpp:observed_2D}.  These imply the following limits on $m_{\mathrm{KK}}$:
in the symmetric case $(R_4=R_5)$, the observed limit coming from tier $(1,1)$ is $0.96$~\TeV{} (where the expected limit is 1.05~\TeV{}). The observed limit coming from tiers $(2,0)+(0,2)$ alone (BR$(A^{(1,1)}\mapsto t\bar{t}t\bar{t}) = 0)$ is $0.50$~TeV (where the expected limit is 0.55~\TeV{}). In the highly asymmetric case $(R_4>R_5)$, tier $(0,2)$ does not contribute any longer and the observed limit on $m_{\mathrm{KK}}$ from tier $(2,0)$ alone is $0.45$~TeV (where the expected limit is 0.51~\TeV{}).  Figure~\ref{fig:rpp:observed_2D} shows the limits in the $m_{\mathrm{KK}}$--$\xi$ plane, with the constraints from cosmological considerations superimposed.

\begin{figure}  
   \begin{center}
  	\includegraphics[width=\linewidth]{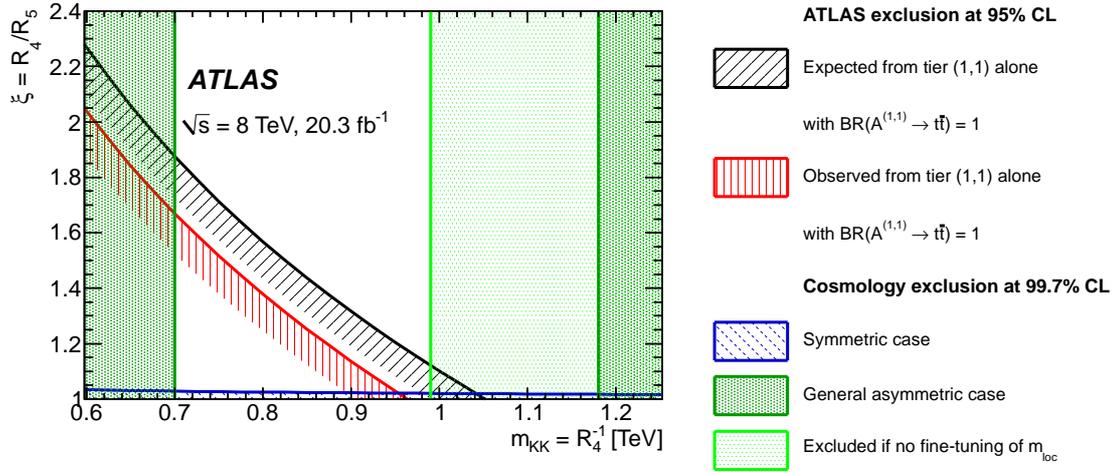}
	\caption{Expected and observed limits in the $(m_{\mathrm{KK}} = 1/R_4, \xi = R_4/R_5)$ plane for the 2UED/RPP model. Cosmological constraints are also shown.  They are due both to direct searches for dark matter and from constraints on its relic density (see ref.~\cite{Lyon12} for more details). The $m_{\mathrm{loc}}$ parameter from ref.~\cite{Lyon12} affects the dark-matter relic density predicted by the 2UED/RPP model.}
	\label{fig:rpp:observed_2D}
   \end{center}
\end{figure}

\section{Checks of the background estimate}
\label{sec:crosschecks}

Several checks were performed to assess the validity of the background estimate. The most important tests are summarized here. For the simulation-based background estimates: variations in the cross section, in the generators, and their settings that span the range consistent with theoretical expectations or direct measurements were applied. Variations in the expected yield are part of the systematic uncertainty.  The data-driven estimates (for the charge misidentification and fake/non-prompt lepton backgrounds) were checked in several ways.  The particular leptons observed in SRVLQ6 and SRVLQ7 were scrutinized, and their quality was  found to be consistent with a sample dominated by real leptons.  Similarly, the multivariate discriminant for $b$-tagging is well above the required threshold for tagged jets found in the sample.  In addition, the expected contribution from charge misidentification and from fake/non-prompt leptons was assessed using samples of simulated events.  It is found that the yields are consistent within uncertainties with the expectations from the data-driven estimates.  To further investigate whether the matrix method accurately predicts the number of fake/non-prompt leptons in $t\bar{t}$ events, the entire procedure was repeated with samples of simulated events, with $r$ and $f$ measured using simulated single-lepton and multijet samples respectively.  The predicted number of fake/non-prompt leptons in simulated $t\bar{t}$ samples is consistent with the actual number present in the MC samples.

\section{Features of events in signal regions with most significant excesses}
\label{sec:eventfeatures}

Information about the lepton charges and flavours, as well as some key kinematic information, for the observed events in signal regions SR4t3/SRVLQ6 and SR4t4/SRVLQ7 are presented in tables~\ref{tab:dump:cat6}--\ref{tab:dump:cat7}.  One unexpected feature is the dominance of one electric charge over another: in SR4t3/SRVLQ6 there are 10 negatively charged and 16 positively charged leptons, and in SR4t4/SRVLQ7 there are only 2 negatively charged leptons and 11 positively charged leptons.  Similar effects are observed in two other signal regions, with SRVLQ3 being dominated by negatively charged leptons and SR4t2/SRVLQ5 being dominated by positively charged leptons.  These charge asymmetries are interpreted as statistical fluctuations. This interpretation is bolstered by the fact that $i$) there is no known mechanism for the selection to favour one electric charge over the other; $ii$) the asymmetry is present only in some of the signal regions (and both negative and positive charges dominate in the various regions); and $iii$) that the asymmetries do not persist in the loose lepton samples selected with the same kinematic criteria as are applied in the signal regions.

\begin{table}
  \begin{center}
    \begin{tabular}{cccc}
      \hline\hline
      Type & $N_{j}$ & \HT{} [\GeV] & \met{} [\GeV] \\
      \hline
       $e^-e^-$ & 3 & 807 & 171 \\
       $e^+e^+$ & 5 & 862 & 268 \\
       $e^+e^+$ & 5 & 868 & 113 \\
      \hline
       $\mu^-e^-$ & 6 & 1346 & 353 \\
       $e^+\mu^+$ & 5 & 810 & 106 \\
       $e^-\mu^-$ & 3 & 707 & 184 \\
       $e^-\mu^-$ & 2 & 706 & 174 \\
     $\mu^+e^+$ & 8 & 882 & 150 \\
      $\mu^+e^+$ & 4 & 860 & 112 \\
      \hline
       $\mu^+\mu^+$ & 5 & 888 & 111 \\
      \hline
      $\mu^-e^+e^+$ & 5 & 773 & 197 \\
      $\mu^-e^+e^+$ & 9 & 968 & 355 \\
      \hline
    \end{tabular}
  \end{center}
    \caption{List of data events in the SR4t3/SRVLQ6 category (by definition, all events have exactly two
      $b$-jets, the \met{} is above 100~\GeV{} and the \HT{} above 700~\GeV).}\label{tab:dump:cat6}
\end{table}

\begin{table}[tb]
  \begin{center}
    \begin{tabular}{ccccc}
      \hline\hline
      Type & $N_{j}$ & $N_{b}$ & \HT{} [\GeV] & \met{} [\GeV] \\
      \hline
      $e^+e^+$ & 4 & 3 & 709 & 298 \\
       $e^+e^+$ & 6 & 3 & 800 & 137 \\
      \hline
   $e^+\mu^+$ & 5 & 3 & 744 & 216 \\
 $e^+\mu^+$ & 4 & 3 & 888 & 155 \\
     $\mu^+e^+$ & 3 & 3 & 1439 & 239 \\
      \hline
     $\mu^-\mu^+\mu^-$ & 4 & 4 & 1072 & 176 \\
      \hline
    \end{tabular}
  \end{center}
    \caption{List of data events in the SR4t4/SRVLQ7 category (by definition, the \met{} is above 40~\GeV{} and the \HT{} above 700~\GeV).}\label{tab:dump:cat7}
\end{table}

\FloatBarrier

\section{Conclusion}

A search for BSM physics has been performed using $pp$ collisions at $\sqrt{s} =8$~\TeV{} corresponding to an integrated luminosity of 20.3~\ifb{} recorded by the ATLAS detector at the LHC, where events with at least two leptons, including a pair of the same electric charge, at least one $b$-tagged jet,  sizeable missing transverse momentum, and large \HT were considered.  Several BSM physics effects could enhance the yield of such events over the small SM expectation.  The search was performed in the context of several BSM physics models, with signal regions defined for different models.  The regions of parameter space excluded by the data are quantified by setting 95\% CL limits.  The observed yield in the signal region for positively charged top quark pair production is consistent with the expected background, resulting in limits of 8.4--62~fb on the cross section of this process (depending on the model considered) and a limit BR($t\rightarrow uH)<1\%$.  In the set of signal regions defined for vector-like quark, four-top-quark,  and chiral $b^\prime$-quark searches there is an excess of observed events over the SM prediction, particularly in the subset of those signal regions that require at least two $b$-tagged jets (and thus are relevant to the search for four-top-quark production). The significance of the excess varies with the signal being considered, reaching 2.5 standard deviations for hypotheses involving heavy resonances decaying to four top quarks.   Nonetheless the data can still constrain some of the BSM physics models considered; 95\% CL limits are set as follows:  the mass of the chiral $b^{\prime}$-quark is constrained as a function of the branching ratio to $Wt$, the masses of vector-like $B$ and $T$ quarks are constrained to $m_B > 0.62$ \TeV{}, $m_T > 0.59$~\TeV{} (assuming  branching fractions to the $W$, $Z$, and $H$ decay modes arising from a singlet model), the mass of the $T_{5/3}$ quark is  greater than 0.75~\TeV{}, the SM four-top production cross section is  less than 70~fb, the sgluon mass is greater than 0.83~\TeV{} and the Kaluza--Klein mass (in the context of models with two universal extra dimensions) is greater than 0.96~\TeV{}.  

\section*{Acknowledgements}


We thank CERN for the very successful operation of the LHC, as well as the
support staff from our institutions without whom ATLAS could not be
operated efficiently.

We acknowledge the support of ANPCyT, Argentina; YerPhI, Armenia; ARC,
Australia; BMWFW and FWF, Austria; ANAS, Azerbaijan; SSTC, Belarus; CNPq and FAPESP,
Brazil; NSERC, NRC and CFI, Canada; CERN; CONICYT, Chile; CAS, MOST and NSFC,
China; COLCIENCIAS, Colombia; MSMT CR, MPO CR and VSC CR, Czech Republic;
DNRF, DNSRC and Lundbeck Foundation, Denmark; EPLANET, ERC and NSRF, European Union;
IN2P3-CNRS, CEA-DSM/IRFU, France; GNSF, Georgia; BMBF, DFG, HGF, MPG and AvH
Foundation, Germany; GSRT and NSRF, Greece; RGC, Hong Kong SAR, China; ISF, MINERVA, GIF, I-CORE and Benoziyo Center, Israel; INFN, Italy; MEXT and JSPS, Japan; CNRST, Morocco; FOM and NWO, Netherlands; BRF and RCN, Norway; MNiSW and NCN, Poland; GRICES and FCT, Portugal; MNE/IFA, Romania; MES of Russia and NRC KI, Russian Federation; JINR; MSTD,
Serbia; MSSR, Slovakia; ARRS and MIZ\v{S}, Slovenia; DST/NRF, South Africa;
MINECO, Spain; SRC and Wallenberg Foundation, Sweden; SER, SNSF and Cantons of
Bern and Geneva, Switzerland; NSC, Taiwan; TAEK, Turkey; STFC, the Royal
Society and Leverhulme Trust, United Kingdom; DOE and NSF, United States of
America.

The crucial computing support from all WLCG partners is acknowledged
gratefully, in particular from CERN and the ATLAS Tier-1 facilities at
TRIUMF (Canada), NDGF (Denmark, Norway, Sweden), CC-IN2P3 (France),
KIT/GridKA (Germany), INFN-CNAF (Italy), NL-T1 (Netherlands), PIC (Spain),
ASGC (Taiwan), RAL (UK) and BNL (USA) and in the Tier-2 facilities
worldwide.

\bibliographystyle{elsarticle-num}
\bibliography{paper}

\newpage 
\begin{flushleft}
{\Large The ATLAS Collaboration}

\bigskip

G.~Aad$^{\rm 85}$,
B.~Abbott$^{\rm 113}$,
J.~Abdallah$^{\rm 152}$,
O.~Abdinov$^{\rm 11}$,
R.~Aben$^{\rm 107}$,
M.~Abolins$^{\rm 90}$,
O.S.~AbouZeid$^{\rm 159}$,
H.~Abramowicz$^{\rm 154}$,
H.~Abreu$^{\rm 153}$,
R.~Abreu$^{\rm 30}$
% ...
\bigskip
\\
% ...
$^{11}$ Institute of Physics, Azerbaijan Academy of Sciences, Baku, Azerbaijan\\
% ...
$^{30}$ CERN, Geneva, Switzerland\\
% ...
$^{85}$ CPPM, Aix-Marseille Universit{\'e} and CNRS/IN2P3, Marseille, France\\
% ...
$^{90}$ Department of Physics and Astronomy, Michigan State University, East Lansing MI, United States of America\\
% ...
$^{107}$ Nikhef National Institute for Subatomic Physics and University of Amsterdam, Amsterdam, Netherlands\\
% ...
$^{113}$ Homer L. Dodge Department of Physics and Astronomy, University of Oklahoma, Norman OK, United States of America\\
% ...
$^{152}$ Institute of Physics, Academia Sinica, Taipei, Taiwan\\
$^{153}$ Department of Physics, Technion: Israel Institute of Technology, Haifa, Israel\\
$^{154}$ Raymond and Beverly Sackler School of Physics and Astronomy, Tel Aviv University, Tel Aviv, Israel\\
% ...
$^{159}$ Department of Physics, University of Toronto, Toronto ON, Canada\\
% ...
\end{flushleft}

\end{document}